\pdfoutput=1
\documentclass[
    aps,pre,twocolumn,
    reprint,
    superscriptaddress,
    nofootinbib,
    floatfix,
    amssymb,
    longbibliography
]{revtex4-2}

\usepackage[english]{babel}

\usepackage{amsmath}
\usepackage{amsfonts}
\usepackage{amssymb}
\usepackage{graphicx}

\usepackage[utf8]{inputenc}

\usepackage{nicefrac}

\usepackage[normalem]{ulem}

\usepackage{xcolor}

\definecolor{linkColor}{rgb}{0,0.3,0.7}
\usepackage[colorlinks=true,
            allcolors=linkColor,
            pdfborder={0 0 0},
            pdfencoding = auto
            ]{hyperref}
            
\usepackage{url}

\definecolor{myGreen}{rgb}{0.1,0.5,0.1}

\def\HSS{\mathrm{HSS}}

\newcommand{\rhostat}{\rho_\mathrm{stat}}
\newcommand{\etastat}{{\eta_\mathrm{stat}}}
\newcommand{\rhostateps}{\rho_\mathrm{stat}^\varepsilon}
\newcommand{\etastateps}{\eta_\mathrm{stat}^\varepsilon}
\newcommand{\order}[1]{\mathcal{O}\!\left(#1\right)}

\begin{document}
\title{
Coarsening and wavelength selection far from equilibrium:
a unifying framework based on singular perturbation theory
}

\author{Henrik Weyer}
\author{Fridtjof Brauns} 
\email{fbrauns@gmail.com}
\thanks{Present address: Kavli Institute for Theoretical Physics, University of California Santa Barbara, Santa Barbara, CA 93106, USA}
\affiliation{Arnold Sommerfeld Center for Theoretical Physics and Center for NanoScience, Department of Physics, Ludwig-Maximilians-Universit\"at M\"unchen, Theresienstra\ss e 37, D-80333 M\"unchen, Germany}
\author{Erwin Frey}
\email{frey@lmu.de}
\affiliation{Arnold Sommerfeld Center for Theoretical Physics and Center for NanoScience, Department of Physics, Ludwig-Maximilians-Universit\"at M\"unchen, Theresienstra\ss e 37, D-80333 M\"unchen, Germany}
\affiliation{Max Planck School Matter to Life, Hofgartenstraße 8, D-80539 Munich, Germany}

\date{\today}

\begin{abstract}
    Intracellular protein patterns are described by (nearly) mass-conserving reaction--diffusion systems.
    While these patterns initially form out of a homogeneous steady state due to the well-understood Turing instability, no general theory exists for the dynamics of fully nonlinear patterns.
    We develop a unifying theory for nonlinear wavelength-selection dynamics in (nearly) mass-conserving two-component reaction--diffusion systems independent of the specific mathematical model chosen.
    Previous work has shown that these systems support an extremely broad band of stable wavelengths, but the mechanism by which a specific wavelength is selected has remained unclear. We show that an interrupted coarsening process selects the wavelength at the threshold to stability. Based on the physical intuition that coarsening is driven by competition for mass and interrupted by weak source terms that break strict mass conservation, we develop a singular perturbation theory for the stability of stationary patterns.
    The resulting closed-form analytical expressions enable us to quantitatively predict the coarsening dynamics and the final pattern wavelength.
    We find excellent agreement with numerical results throughout the diffusion- and reaction-limited regimes of the dynamics, including the crossover region.
    Further, we show how, in these limits, the two-component reaction--diffusion systems map to generalized Cahn--Hilliard and conserved Allen--Cahn dynamics, therefore providing a link to these two fundamental scalar field theories.
    The systematic understanding of the length-scale dynamics of fully nonlinear patterns in two-component systems provided here builds the basis to reveal the mechanisms underlying wavelength selection in multi-component systems with potentially several conservation laws.
\end{abstract}
\maketitle

\section{Introduction}
\label{sec:Introduction}
Across many non-equilibrium systems, small constituents self-organize into macroscopic patterns on much larger length scales.
One of the most intriguing aspects of such patterns far from equilibrium is the emergence of intrinsic length scales independent of the system size or other spatial cues but entirely determined by the local interaction of the constituents.
Examples include chemical systems far from equilibrium \cite{Turing1952, Maini.etal1997, Epstein.Pojman1998, Cross.Greenside2009}, especially intracellular protein patterns \cite{Halatek.Frey2012,Murray.Sourjik2017,Halatek.etal2018,Chiou.etal2017}, collective states in active matter \cite{Caussin.etal2014,Solon.etal2015,Chate2020,Tjhung.etal2018,Schaller.etal2010,Huber.etal2021,Peruani.etal2012,Shi.etal2020}, and phase separation of chemically active species \cite{Glotzer.etal1995,Weber.etal2019,Zwicker2022} or species undergoing population dynamics \cite{Cates.etal2010,Li.Cates2020,Hillen.Painter2009}.
In contrast, phase-separation processes approaching thermodynamic equilibrium usually develop toward full separation via a continuous growth of the average domain size, a process termed ``coarsening''.
The coarsening process can be interrupted, and a wavelength selected in close-to-equilibrium processes if the system allows for long-range interactions~\cite{Goldenfeld.etal1989,Desai.Kapral2009,Politi.Torcini2015}.
Recently, the mechanisms underlying length-scale selection were explained for non-equilibrium extensions of classical phase separation, advancing the understanding of active phase separation important for intracellular condensates and active matter systems \cite{Tjhung.etal2018,Weber.etal2019,Li.Cates2020,Zwicker2022}.

Reaction--diffusion systems are governed by (chemical) reactions that induce transitions between particle states with different diffusivities, instead of physical interactions that induce the demixing of phases.
Originally, Alan Turing proposed such systems to explain the patterning during morphogenesis, i.e., the development of organisms~\cite{Turing1952}.
By now, reaction--diffusion systems are found to describe diverse biological processes, for example, intracellular protein pattern formation \cite{Lutkenhaus2007,Goehring2014,Chiou.etal2017,Halatek.etal2018,Burkart.etal2022a} or signaling via trigger waves \cite{Gelens.etal2014}.
In the intracellular context, the timescale of protein production and degradation is long compared to the timescale of pattern formation, and the total number of proteins can be assumed to be (approximately) constant.
The proteins just switch between different states, for example, a membrane-bound and a cytosolic state.
These dynamics are captured by mass-conserving reaction--diffusion (McRD) systems which describe the concentrations of the different protein states~\cite{Otsuji.etal2007, Mori.etal2008, Altschuler.etal2008, Goryachev.Pokhilko2008, Jilkine.Edelstein-Keshet2011, Halatek.Frey2012, Trong.etal2014, Chiou.etal2018, Halatek.etal2018,Halatek.Frey2018,Brauns.etal2020};
for an introduction to the theory of McRD systems, see the lecture notes Ref.~\cite{Frey.Brauns2022}.
Importantly, McRD systems do not conserve each individual component but the total number density of each protein summing over the concentrations of its conformational states.
McRD systems also describe, for example, granular media \cite{Aranson.Tsimring2008}, precipitation patterns \cite{Scheel2009}, and braided polymers \cite{Forte.etal2019}.

Two-component reaction--diffusion (2cRD) systems \emph{lacking} mass conservation generally exhibit patterns with an intrinsic length scale \cite{Cross.Hohenberg1993, Maini.etal1997, Epstein.Pojman1998, Cross.Greenside2009, Wei.Winter2014}.
In contrast, the fully nonlinear patterns in two-component \emph{mass-conserving} reaction--diffusion (2cMcRD) systems have been observed to undergo coarsening until completion, that is, the pattern length scale grows until the system size is reached \cite{Otsuji.etal2007, Ishihara.etal2007, Chiou.etal2018, Brauns.etal2021}.
This distinct behavior of two-component reaction--diffusion systems without and with mass conservation calls for a detailed analysis of the length-scale dynamics in nearly mass-conserving 2cRD systems.
We started to investigate this question in Ref.~\cite{Brauns.etal2021} by proposing that wavelength selection in 2cRD systems can be understood as an interruption of the coarsening process due to (weak) source terms that break strict mass conservation.
Together with the splitting of pattern domains at larger wavelengths, this gives rise to a broad range of stable wavelengths.
Here, we systematically analyze under what conditions this reasoning applies in general 2cMcRD systems and how the final pattern wavelength is selected dynamically.

The coarsening process observed in mass-conserving 2cRD systems is reminiscent of the typical dynamics found in phase-ordering and phase-separation kinetics.
In these systems, which are close to thermal equilibrium, the dynamics is governed by gradient flows in a free energy landscape toward the minimum of the respective free energy functional.
The free-energy cost due to interfaces between the different phases leads to dynamics that continuously minimizes the surface area of the interfaces (Model A/B dynamics \cite{Hohenberg.Halperin1977}) \cite{Pismen2006}.
Thus, small domains with a high surface-to-bulk ratio collapse in favor of larger domains,
and the characteristic pattern length scale grows uninterruptedly until the fully phase-ordered or phase-separated state is reached \cite{Lifshitz.Slyozov1961, Wagner1961, Langer1971, Bray2002, Rubinstein.Sternberg1992}.
Intriguingly, the uninterrupted coarsening in (inherently far-from-equilibrium) 2cMcRD systems exhibits regimes where the coarsening resembles either Cahn--Hilliard dynamics (bulk-diffusion-controlled phase separation \cite{Cahn.Hilliard1958}) \cite{Brauns.etal2021, Tateno.Ishihara2021} or conserved Allen--Cahn dynamics (interface-controlled kinetics \cite{Rubinstein.Sternberg1992, Wagner1961,Conti.etal2002}) \cite{Nishiura1982, Ni.etal2001, McKay.Kolokolnikov2012}, two classical models of phase separation.
Moreover, stationary states of 2cMcRD systems can be analyzed similarly to a Maxwell construction \cite{Brauns.etal2020}, and close to the supercritical onset of pattern formation 2cMcRD systems reduce to an amplitude equation which agrees with the Cahn--Hilliard equation \cite{Bergmann.etal2018}.
While some 2cMcRD systems with a specific mathematical form of the reaction term allow for an abstract mapping onto a gradient flow for an effective free-energy functional \cite{Morita.Ogawa2010, Jimbo.Morita2013}, a general and unifying understanding of the similarities between reaction--diffusion and phase-separation dynamics is lacking.

Similarly, the changed phenomenology giving rise to wavelength selection in the presence of source terms is reminiscent of interrupted coarsening observed if non-equilibrium extensions are included in classical phase-separation dynamics.
In binary phase separation, for example, the coarsening process can be arrested if chemical reactions are introduced that convert particles from one of the phase-separating species into the other \cite{Glotzer.etal1995}.
This mechanism has been of increasing interest in recent years to describe intracellular condensates~\cite{Zwicker.etal2015, Weber.etal2019}.
Moreover, in the fields of active matter and nonreciprocal systems, wavelength selection in non-equilibrium settings is of increasing interest and interrupted coarsening has been frequently studied as well~\cite{Tjhung.etal2018, Saha.etal2020, Grafke.etal2017, Li.Cates2020, Frohoff-Hulsmann.etal2021}.
It would therefore be telling to find out whether common principles underlie coarsening and wavelength selection in these systems taking the form of active phase separation and 2cRD models.

Several approaches have been developed to analyze the length-scale dynamics in systems of (active) phase separation and reaction--diffusion systems.
Close to a \emph{supercritical} onset of pattern formation, the \emph{amplitude equation} formalism provides a powerful tool to study the pattern properties and dynamics, including the question of wavelength selection \cite{Cross.Hohenberg1993, Cross.Greenside2009, Desai.Kapral2009}.
As this approach critically depends on a small pattern amplitude, it is not applicable to fully developed patterns of large amplitude.
If the patterns only evolve on length scales much larger than the typical pattern wavelength, an approach based on \emph{phase equations}, which are conceptually closely related to amplitude equations, can be applied to obtain the long-time dynamics in the highly nonlinear regime~\cite{Cross.Newell1984,Politi.Misbah2004,Politi.Misbah2006}.
In both the amplitude and phase equation approach, conservation laws play a critical role and must be accounted for explicitly  ~\cite{Cross.Hohenberg1993,Matthews.Cox2000}.

(Nearly) mass-conserving two-component reaction--diffusion systems as well as thermodynamic systems exhibiting phase separation generally show a \emph{subcritical} onset of pattern formation such that the regime of (spontaneous) lateral instability (termed as \emph{spinodal} regime in the language of phase separation) is surrounded by a multistable regime which shows stimulus-induced pattern formation (\emph{nucleation-and-growth} regime) \cite{Brauns.etal2020,Desai.Kapral2009}.
In these reaction--diffusion systems, the coarsening process proceeds by mass redistribution, which is fastest on the shortest distances, that is, between neighboring pattern domains \cite{Brauns.etal2021}.
The dominant dynamic process thus acts on the length scale of the patterns.
As a result, amplitude and phase equations cannot capture the dynamics of the highly nonlinear patterns in nearly mass-conserving reaction--diffusion systems.
The same holds for systems describing (close-to-equilibrium) phase separation if the dynamics is not analyzed close to the critical point.
Therefore, different methods have been developed and have been first used to describe the long-time dynamics of close-to-equilibrium phase separation.
In the mathematical literature, \emph{slow-manifold theory} is applied to phase-separating systems~\cite{Kawasaki.Ohta1982, Kawasaki1984, Fusco.Hale1989, Alikakos.etal1991, Argentina.etal2005}.
Moreover, Lifshitz, Slyozov, and Wagner \cite{Lifshitz.Slyozov1961,Wagner1961} 
developed their classical theory of Ostwald ripening (LSW theory), building on the physical properties of single (quasi-)stationary droplets [quasi-steady-state (QSS) approximation].
The resulting theory describes the particle exchange between droplets of different sizes of an immiscible minority phase sparsely distributed in the majority phase.
LSW theory was extended to analyze wavelength selection in systems introducing chemical reactions between the species undergoing phase separation~\cite{Zwicker.etal2015, Weber.etal2019, Li.Cates2020}.
In contrast, this physical reasoning has not been employed to describe reaction--diffusion dynamics.
Instead, in 2cRD systems \emph{singular perturbation theory} \cite{Nishiura1982} was adopted to analyze wavelength selection
\cite{Kerner.Osipov1994,Kolokolnikov.etal2006, McKay.Kolokolnikov2012, Ni.etal2001, Wei.Winter2014}.
Also, numerical bifurcation analysis was applied to discuss the wavelength of stable stationary patterns when tuning parameters \cite{Fujii.etal1982, Frohoff-Hulsmann.etal2021}.
In Ref.~\cite{Brauns.etal2021}, we have employed a QSS approximation inspired by LSW theory in (nearly) mass-conserving reaction--diffusion systems, gaining an in-depth understanding of the coarsening dynamics and the stabilization of finite pattern wavelengths in these systems.
However, a systematic justification for the QSS approximation in reaction--diffusion systems is still missing. Providing such a justification in terms of singular perturbation theory supported by physical arguments is the central goal of the present work.

Our theoretical analysis in Ref.~\cite{Brauns.etal2021} has shown that the length-scale dynamics in reaction--diffusion systems can be understood---analogous to active phase separation---as a coarsening process which is arrested by counteracting processes above a particular length scale.
Coarsening is explained on the basis of a mass-competition instability between neighboring pattern domains (the droplets in LSW theory) and wavelength selection by suppression of this instability.
Here, we complement the physical reasoning by a singular perturbation analysis for (weakly) mass-conserving 2cRD systems, which provides explicit expressions for the growth rates of the different processes involved in the mass-competition instability.
Previous works have focused on specific mathematical forms of the reaction term \cite{Kolokolnikov.etal2006, Wei.Winter2014, Kolokolnikov.Wei2018, Gai.etal2020} and the reaction-limited regime \cite{McKay.Kolokolnikov2012,Ni.etal2001}.
Our results are independent of the specific mathematical form of the reaction terms as they build on geometric reasoning in phase space \cite{Brauns.etal2020}.
Moreover, our results apply in both diffusion- and reaction-limited regimes, including the crossover between them.
These mathematical results are explained by a detailed analysis of how mass is exchanged between pattern domains, revealing a systematic link of the reaction--diffusion dynamics to diffusion-limited (bulk-diffusion-controlled) and reaction-limited (interface-controlled) phase-separation kinetics.
We then derive the wavelength dynamics of patterns in large systems from the rate expressions for the competition between neighboring peaks.
This dynamics reveals that the \emph{threshold} of interrupted coarsening selects the final pattern wavelength.
Taken together, our results underline that wavelength selection in both the reaction--diffusion and phase-separating systems is driven by the mass exchange between domains and the mass exchange with a reservoir described by the source terms.
Consequently, we expect that the analysis provides the basis to analyze wavelength-selection dynamics in multi-component reaction--diffusion and active matter systems governed by one, or possibly multiple, (approximate) conservation laws.

The remainder of this paper is organized as follows.
We first describe in Section~\ref{sec:phenomenology} the phenomenology of the pattern dynamics we set out to explain, that is, the phenomenology of coarsening and its arrest.
In Section~\ref{sec:models}, we then focus on the approximate conservation law governing 2cRD system with weak source terms. 
We also introduce two standard models exhibiting phase separation---(generalized) \textit{Cahn--Hilliard} (CH) and \textit{conserved Allen--Cahn} (cAC) models---which account for mass conservation by either a local or a global constraint. 
These classical systems also serve to compare the reaction--diffusion with (active) phase-separation dynamics.
Section~\ref{sec:HSS-instability} discusses the stability properties of the homogeneous steady state which highlights the connections found between these three models in their initial pattern-forming instabilities.
Afterward, we construct the stationary patterns of all three systems in a unifying picture (Sec.~\ref{sec:stat-patterns}).
On the basis of this classification of stationary patterns, we then focus in  section~\ref{sec:MC-mass-competition} on strictly mass-conserving systems.
The mass-competition instability is described, which underlies the self-amplifying mass transport between neighboring pattern domains, and which causes the growth of larger droplets at the expense of smaller droplets.
This instability is the elementary motive underlying the uninterrupted coarsening process observed in the mass-conserving systems, and we use the derived growth rate of the instability in section~\ref{sec:coarsening-scaling} to obtain a scaling law for the time evolution of the coarsening process.
This completes the description of uninterrupted coarsening in mass-conserving systems, and we turn in
section~\ref{sec:nMC-mass-competition} to discuss the influence of weak source terms that break the strict conservation law. 
From the growth rate of the mass-competition instability under the influence of weak source terms, we explain in this section the central criterion for the wavelength selected by interrupted coarsening.
In particular, we discuss how the suppression of the mass-competition instability by weak source terms above a threshold pattern wavelength determines the wavelength-selection dynamics.
After explaining the underlying processes, the comparison with several examples analyzed numerically verifies the found relations.
The discussion of our findings and future applications are found in Sec.~\ref{sec:discussion}.

Throughout this work, the main text focuses on the discussion of the results from the singular perturbation analysis, i.e., their implications and physical interpretations while the formal mathematical derivations are deferred to the appendices.
Moreover, we restrict the analysis to one-dimensional systems to avoid mathematical complications resulting from the system's geometry that would hamper the analysis of the underlying mechanisms.
Additional effects appearing in higher dimensions are mentioned in the discussion.

\begin{figure*}
    \centering
    \includegraphics{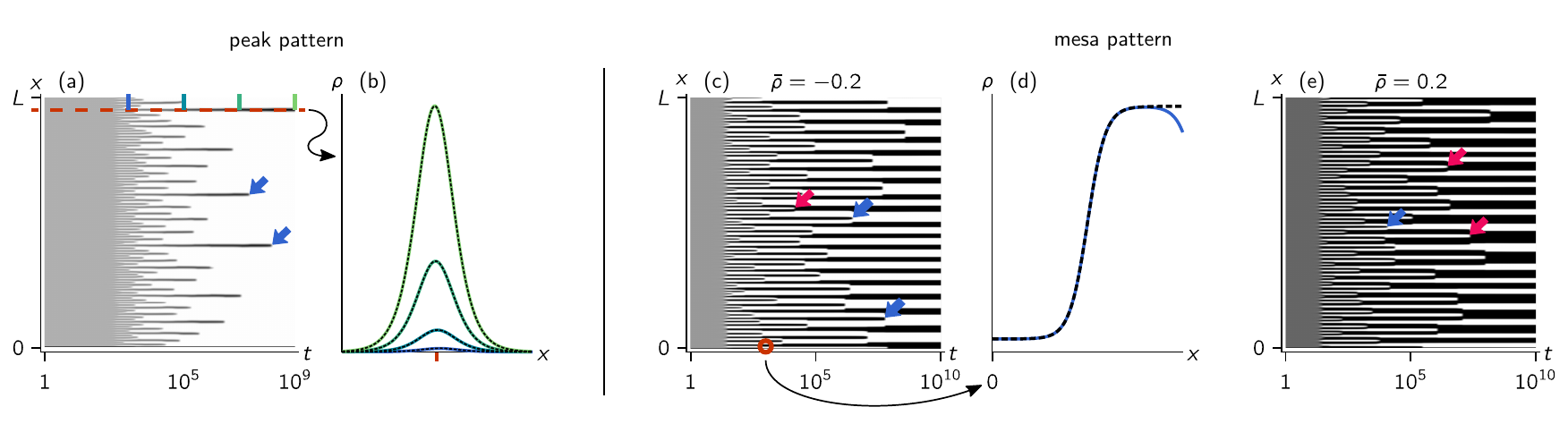}
    \caption{
    The phenomenology of coarsening.
    (a) The kymograph shows the density profile $\rho$ (grayscale, black corresponds to large densities) for the simulation of the second model of Ref.~\cite{Otsuji.etal2007}, a peak-forming 2cMcRD system defined by ${f = \rho^2\eta-\rho}$, where $\rho=u+v$ is the total density and $\eta=v + d u$ is the mass-redistribution potential (see Sec.~\ref{sec:2cRD-mass-conservation}). 
    The parameters are ${D_u = 1}$, ${D_v = 10}$ and ${\bar{\rho} = 10}$.
    (b) A single peak participating in the coarsening process [(red dashed line in panel (a); colored (dark to light gray) profiles in (b) correspond to the times indicated in (a)] is well described by the stationary peak profile (black dashed, analytical form was derived in Ref.~\cite{Otsuji.etal2007}).
    (c,e) A mesa-forming model shows much slower coarsening [cubic model ${f = (\eta - \rho^3+\rho)/(1-d)}$; see Sec.~\ref{sec:cubic-model} and Appendix~\ref{app:cubic-model}].
    At low average total density $\bar{\rho}$ [panel (c)], coarsening proceeds mainly via competition [blue (dark gray) arrows] while coalescence [red (light gray) arrows] dominates at large average density [panel (e)].
    The parameters are $D_u =1$, $D_v=10$ and $\bar{\rho}$ as indicated.
    (d) Again, the single interface profiles are well approximated by the stationary profile.
    All simulations employ a system length $L=1000$ and periodic boundary conditions.
    }
    \label{fig:coarsening-phenomenology}
\end{figure*}

\section{Phenomenology of the coarsening process and its arrest}
\label{sec:phenomenology}
Let us start with an overview of the typical phenomenology of (nearly mass-conserving) 2cRD systems before getting into their detailed mathematical analysis.
Consider the general reaction--diffusion dynamics of two species $u$ and $v$.
We decompose the governing equations into a mass-conserving core system and source terms of strength $\varepsilon$ (cf.\ \cite{Kuwamura.Morita2015, Kuwamura.Izuhara2017, Brauns.etal2021})
\begin{subequations}
\label{eq:2cRD}
\begin{align}
    \partial_t u(x,t) &= D_u\, \nabla^2 u + f(u,v) +\varepsilon s_1(u,v)
    \, ,\label{eq:2cRD-u}\\
    \partial_t v(x,t) &= D_v\, \nabla^2 v - f(u,v) +\varepsilon s_2(u,v)
    \, ,\label{eq:2cRD-v}
\end{align}
\end{subequations}
defined on a $D$-dimensional spatial domain $\Omega$ with no-flux boundary conditions for both $u$ and $v$.
Although we here introduce the general system, our analysis in the following chapters focuses on the one-dimensional case ${D=1}$. 
For specificity, we choose the relative diffusion constant as ${d := D_u/D_v < 1}$.

Thus, $u(x,t)$ describes the density of a slowly diffusing species, which we will interpret in the context of intracellular pattern formation as proteins attached onto a cell membrane.
In contrast, $v(x,t)$ describes the density of the fast-diffusing species, which may be a different conformational state of the proteins that has detached from the membrane and undergoes fast diffusion in the cytosol of the cell (${D_v \gg D_u}$ for intracellular pattern formation).
The reaction term $f(u,v)$ accounts for the mass-conserving conversion between these two species, that is, between the two conformational states.
In this intracellular context, $f$ is typically given as attachment-detachment kinetics ${f(u,v) = a(u)v - b(u)u}$, where the attachment rate $a(u)$ and detachment rate $b(u)$ depend on the membrane concentration $u$. These terms model, for instance, nonlinear recruitment and enzyme-driven detachment \cite{Jilkine.Edelstein-Keshet2011, Frey.Brauns2022, Goryachev.Leda2020}.

As the chemical reactions represented by $f(u,v)$ correspond to conversions between the two protein states, they locally conserve  the total density ${\rho(x,t) = u + v}$ of proteins at each point $x$ in the domain $\Omega$.
Consequently,  in the case of mass conservation (${\varepsilon=0}$), the density $\rho(x,t)$  changes only due to the redistribution of proteins within the system.
We stress that---different from conserving systems of Cahn--Hilliard type---mass conservation in the 2cMcRD system [Eq.~\eqref{eq:2cRD} with ${\varepsilon=0}$] only holds for the total density $\rho(x,t)$ while the dynamics of the single protein states $u$ and $v$ is not conserving.
In contrast, $s_1$ and $s_2$ are source terms that break strict mass conservation of the total density $\rho$; their strength is given by the dimensionless parameter $\varepsilon$.
These source terms, for instance, account for the production or degradation of proteins, and they change the total number of proteins.

\subsection{Long-time dynamics of 2cMcRD systems}
\label{sec:coarsening-phenomenology}
We study two-component mass-conserving reaction--diffusion (2cMcRD) systems for their ability to form spatially heterogeneous protein concentration patterns.
Such patterns form if homogeneous steady states (HSS) with uniform protein concentrations within the whole domain exhibit a lateral instability.
This instability drives the exponential growth of small fluctuations in the protein concentration fields around the homogeneous steady state.
In 2cMcRD systems, this process results in the formation of peak or mesa patterns (Fig.~\ref{fig:coarsening-phenomenology}).
Both types of patterns subsequently undergo a coarsening process, but typically on a much larger timescale, during which smaller peaks or mesas vanish while larger domains grow.

Figure~\ref{fig:coarsening-phenomenology}(a) exemplifies this dynamics for peak-forming patterns.
Small perturbations around the HSS grow exponentially and rapidly form a series of density peaks.
This is followed by a slow coarsening process which continues until just one peak is left.
During the coarsening process, the dynamics of each individual peak is well described by a (quasi-)stationary profile that slowly evolves over time [Fig.~\ref{fig:coarsening-phenomenology}(b)].
The elementary motif of the coarsening process is the competition for mass between peaks:
We observe that between two peaks of almost equal size, mass is redistributed from the smaller peak toward the larger peak.
This process destabilizes the symmetric state of equally sized peaks and leads to the collapse of the smaller peak.
Because diffusive mass redistribution from one to the other peak is slower over longer length scales, the competition is strongest between \emph{neighboring} peaks.
We call this elementary instability between neighboring peaks the \emph{mass-competition instability} \cite{Brauns.etal2021}.

\begin{figure}
    \centering
    \includegraphics{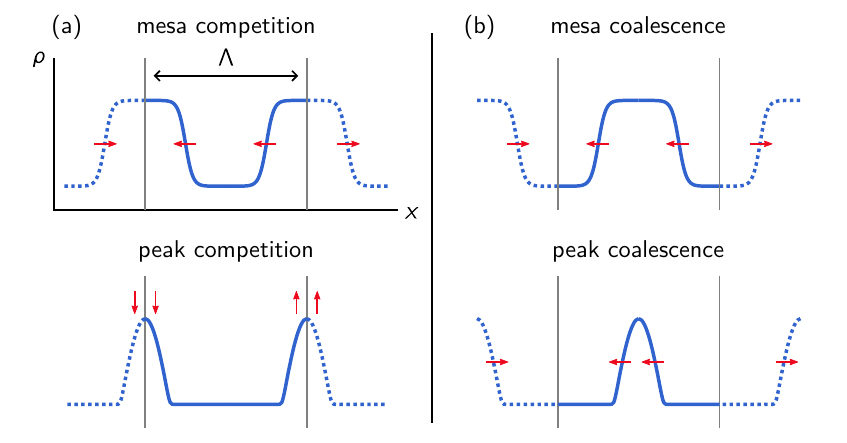}
    \caption{
    The coarsening scenarios.
    (a) The \textit{competition scenario} is illustrated for mesa (top) and peak (bottom) patterns: One of the two mesas/peaks is growing, and the other is shrinking [indicated by (red) arrows] due to competition for mass. 
    (c) Coarsening can also proceed by \textit{coalescence} of mesas (top) or peaks (bottom): Neighboring high-density domains move toward each other until they merge. This coalescence scenario can also be understood as competition for mass between low-density regions (`troughs') of the mesa/peak patterns.
    }
    \label{fig:coarsening-modes}
\end{figure}

Similar dynamics as for peak patterns occurs in mesa-forming systems [Fig.~\ref{fig:coarsening-phenomenology}(c,d)].
However, the coarsening process is much slower, and at high average protein densities, it is dominated by the \emph{coalescence}, i.e.\ relative motion, of mesas rather than their competition for mass [examples marked by red (light gray) arrows in Fig.~\ref{fig:coarsening-phenomenology}(e)]. Competition for mass, in contrast, is (mainly) observed in peak-forming systems as well as at lower densities in mesa-forming systems [blue (dark gray) arrows in Fig.~\ref{fig:coarsening-phenomenology}(e)].
At coalescence events, the mesa number decreases not because a mesa loses all its mass and vanishes but because mesas shift their positions and merge.
The interaction of fronts resulting in coarsening dynamics has been analyzed in diverse mesa-forming systems \cite{Langer1971, Kawasaki.Ohta1982, Kawasaki1984, Fusco.Hale1989, Alikakos.etal1991, Argentina.etal2005, Nepomnyashchy2010}.
We will use a linear stability analysis of periodic patterns to discuss the mass-competition instability and the resulting dynamics in 2cMcRD systems.

As in the case of competition for mass between the mesas/peaks, the coalescence process is also based on an instability of the periodic patterns:
A high-density mesa which is, say, closer to its left neighbor than to its right neighbor moves even farther to the left until it merges with its left neighbor.
Figure~\ref{fig:coarsening-modes} compares the
different coarsening scenarios where coarsening is driven by competition for mass [panel (a)] and coalescence [panel (b)], respectively.
This comparison illustrates that the dynamics in the coalescence scenario is actually driven by (inverted) mass competition of the low-density domains, that is, competition for (negative) mass between the `troughs' of the pattern. 
For example, if a peak moves toward a neighboring peak [Fig.~\ref{fig:coarsening-modes}(b), bottom], the trough on one side grows, and the trough on the other side collapses similarly as the larger high-density mesa grows while the smaller one collapses during the mesa-competition process [Fig.~\ref{fig:coarsening-modes}(a), top].
Consequently, both competition and coalescence are driven by destabilizing mass redistribution between pattern domains, i.e., a mass-competition instability between domains of high or low density.

In the next sections, we analyze mass competition in an isolated compartment containing only two `half' peaks or mesas, that is, one period of the stationary pattern (gray vertical lines in Fig.~\ref{fig:coarsening-modes}).
This corresponds to the elementary motif of nearest-neighbor competition and it allows us to isolate each of the two coarsening scenarios by placing the no-flux boundaries of the compartment such that they reflect the symmetry of the respective perturbation mode of the pattern [cf.\ (red) arrows in Fig.~\ref{fig:coarsening-modes}.

Below, we will use linear stability analysis to describe the evolution of small perturbations from the periodic patterns.\footnote{
Note that the linear stability analysis is performed for the dynamics linearized around fully nonlinear periodic patterns.
This is different from the typical Turing analysis where the dynamics is linearized around the HSS.}
\textit{Mass competition} and \textit{coalescence} then correspond to distinct unstable eigenmodes of the linearized dynamics close to the periodic patterns.
In this linear regime, deviations from the periodic pattern grow exponentially ${\sim \operatorname{e}^{\sigma t}}$ in time, where $\sigma$ is the growth rate of the corresponding eigenmode.
The sign of $\sigma$ determines whether the associated coarsening mode is stable (${\sigma <0}$) or unstable (${\sigma>0}$), i.e., whether the perturbations decay or grow.\footnote{
In general, $\sigma$ is complex, and the sign of the real part of the growth rate $\Re[\sigma]$ determines the stability.
For the competition and coalescence modes, it will turn out that $\sigma$ is real.}
We will learn from this  analysis why 2cMcRD systems always exhibit uninterrupted coarsening.
Moreover, the magnitudes of the growth rates $\sigma$ for the different coarsening scenarios illustrated in Fig.~\ref{fig:coarsening-modes} will explain that (almost) no coalescence events occur for peak patterns because peak competition is much faster than peak coalescence.
Moreover, for mesa patterns, the magnitude of both growth rates for mesa competition and mesa coalescence will turn out to be strongly reduced compared to the rate of peak competition, revealing why the coarsening process is much slower for mesa than for peak patterns.
In addition, we will show that the relative strength of mesa competition and mesa coalescence depends on the average density $\bar{\rho}$, explaining why the coarsening process for mesa patterns is dominated by competition at low average densities $\bar{\rho}$ and coalescence at high average densities [cf.\ Fig.~\ref{fig:coarsening-phenomenology}(c,e)].
To describe the coarsening dynamics in a large system containing many peaks or mesas (cf.\ Fig.~\ref{fig:coarsening-phenomenology}), we use the growth rate $\sigma$ to determine the temporal law of the coarsening dynamics, i.e., the time evolution of the average peak (mesa) separation or of the peak (mesa) number (see Ref.~\cite{Brauns.etal2021} and Sec.~\ref{sec:coarsening-scaling}).

\begin{figure}
    \centering
    \includegraphics{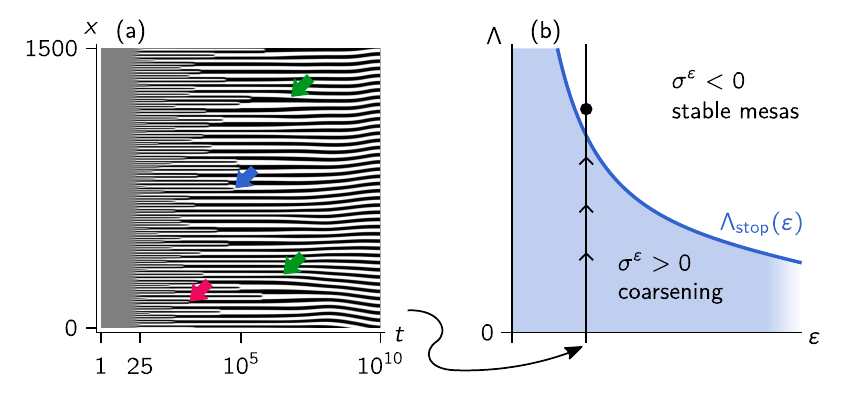}
    \caption{Weak source terms interrupt the coarsening process by suppression of the mass-competition instability at large wavelengths.
    (a) The kymograph obtained from numerical simulations shows mesa patterns (grayscale, $-1$ to $1$) that collapse and coalesce for an intermediate period of time [blue (dark gray) and red (light gray) arrow].
    At later times ${t\gtrsim 10^6}$, the domains merely rearrange into the final periodic stationary state [green (two right-most) arrows].
    As an example, the cubic model ${f=(\eta-\rho^3+\rho)/(1-d)}$ ($\rho$ and $\eta$ defined as in the caption of Fig.~\ref{fig:coarsening-phenomenology}) was simulated with the source terms ${(s_1,s_2) = (0,-\rho)}$ on a domain of length ${L=20 \, 000}$ (see Sec.~\ref{sec:cubic-model} and Appendix~\ref{app:cubic-model}).
    The parameters are ${D_u = 1}$, ${D_v=10}$ and ${\varepsilon = 10^{-6}}$.
    (b) A given source strength $\varepsilon$ is strong enough to stabilize patterns at length scales ${\Lambda>\Lambda_\mathrm{stop}(\varepsilon)}$ (top edge of the shaded region, blue line).
    Patterns with a shorter characteristic length undergo coarsening [(blue-) shaded region].
    }
    \label{fig:interrupted-coarsening-phenomenology}
\end{figure}

\subsection{Weak source terms interrupt coarsening}
\label{sec:interrupted-coarsening-phenomenology}
As we have just demonstrated phenomenologically, strictly mass-conserving two-component reaction--diffusion systems exhibit uninterrupted coarsening:
The pattern length scale grows until it reaches the system size, and no intrinsic length scale is selected.
In contrast, classical non-mass-conserving 2cRD systems, including production and degradation terms, result in patterns with a fixed length scale \cite{Epstein.Pojman1998, Cross.Greenside2009, Wei.Winter2014}.
To bridge the phenomenological gap between mass-conserving 2cRD systems and those without mass conservation, we analyze systems with weak source terms.
Such weak coupling to a reservoir is also a first step to generalize the analysis toward reaction--diffusion systems with more than two components, which are important as models of pattern formation in complex biochemical reaction networks \cite{Halatek.Frey2012,Halatek.etal2018,Murray.Sourjik2017,Chiou.etal2021}.

Figure~\ref{fig:interrupted-coarsening-phenomenology}(a) shows the time evolution of a mesa-forming model with weak source terms.
Initially, the mesa pattern develops out of perturbations around the homogeneous steady state.
These mesas then undergo coarsening [blue (second from the left) and red (left-most) arrow in Fig.~\ref{fig:interrupted-coarsening-phenomenology}(a)], which halts after some time.
After that, no more mesas (and troughs) collapse, and the remaining mesas rearrange themselves into a periodic pattern.
During this rearrangement, smaller mesas grow, and larger mesas shrink so that their masses balance out [green (two right-most) arrows in Fig.~\ref{fig:interrupted-coarsening-phenomenology}(a)].
These observations suggest that the mass-competition instability must be suppressed by the weak source terms, and the mass competition between two neighboring peak/mesas no longer results in the collapse of the smaller peak/mesa.
Rather, the mass-competition process is reversed and stabilizes the symmetrical configuration of two equally sized domains.

The slowdown of the coarsening process as it progresses [see Fig.~\ref{fig:coarsening-phenomenology}] indicates that the mass-competition instability weakens as the typical distance $\Lambda$ between peaks/mesas increases.
Moreover, we will show below that source terms of strength $\varepsilon$ [cf.\ Eqs.~\eqref{eq:2cRD}] lead to stabilizing effects ${\sim \varepsilon}$.
Thus, a critical pattern length scale ${\Lambda_\mathrm{stop}(\varepsilon)}$ exists where the source terms are sufficiently strong to suppress the mass-competition instability [see Fig.~\ref{fig:interrupted-coarsening-phenomenology}(b)].
For ${\Lambda < \Lambda_\mathrm{stop}(\varepsilon)}$ the growth rate ${\sigma^\varepsilon(\Lambda)}$ of at least one coarsening mode under the influence of weak source terms stays positive, giving rise to a mass-competition instability, and driving coarsening.
For patterns of larger length scale ${\Lambda>\Lambda_\mathrm{stop}(\varepsilon)}$  the growth rates for all coarsening scenarios will fulfill ${\sigma^\varepsilon <0}$ which signifies that deviations from the symmetric pattern of equally sized peaks or mesas are not amplified any longer but relax back toward the symmetric state.
Below, we determine this threshold $\Lambda_\mathrm{stop}(\varepsilon)$ where $\sigma^\varepsilon = 0$.

\section{Models}
\label{sec:models}
In the last section, we introduced the general form of (nearly) mass-conserving two-component reaction--diffusion dynamics.
Here, we will rewrite this dynamics to make the role of mass conservation explicit.
Then, we discuss the Cahn--Hilliard and conserved Allen--Cahn models for phase separation and recast these models into a form that underlines their structural similarity to mathematical descriptions of reaction--diffusion dynamics.
Importantly, by pointing out how the three models are akin at a mathematical level, we are not claiming that these models describe the same physical processes, but rather we are trying to explain in what sense the phenomena they exhibit are related.

\subsection{Two-component reaction--diffusion system}
\label{sec:2cRD-mass-conservation}
We consider the general 2cRD dynamics Eqs.~\eqref{eq:2cRD}.
If the source terms are switched off by setting ${\varepsilon=0}$, the system locally conserves the total density ${\rho(x,t) = u + v}$ of proteins.
Including the source terms one has in a well-mixed reaction compartment
\begin{equation}
    \partial_t \rho(t) = \varepsilon s_\mathrm{tot}(u,v)
    \, ,
\end{equation}
with the total source term ${s_\mathrm{tot} := s_1 + s_2}$.

In contrast, in a spatially extended system the total density $\rho(x,t)$ additionally changes by redistribution of proteins, and its time evolution is governed by the (modified) continuity equation [add up Eqs.~\eqref{eq:2cRD}, and, for ease of notation denote $s_\mathrm{tot}(u(\rho,\eta),v(\rho,\eta))$ by $s_\mathrm{tot}(\rho,\eta)$]
\begin{equation}
    \partial_t \rho 
    = 
    D_v \nabla^2 \eta + \varepsilon s_\mathrm{tot}(\rho,\eta)
    \, .
    \label{eq:cont-eq}
\end{equation}
Here we defined ${\eta(x,t) := v(x,t) + d\, u(x,t)}$, termed the \emph{mass-redistribution potential} \cite{Otsuji.etal2007,Ishihara.etal2007,Forte.etal2019,Brauns.etal2020}.
In the absence of source terms (${\varepsilon = 0}$) the spatially averaged density ${\bar{\rho} = 1/|\Omega|\int_\Omega \mathrm{d} x \, \rho}$ is conserved.
Thus, the mass-conserving chemical reactions entail that the \emph{total} protein density $\rho$ follows a locally conserved dynamics given by a continuity equation, akin to the Cahn--Hilliard equation and gradient dynamics of scalar field theories for conserved order parameters (``Model B'') \cite{Cahn.Hilliard1958,Hohenberg.Halperin1977,Desai.Kapral2009}.
The mass-redistribution potential here plays a similar role as the chemical potential in the Cahn--Hilliard equation.

Unlike the chemical potential in thermal equilibrium systems, however, the mass redistribution potential is \emph{not} given by the gradient of a free energy functional but follows its own time evolution [using Eqs.~\eqref{eq:2cRD} and the definition $\eta = v + d u$]
\begin{equation}
    \partial_t \eta 
    = 
    \big(
    D_v {+} D_u
    \big)
    \nabla^2 \eta 
    -
    D_u 
    \nabla^2 \rho 
    - 
    \tilde{f}(\rho,\eta) 
    + 
    \varepsilon 
    \big(
    d s_1 {+} s_2
    \big)
    \, ,
    \label{eq:eta-evol}
\end{equation}
with ${\tilde{f}(\rho,\eta) := (1-d) \, f(u(\rho,\eta),v(\rho,\eta))}$.
While the continuity equation, Eq.~\eqref{eq:cont-eq}, describes the redistribution of the total density and its production or degradation, Eq.~\eqref{eq:eta-evol} describes the local reactions between $u$ and $v$ that adjust their ratio given a prescribed total-density profile $\rho$.
These reactions induce the relaxation of the mass-redistribution potential $\eta = v + d u$ toward the reactive equilibrium $\tilde{f}(\eta^*,\rho)=0$.\footnote{
Consequences of bistability of the reaction kinetics are discussed in~\cite{Brauns.etal2020}.
}
Here, the nullcline $\eta^*(\rho)$ gives the family of reactive equilibria for different total densities $\rho$.
If the density profile is not uniform, the densities $u$ and $v$ show gradients that lead to particle diffusion.
The (differential) diffusion of $u$ and $v$ is accounted for by the gradient terms in Eq.~\eqref{eq:eta-evol}.

\subsection{Generalized Cahn--Hilliard equation}
The classical Cahn--Hilliard (CH) equation \cite{Cahn.Hilliard1958} describes the equilibrium dynamics of an incompressible binary mixture of two types of particles A and B which undergo phase separation due to a stronger affinity between particles of the same type than between particles of distinct types.
The thermodynamics of such phase-separating systems are determined by the free-energy functional \cite{Cahn.Hilliard1958,Desai.Kapral2009}
\begin{equation}\label{eq:free-energy}
    \mathcal{F}[\phi] = \int\mathrm{d}x
    \bigg[
    \frac{\kappa}{2}(\nabla\phi)^2 
    + 
    g(\phi)
    \bigg]
    \, ,
\end{equation}
where the scalar field $\phi (x)$ corresponds to the local composition of the binary mixture and is linearly related to the density of the A particles.
The square-gradient term in $\mathcal{F}[\phi]$ accounts for the free energy costs of gradients in the composition.
We denote the local free-energy density by $g(\phi)$ to avoid confusion with the term $f(u,v)$ describing chemical reactions in the 2cRD model.
This free-energy density $g(\phi)$ can be obtained from symmetry arguments \cite{Chaikin.Lubensky1995,Desai.Kapral2009} or, alternatively, derived, for example, from a lattice model \cite{Flory1942,Huggins1942,Doi2013}.
Below the critical point at sufficiently low temperatures, the free-energy density is a double-well potential that models both entropic effects and the effectively repulsive interactions between particles of different types.
For symmetric binary mixtures,  one often takes the simple quartic form ${g(\phi) = -\frac{r}{2}\phi^2 + \frac{u}{4}\phi^4}$ as obtained from a  Ginzburg-Landau expansion close to the critical point at ${r=0}$ and ${\phi=0}$.
This simple form also captures the qualitative shape of the free-energy density further away from the onset of phase separation.

To describe the dynamics of phase separation, we need an evolution equation associated with the free-energy functional $\mathcal{F}[\phi]$.
Since the total number of A and the total number of B particles are each conserved, the local composition $\phi(x,t)$ must obey a continuity equation ${\partial_t \phi(x,t) = - \nabla J(x,t)}$, and for the system to relax into thermal equilibrium, the particle current $J(x,t)$ must be proportional to the gradient of a corresponding chemical potential $\mu(x,t)$ \cite{Groot.Mazur1984,Balian2007,Desai.Kapral2009}.
These general concepts of non-equilibrium thermodynamics lead to the following equations\footnote{
We consider only the deterministic dynamics here.
Effects of noise during the pattern-formation process (phase-separation process) are discussed shortly in the discussion Sec.~\ref{sec:discussion}.}
\begin{subequations}
\begin{align}
    \partial_t \phi(x,t) 
    &= 
    \nabla 
    \big[ 
    \Gamma(\phi) \nabla \mu 
    \big] 
    \, , \\
    \mu(x,t) 
    &= \frac{\delta \mathcal{F}[\phi]}{\delta \phi} 
    = 
    - 
    \kappa \nabla^2\phi 
    + 
    \partial_\phi g(\phi)
    \, , 
    \label{eq:Model-B-mu}
\end{align}
\end{subequations}
where $\Gamma(\phi)$ denotes the mobility, i.e., an Onsager coefficient that may depend on the composition $\phi$.
These equations correspond to Model B in the classification scheme of Hohenberg and Halperin \cite{Hohenberg.Halperin1977}. 
Beyond phase separation, equations of this form are also used to describe, for example, the dynamics of fluid thin films \cite{Thiele2018}.

The classical CH equation follows for constant mobility ${\Gamma(\phi) = \Gamma}$ and the quartic free-energy density $g(\phi)$.
It reads
\begin{subequations}
\begin{align}
    \partial_t \phi 
    &= 
    \Gamma \nabla^2\mu
    \, , \\
    \mu 
    &= 
    -\kappa \nabla^2 \phi + \phi^3-\phi
    \, ,
\end{align}
\end{subequations}
where we eliminated the parameters in the free-energy density by rescaling.
This equation yields phase separation into A- and B-rich domains with ${\phi_\pm \approx \pm 1}$ which undergo coarsening and grow until the fully phase-separated state is reached~\cite{Bray2002}.

As an extension of the equilibrium phase-separation dynamics, reactions $R(\phi)$ between the particles involved in the phase separation have been considered \cite{Glotzer.etal1995,Weber.etal2019,Li.Cates2020}.
The resulting model couples Model A (reactions) and Model B (phase separation) dynamics
\begin{equation}
\label{eq:Model-A-plus-B}
    \partial_t \phi 
    = 
    \nabla 
    \big[
    \Gamma(\phi) \nabla \mu
    \big]
    + 
    R(\phi) 
    \, ,
\end{equation}
with the chemical potential $\mu$ again given by Eq.~\eqref{eq:Model-B-mu}.
For example, a linear reaction that allows for conversion between A and B particles yields ${R(\phi) = k(\phi_0 - \phi)}$ where $k$ is the reaction rate, and $\phi_0$ denotes the chemical equilibrium composition balancing the forward and backward reactions.
It was shown that such reactive dynamics may lead to an interruption of the coarsening process found in pure Model B dynamics and thus a selection of a finite domain size \cite{Glotzer.etal1995}.
This motivates the study of such models to understand the size control of intracellular condensates which are thought to compartmentalize biochemical reactions in living cells~\cite{Weber.etal2019}.
This system also captures models for active matter considering birth and death \cite{Cates.etal2010} and the Oono--Shiwa equation \cite{Oono.Shiwa1987}.
Overall, Eq.~\eqref{eq:Model-A-plus-B} defines a paradigmatic non-equilibrium field theory if the free energies for the Model A and Model B dynamics are chosen independently \cite{Li.Cates2020}. 
Allowing for the interaction with different diffusing species through the reaction term, a related model was considered to analyze phase separation in cell membranes coupled to protein-pattern formation \cite{John.Bar2005a}.

To work out the similarities with the 2cRD system,
we rewrite the generalized CH equation, Eq.~\eqref{eq:Model-A-plus-B}, in the form
\begin{subequations}
\label{eq:gen-Cahn-Hilliard}
\begin{align}
    \partial_t \rho 
    &=  
    \nabla 
    \big[
    D_v(\rho)\nabla \eta
    \big]
    + 
    \varepsilon s_\mathrm{tot}(\rho,\eta)
    \, ,
    \label{eq:gen-CH-cont-eq}
    \\
    0 
    &= 
    - 
    D_u \nabla^2 \rho 
    - 
    \big[
    \eta-\eta^*(\rho)
    \big]
    \, .
    \label{eq:gen-CH-eta}
\end{align}
\end{subequations}
To underline the close similarities, we are using the same nomenclature as for the reaction--diffusion system [Eqs.~\eqref{eq:cont-eq},~\eqref{eq:eta-evol}]:
${\phi \leftrightarrow \rho}$, ${\mu \leftrightarrow \eta}$, ${\Gamma(\phi) \leftrightarrow D_v(\rho)}$, ${R(\phi)\leftrightarrow\varepsilon s_\mathrm{tot}(\rho,\eta)}$, ${\kappa\leftrightarrow D_u}$, and ${\partial_\phi g(\phi) \leftrightarrow \eta^*(\rho)}$.

Comparing the 2cRD dynamics, Eqs.~\eqref{eq:cont-eq},~\eqref{eq:eta-evol}, with the rewritten CH equation including chemical reactions, Eqs.~\eqref{eq:gen-Cahn-Hilliard}, one notices the following similarities and differences.
First, in the 2cRD system, $\rho$ describes the total molecule density which may change only due to redistribution of molecules within the system (current $-D_v\nabla\eta$) or production and degradation via $\varepsilon s_\mathrm{tot}$ [cf.\ Eq.~\eqref{eq:cont-eq}].
Similarly, $\phi$ in the generalized CH equation is a measure for the local fraction of A particles, and the local amount of A particles can also only change by redistribution of A particles [and ensuing redistribution of B particles due to the incompressibility constraint; driven by the current $-(\Gamma\nabla\mu)$] or by conversion of particles between A and B type [described by $R(\phi)$].
Therefore, $\rho$ and $\phi$ follow analogous dynamics with the mass-redistribution potential $\eta$ corresponding to the chemical potential $\mu$ and the total source term $\varepsilon s_\mathrm{tot}$ taking the place of the conversion reactions $R(\phi)$.
The timescale of redistribution is set by the, possibly density-dependent, mobility $\Gamma$ in the extended CH system. In the 2cRD system, it is set by $D_v$, the larger of the two diffusion coefficients, which is typically density-independent.

Rearranging the defining equation for the chemical potential, Eq.~\eqref{eq:Model-B-mu}, into Eq.~\eqref{eq:gen-CH-eta} suggests that the value of the chemical potential for the CH equation is fixed by the steady-state balance of a diffusion [$D_u\nabla^2\rho$] and a reaction term. 
One can choose ${\tilde{f}=\eta-\eta^*(\rho)}$ as reaction term in the 2cRD system [Eq.~\eqref{eq:eta-evol}; for the classical CH equation we have ${\eta-\eta^*(\rho)=\eta+r\rho-u\rho^3}$].
Then, Eq.~\eqref{eq:gen-CH-eta} corresponds to the (quasi-)stationary equation for Eq.~\eqref{eq:eta-evol}, that is, the equation following from ${\partial_t\eta=0}$, up to the additional diffusion term ${{\sim}\nabla^2\eta}$.
The continuity equation, Eq.~\eqref{eq:cont-eq}, shows that the strength of the diffusion term agrees with the strength of mass transport. 
Consequently, the diffusion term is negligible if mass transport is slow in comparison to the local reactions.
This indicates that a connection between the reaction--diffusion and CH dynamics can be found under this condition.
Moreover, the identification ${\tilde{f}=\eta-\eta^*(\rho)}$ shows that the steady-state mass-redistribution potential of the homogeneous (well-mixed) 2cMcRD system determined by the reactive equilibria ${\tilde{f}=0}$ corresponds to the chemical potential ${\mu = \partial_\phi g=\eta^*}$ of a uniform composition.

Finally, this comparison shows that the rigidity $\kappa$ plays the role of an effective diffusion constant, a correspondence well known for non-conserved dynamics where the Allen--Cahn system (Model A) is compared to the Schlögl reaction--diffusion model \cite{Schlogl1972, Desai.Kapral2009}.
In phase separation, the rigidity $\kappa$ determines the width of the interface between the two separated phases.
Similarly, the diffusion constant $D_u$ determines the width of the transition region between high- and low-density domains in the (nearly) mass-conserving 2cRD system \cite{Brauns.etal2020}.

\subsection{Conserved Allen--Cahn equation}
If the redistribution of material between different droplets is fast during a phase-separation process, the dynamics can be limited by the growth of individual droplets and not by the redistribution of matter between the different phase-separating domains \cite{Wagner1961}.
Examples are sublimation--deposition processes where diffusion through the gaseous phase is orders of magnitude faster than the growth of individual domains \cite{Conti.etal2002}.
Model B, describing only the mass-redistribution dynamics, cannot account for this process.
Instead, the growth of individual domains has to be described.

Because one again describes phase separation, the model is based on the same square-gradient free energy functional Eq.~\eqref{eq:free-energy} as Model B.
However, the time evolution is different.
The growth and shrinking of single domains proceeds to minimize the free-energy functional $\mathcal{F}[\phi]$.
During this process, the domains are coupled to a global pool which represents  the  fast-diffusing  phase  (the  gaseous  phase).
Thereby, mass conservation is enforced as a global conservation law.
Consequently, the conserved Allen--Cahn (cAC) dynamics is derived using relaxational Model A dynamics restricted to the mass-conserving (MC) submanifold in phase space \cite{Rubinstein.Sternberg1992,Conti.etal2002}
\begin{align}
\label{eq:cAC-classical}
    \partial_t \phi(x,t)
    &=
    - \left[\Gamma(\phi) \frac{\delta \mathcal{F}[\phi]}{\delta \phi} \right]_\mathrm{MC} \\
    &=
    - \left[\Gamma(\phi) \frac{\delta \mathcal{F}[\phi]}{\delta \phi} - \frac{1}{|\Omega|}\int_\Omega\mathrm{d}x\, \Gamma(\phi) \frac{\delta \mathcal{F}[\phi]}{\delta \phi} \right]\nonumber\\
    &=
    \Gamma(\phi) 
    \big(
    \kappa \nabla^2\phi - \partial_\phi g
    \big)
    + \frac{1}{|\Omega|}\int_\Omega\mathrm{d}x\, \Gamma(\phi) \partial_\phi g 
    \, , \nonumber
\end{align}
again implemented on a system domain $\Omega$ with no-flux boundary conditions for $\phi$.
In the second line, the global mass conservation constraint is written out explicitly as a Lagrange-multiplier term.
Originally, the model was introduced in Ref.~\cite{Rubinstein.Sternberg1992} to describe phase separation dominated by viscous effects.
The cAC model also finds application in granular media \cite{Aranson.etal2002} and the simulation of incompressible two-fluid flow \cite{Yang.etal2006,Zhang.Tang2007}.

We are interested in the cAC system because in the 2cRD system, the cytosolic density ${v(x,t)=v(t)}$ is spatially constant in the limit of fast cytosolic diffusion ${D_v \to \infty}$.
The species $v$ then effectively acts as a global pool that ensures instantaneous redistribution of molecules over the whole domain (like the gaseous phase in a sublimation/deposition process).
For reaction--diffusion systems this limit is called the \textit{shadow limit} \cite{Nishiura1982}.
In the limit ${D_v \to \infty}$, the time evolution of the global pool ${v(t)=\eta(t)}$ (because ${d\to 0}$) follows from integration of Eq.~\eqref{eq:eta-evol} (${\varepsilon = 0}$) over the whole domain $\Omega$ with no-flux or periodic boundary conditions and yields
\begin{equation}
\label{eq:cAC-reservoir}
    \partial_t v 
    = 
    \partial_t \eta 
    = 
    - 
    \frac{1}{|\Omega|}
    \int_\Omega\mathrm{d}x \, \left(\tilde{f} -\varepsilon s_2\right)
    .
\end{equation}
Inserting the continuity equation, Eq.~\eqref{eq:cont-eq}, into Eq.~\eqref{eq:eta-evol}, and identifying $f_\mathrm{cAC}= \tilde{f} - \varepsilon s_2$, we find the time evolution of the density profile as
\begin{align}\label{eq:cAC-rho}
    \partial_t \rho 
    &= 
    D_u \nabla^2\rho 
    + 
    f_\mathrm{cAC}(\rho,\eta)
    - 
    \frac{1}{|\Omega|}
    \int_\Omega\mathrm{d}x \, 
    f_\mathrm{cAC}(\rho,\eta)\nonumber\\
    &\quad +
    \varepsilon s_\mathrm{tot}(\rho,\eta)
    \, ,
\end{align}
which has the standard form of the cAC equation, Eq.~\eqref{eq:cAC-classical}, with an additional source term $\varepsilon s_\mathrm{tot}$ and the density $\eta$ in the global pool as an additional `parameter'.
If the 2cRD system is mass-conserving, the source term drops out.
If, additionally, one sets ${\tilde{f} = \eta - \eta^*(\rho)}$ (as for the extended CH dynamics), the dependence on $\eta$ drops out and the shadow limit of the 2cMcRD system, Eq.~\eqref{eq:cAC-rho}, agrees exactly with the cAC model, Eq.~\eqref{eq:cAC-classical}.

\section{Instability of the homogeneous steady-state}
\label{sec:HSS-instability}
To start the comparison of the 2cRD dynamics with the CH and cAC dynamics, we begin with the analysis of the homogeneous steady states (HSS) and their stability properties.
The instability of the homogeneous steady state against spatial modulations describes the onset of pattern formation due to the growth of small fluctuations around the uniform density distribution.\footnote{
Outside the regime of lateral instability, a multistable regime allows for pattern formation through a finite perturbation of a stable homogeneous steady state in (nearly) mass-conserving 2cRD systems \cite{Brauns.etal2020} (nucleation-and-growth regime in phase-separation dynamics \cite{Desai.Kapral2009}).}
We begin our analysis with mass-conserving 2cRD systems, compare these with the CH and cAC systems, and then discuss the implications of broken mass conservation.

The homogeneous steady states $(\rho_\HSS^{},\eta_\HSS^{})$ of the 2cMcRD system depend on the average density ${\bar{\rho} = \rho_\HSS^{}}$ as a control parameter because the total mass in the system is not determined by the system dynamics but fixed by the initial condition.
With the nullcline $\eta^*(\rho)$ defined as the set of reactive equilibria ${\tilde{f}(\rho,\eta^*(\rho)) = 0}$, the homogeneous steady state at average total density $\bar{\rho}$ can be written as $[\bar{\rho},\eta^*(\bar{\rho})]$ [see Eqs.~\eqref{eq:cont-eq},~\eqref{eq:eta-evol} with ${\varepsilon=0}$].
By the identification ${\partial_\phi g \leftrightarrow \eta^*(\rho)}$ introduced above [cf.\ Eqs.~\eqref{eq:gen-Cahn-Hilliard}],
the same expression describes the homogeneous steady states in the mass-conserving CH and cAC systems [see Eqs.~\eqref{eq:gen-Cahn-Hilliard} and~\eqref{eq:cAC-rho}].
We assume that these homogeneous steady states are linearly stable against homogeneous perturbations, i.e., we demand that ${\partial_\eta \tilde{f}(\rho_\HSS^{},\eta_\HSS^{}) > 0}$ [cf.\ Eqs.~\eqref{eq:eta-evol},~\eqref{eq:gen-Cahn-Hilliard},~\eqref{eq:cAC-reservoir}].
Otherwise, already the dynamics of the well-mixed system would leave the considered, unstable steady state under an infinitesimal perturbation, and the system would evolve into a different stable steady state.

\begin{figure}
    \centering
	\includegraphics{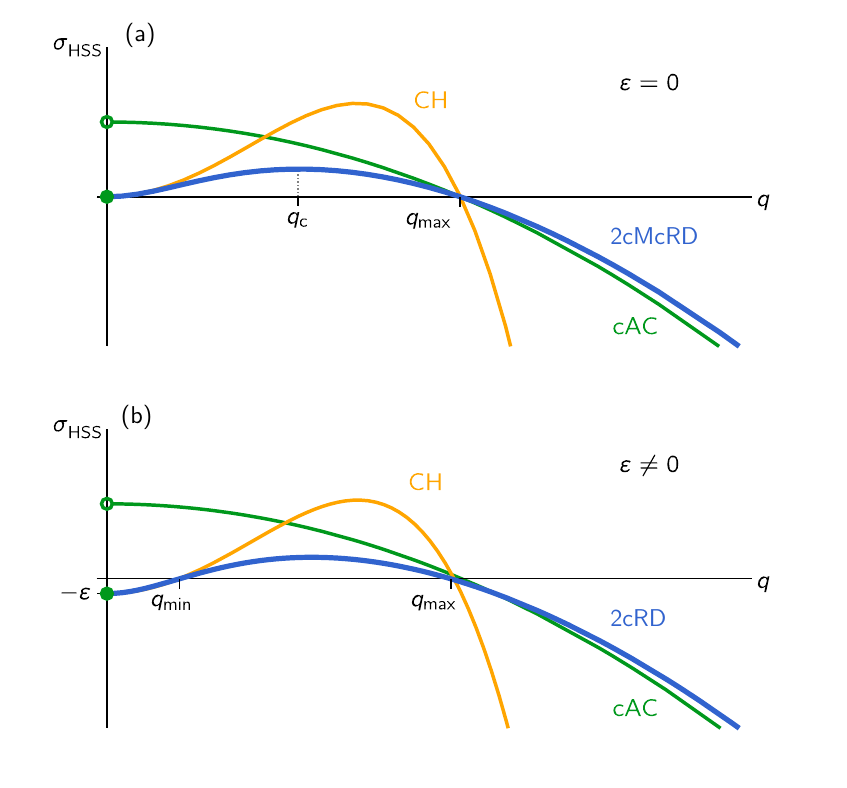}
	\caption{Dispersion relations obtained from a linear stability analysis of the homogeneous steady state  (HSS).
	(a) The dispersion relations $\sigma_\HSS^{} (q)$ as a function of wavenumber $q$ for the mass-conserving 2cMcRD (${\varepsilon = 0}$), CH and cAC models show a band of linearly unstable modes (${\sigma_\HSS^{}>0}$ for ${0<q<q_\mathrm{max}}$).
	The 2cMcRD model has two eigenvalue branches of which only the unstable branch is shown (see Appendix~\ref{app:dispRel}).
	For small wavenumbers $q$, the dispersion relation for the 2cMcRD dynamics is well approximated by the dispersion relation of the CH equation, while in the regime of large wavenumbers, the dispersion relation of the cAC model is found to be a good approximation.
	For the 2cMcRD system, the fastest-growing mode $q_\mathrm{c}$ is indicated in the graph.
	(b) Weak source terms (${\varepsilon \neq 0}$) stabilize the homogeneous mode [${\sigma_\HSS^{}(q=0) < 0}$] in the 2cRD, the CH and the cAC model.
	As a result, in the 2cRD and CH models only modes with wave numbers larger than ${q_\mathrm{min}\sim \sqrt{\varepsilon/D_v}}$ are unstable.
	}
	\label{fig:dispersion-relations}
\end{figure}

Performing a linear stability analysis of Eqs.~\eqref{eq:cont-eq},~\eqref{eq:eta-evol} in terms of Fourier modes for small perturbations around the homogeneous steady state  $(\rho_\HSS^{},\eta_\HSS^{})$ yields the dispersion relation $\sigma_\HSS^{}(q)$ that determines the growth rates for each mode with wavenumber $q$ (see Appendix~\ref{app:dispRel}).
One finds that 2cMcRD models show a lateral instability if and only if \cite{Brauns.etal2020}
\begin{equation}\label{eq:lateral-instability}
    \partial_\rho\eta^*(\rho)
    {\big|}_{\rho=\rho_\HSS^{}} 
    < 
    0
    \, .
\end{equation}
Then the dispersion relation shows a band of unstable modes ${0<q<q_\mathrm{max}}$ with growth rates ${\sigma_\HSS^ {} (q)>0}$ and a fastest-growing mode $q_\mathrm{c}$ (Fig.~\ref{fig:dispersion-relations}).

The condition Eq.~\eqref{eq:lateral-instability} can be heuristically understood as follows: If it is satisfied, the mass-redistribution potential $\eta$ is lowered in regions of higher total density $\rho$.
As gradients in $\eta$ drive mass redistribution, the decrease of $\eta$ in regions of high total density $\rho$ leads to additional mass transport toward these regions, thus amplifying the initial perturbation.
The same condition holds for the (mass-conserving) CH model.
For these thermal-equilibrium systems, the `nullcline' $\eta^*(\phi)$ is the derivative $\partial_\phi g(\phi)$ of the free-energy density $g$ such that the above criterion agrees with the well-known curvature criterion $\partial_\phi^2 g < 0$ which defines the spinodal regime \cite{Doi2013}.

For the cAC model, the dispersion relation agrees with the one obtained for the standard Allen--Cahn model for all modes with wavenumbers $q>0$ because the (linearized) integral terms in Eqs.~\eqref{eq:cAC-reservoir},~\eqref{eq:cAC-rho} vanish for modes with a finite wavenumber:
\begin{equation}\label{eq:disp-rel-cAC}
    \sigma_\HSS^{}(q)
    = 
    -D_u q^2 
    + 
    \partial_\rho \tilde{f}(\rho_\HSS^{},\eta_\HSS^{})
    \, .
\end{equation}
Since the diffusion term is negative, a lateral instability (${\sigma_\HSS^{}>0}$) can only be induced by the reaction term, and the instability condition reads ${\partial_\rho \tilde{f}(\rho_\HSS^{},\eta_\HSS^{}) > 0}$.
With ${\partial_\rho \eta^* = -\partial_\rho \tilde{f} / \partial_\eta \tilde{f}}$, which follows from the definition ${\tilde{f}=0}$ of the nullcline, and the stability of the homogeneous steady state against homogeneous perturbations [${\partial_\eta \tilde{f}(\rho_\HSS,\eta_\HSS) > 0}$] we recover the same instability criterion Eq.~\eqref{eq:lateral-instability} also for the cAC system.

Further analysis shows that not only do the instability criteria agree, but that the mass-conserving CH and cAC models also approximate the dispersion relation of the 2cMcRD system for small and large wavenumbers, respectively (see Fig.~\ref{fig:dispersion-relations}).
We refer to these limits as the \emph{diffusion-} and \emph{reaction-limited regimes} of the reaction--diffusion dynamics (cf.\ \cite{Brauns.etal2020,Brauns.etal2021}).
The two regimes arise because the 2cMcRD dynamics given by Eqs.~\eqref{eq:cont-eq},~\eqref{eq:eta-evol} (${\varepsilon = 0}$) contains two distinct processes.
At low wavenumbers $q$, the wavelength of the Fourier modes is large and mass is transported over long distances during the dynamics.
Over these large distances, mass transport by gradients in the mass-redistribution potential---the process described by the CH model---is the rate-limiting process and determines the growth rate.
In contrast, at large wavenumbers, mass transport proceeds on short distances and is fast compared to the local reaction dynamics that drives the conversion between the individual species $u$ and $v$.
Thus, for large $q$, the dynamics can be approximated by the shadow limit ${D_v\to\infty}$ where mass redistribution becomes instantaneous, keeping only the dynamics of the local reactions between $u$ and $v$.
This corresponds to the cAC model.
The mathematical analysis of the dispersion relations is provided in Appendix~\ref{app:dispRel} (see also Ref.~\cite{Brauns.etal2020}).
 
In 2cRD models with finite source terms ($\varepsilon>0$), the homogeneous steady states (HSS) have to fulfill the two conditions: ${0 = \tilde{f}+\varepsilon (s_2 + d s_1)}$ and ${0 = s_\mathrm{tot}}$ where the second condition replaces the mass constraint of mass-conserving systems.
The first condition entails that the HSS under the influence of weak source terms $0<\varepsilon \ll 1$ is, at the lowest order in the source strength $\varepsilon$ (exact for the CH model), given by the HSS of the mass-conserving system $[\rho_\HSS^{},\eta^*(\rho_\HSS^{})]$.
The second condition arises since in the absence of mass conservation the total density $\rho_\HSS^{}$ is not determined by the initial condition.
Instead, the total density evolves until production and degradation balance at steady state (${s_\mathrm{tot} = 0}$).

In the dispersion relation of the 2cRD and CH models, small but finite source terms (${\varepsilon>0}$) shift the dispersion relation at long wavelength $2\pi/q$ down such that ${\sigma_\HSS^{}(q{=}0) < 0}$ and modes with wavenumbers ${q < q_\mathrm{min}}$ are stabilized [see Fig.~\ref{fig:dispersion-relations}(b) and Appendix~\ref{app:dispRel}].
Due to the quadratic form of the dispersion relation at small $q$, one obtains ${\lambda_\mathrm{min} = 2\pi/q_\mathrm{min} \sim \varepsilon^{-1/2}}$ [see Fig.~\ref{fig:dispersion-relations}(c)], independent of the specific form of the reaction term $\tilde{f}$.
In contrast, the scaling relationship between $\varepsilon$ and the wavelength selected by interrupted coarsening of fully developed patterns depends on the specific reaction term (see Sec.~\ref{sec:growth-rate-nMC-discussion}).
Due to the singular form of the dispersion relation of the cAC system, weak source terms only shift the growth rate for homogeneous perturbations (${q=0}$) to negative values but do not introduce a band of stable modes at small, finite wavenumbers (see Fig.~\ref{fig:dispersion-relations}).

In summary, we have shown that the linear instabilities in the homogeneous steady state of the 2cRD, CH, and cAC models are closely related.
We identified a diffusion- and a reaction-limited regime in which the dynamics of the 2cRD model is well approximated by corresponding CH or cAC models, respectively.
Later, we will also identify parallels between the instability of the homogeneous steady state that drives pattern formation and the mass-competition instability that drives coarsening, and the two different regimes will re-emerge.
In the following section, we first construct the stationary patterns in the 2cRD system to get an overview of the nonlinear patterns formed.

\section{Stationary patterns}
\label{sec:stat-patterns}
Our focus in this section is a conceptual understanding of the final stationary patterns employing a geometric construction in phase space based on local equilibria theory~\cite{Brauns.etal2020}.
These stationary density profiles will play an important role in the analysis of the coarsening process.
There, they serve as quasi-steady states (QSS) to locally approximate the pattern profile during the dynamics at asymptotically long times (see Sec.~\ref{sec:coarsening-phenomenology}).
Such an approach is commonly used in phase-separating systems where the \emph{shape} of the interface profile between the two phases can be approximated as stationary~\cite{Bray2002,Desai.Kapral2009,Pismen2006}.
A more detailed discussion and analysis of the stationary states, which will later also be used for the mathematical analysis of the mass-competition process, is summarized in Appendix~\ref{app:stat-state-MC} for mass-conserving systems and in Appendix~\ref{app:stat-state-nMC} for systems including weak source terms.

The stationary patterns $[\rhostateps(x),\etastateps(x)]$ of the 2cRD system satisfy the equations [stationarity of Eqs.~\eqref{eq:cont-eq},~\eqref{eq:eta-evol}]
\begin{subequations}\label{eq:gen-stat-pattern}
\begin{align}
    0 
    &= D_v \partial_x^2 \etastateps + \varepsilon s_\mathrm{tot}^\varepsilon
    \, ,
    \label{eq:eta-stat}\\
    0 
    &= D_u \partial_x^2\rhostateps + \tilde{f}(\rhostateps,\etastateps)+\varepsilon (s_1^\varepsilon + d s_2^\varepsilon)
    \, ,
    \label{eq:profile-equation}
\end{align}
\end{subequations}
where the superscript $()^\varepsilon$ indicates that we consider the system in the presence of source terms.
These stationary equations have different types of solutions.
First, they are satisfied by the homogeneous steady state discussed in the previous section.
Second, they allow for spatially periodic stationary patterns, which we analyze further below because we are interested in the stability properties of patterns consisting of many equally sized peaks or mesas (see Sec.~\ref{sec:phenomenology}).
Third, asymmetric stationary patterns can be constructed as well.
We do not consider these asymmetric states further because they are unstable \cite{Ward.Wei2002,Ward.Wei2002a}.

\begin{figure*}
\centering
\includegraphics{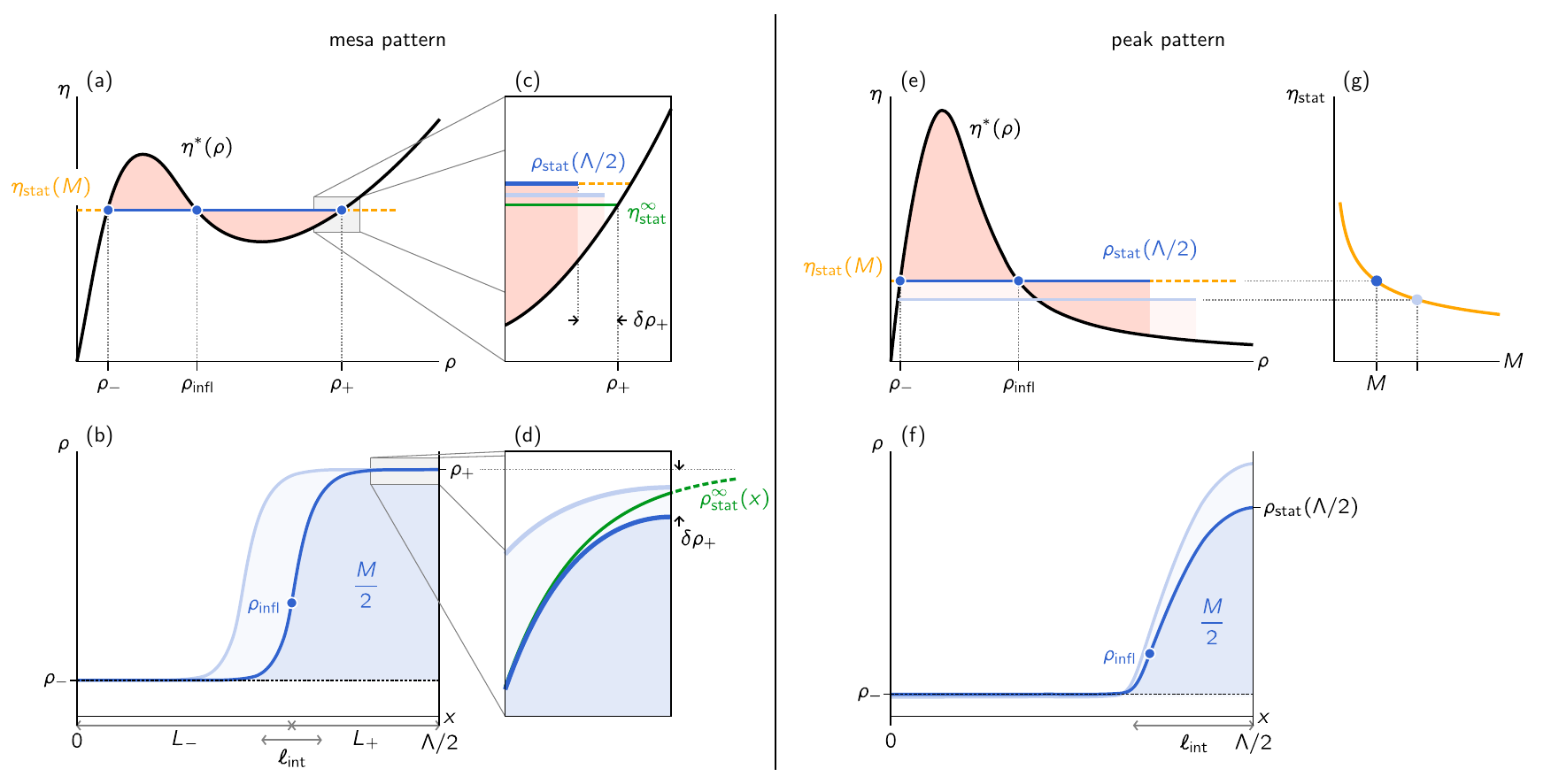}
\caption{
Phase-space construction of the elementary stationary mesa and peak patterns.
(a) An $\mathsf{N}$-shaped nullcline $\eta^*(\rho)$ (NC; black) intersects the flux-balance subspace ${\eta=\etastat(M)}$ [FBS; orange (light gray)] three times, corresponding to the low-density plateau $\rho_-$, the inflection point at density $\rho_\mathrm{infl}$, and the high-density plateau $\rho_+$ [(blue) dots].
The representation of the stationary pattern in phase space [blue (dark gray) line] falls onto the FBS and (approximately) connects the three FBS-NC inflection points.
The position of the FBS is determined by total turnover balance represented approximately by a balance of the (red-) shaded areas.
(b) In real space, the elementary stationary pattern for an $\mathsf{N}$-shaped nullcline has the form of a half-mesa.
An interface of width $\ell_\mathrm{int}$ connects the upper and lower plateaus with lengths $L_\pm$.
The interface position is determined by the mesa mass $M$ [compare the dark (dark gray) and pale blue (light gray) profiles].
(c,d) In more detail, on an infinite domain, the stationary profile $[\rhostat^\infty(x),\eta_\mathrm{stat}^\infty]$ approaches the plateau densities $\rho_\pm$ exponentially [green (thin) line].
Therefore, on a finite domain, the pattern maximum $\rhostat(\Lambda/2)$ and minimum $\rhostat(0)$ deviate from the plateau densities by $\delta\rho_\pm$.
This changes the areas representing total turnover balance, shifting $\etastat(M)$ relative to $\eta_\mathrm{stat}^\infty$.
$\etastat(M)$ approaches $\eta_\mathrm{stat}^\infty$ exponentially as the plateau lengths $L_\pm$ increase (pale blue and red (pale gray) construction).
The same shown here for the high-density plateau applies to the low-density plateau.
(e) A $\mathsf{\Lambda}$-shaped NC $\eta^*(\rho)$ (black) intersects the FBS ${\eta=\etastat(M)}$ [orange (light gray)] only twice.
The FBS-NC intersection points correspond to the low-density plateau $\rho_-$ and the pattern inflection point $\rho_\mathrm{infl}$ [(blue) dots].
Again, the position of the FBS is determined by total turnover balance [represented by the (red-) shaded areas].
An increase in the pattern mass increases the pattern amplitude $\rhostat(\Lambda/2)$, and the FBS shifts downwards (pale blue and red (pale gray) construction).
(f) Systems with $\mathsf{\Lambda}$-shaped nullclines sustain peak-shaped patterns [blue (dark gray)].
The `half-peak' width is denoted by $\ell_\mathrm{int}$.
The peak amplitude grows with the peak mass $M$ [pale blue (light gray) profile].
(g) The stationary mass-redistribution potential $\etastat$ is a decreasing function of the peak mass $M$.
}
\label{fig:MC-stat-construction}
\end{figure*}

The periodic stationary patterns can be constructed from \emph{elementary stationary patterns} which comprise half a period of the periodic pattern.
Because the system is parity symmetric, 
the spatial profiles of the left and right interfaces of peaks or mesas
are mirror images of each other (cf.\ Fig.~\ref{fig:coarsening-modes}).
Thus, the elementary stationary pattern for the periodic pattern with wavelength $\Lambda$ is the monotonic solution to Eqs.~\eqref{eq:gen-stat-pattern} on a domain with no-flux boundary conditions for $\rho$ and $\eta$ and length $\Lambda/2$ [see Figs.~\ref{fig:MC-stat-construction}(b,f)].
\footnote{Next to a stable elementary pattern, an unstable elementary pattern exists in the multistable regime in which both the homogeneous steady state and an elementary stationary pattern are stable~\cite{Brauns.etal2020}.
We do not consider those unstable patterns here.}

For the generalized CH model, the stationary states are determined by the stationarity of the field equations, Eqs.~\eqref{eq:gen-Cahn-Hilliard}.
Thus, they satisfy Eqs.~\eqref{eq:gen-stat-pattern} with ${\tilde{f} = \eta-\eta^*(\rho)}$ and ${s_1 = -d s_2}$.
The first identity shows that the deviation of the chemical potential ${\mu \equiv \eta}$ from its associated fixed point value $\eta^*(\rho)$ for a uniform density profile plays the role of a reaction term.
The choice ${s_1 = -d s_2}$ is required to cancel the source terms in Eq.~\eqref{eq:profile-equation} that do not appear for the generalized CH model [Eqs.~\eqref{eq:gen-Cahn-Hilliard}].
Similarly, the limit ${D_v\to\infty}$ of the stationary equations, Eqs.~\eqref{eq:gen-stat-pattern}, also determines the stationary states of the cAC model, that is, the shadow limit of the 2cMcRD system, Eqs.~\eqref{eq:cAC-reservoir},~\eqref{eq:cAC-rho}.
Therefore, we now focus on the 2cRD system only.

The comparison shows that the uniformity of the mass-redistribution potential in steady state for 2cMcRD systems corresponds to the condition of equal chemical potential in coexisting phases at thermal equilibrium.
The condition of equal pressure will be obtained in a generalized form below by showing that the stationary patterns of 2cMcRD systems fulfill a geometric condition similar to a Maxwell construction.
Importantly, for 2cMcRD systems the geometric construction corresponds to a balance of reactive fluxes instead of following from a common tangent construction for an underlying free-energy potential \cite{Brauns.etal2020,Desai.Kapral2009}.

\subsection{Mass-conserving case}
In the mass-conserving case, the average total density $\bar{\rho}$ of the final stationary pattern is set by the initial conditions which fix the total amount of molecules in the system.
How does $\bar{\rho}$ enter in the stationarity equations Eqs.~\eqref{eq:gen-stat-pattern} of the mass-conserving system (${\varepsilon =0}$)?

We denote the stationary patterns of the mass-conserving system by $[\rhostat(x),\etastat]$.
Balanced mass redistribution, i.e., the \emph{continuity equation}, Eq.~\eqref{eq:eta-stat}, ensures that $\etastat$ is spatially constant and enters only as a parameter in the \emph{profile equation}, Eq.~\eqref{eq:profile-equation}.
For each value of $\etastat$ we can calculate the pattern profile and the average total density $\bar{\rho}$ using Eq.~\eqref{eq:profile-equation}.
This establishes the relation $\bar{\rho}(\etastat)$.
Reciprocally, the average density $\bar{\rho}$ fixes the value of the stationary mass-redistribution potential $\etastat(\bar{\rho})$.
\footnote{For the family of stable elementary stationary patterns, the relation $\bar{\rho}(\etastat)$ is bijective and may therefore be inverted (see supplementary material of Ref.~\cite{Brauns.etal2021}).}

For a fixed average density $\bar{\rho}$ (or, equivalently, fixed $\etastat$) we are now interested in the stationary pattern profile $\rhostat(x)$.
For this, we consider the $(\rho,\eta)$-phase space [see Fig.~\ref{fig:MC-stat-construction}(a,e)].
In phase space, the stationary pattern $[\rhostat(x),\etastat]$ is represented by a curve parametrized by the spatial position $x$.
As noted, the mass-redistribution potential $\etastat$ must be constant because all redistribution processes have to balance.
Thus, a curve corresponding to a stationary pattern $[\rhostat(x),\etastat]$ must be restricted to a horizontal subspace, the \emph{flux-balance subspace} (FBS) ${\eta=\eta_\mathrm{stat}}$  \cite{Brauns.etal2020}.
Now consider the nullcline (NC) $\eta^*(\rho)$ of chemical equilibria ${\tilde{f}=0}$. 
The corresponding curve in phase space represents the homogeneous steady states of the system, and the parts of a pattern which are plateaus, i.e., approximately flat in real space, must lie on (close to) this nullcline.
Thus, FBS-NC intersection points determine the densities of the pattern plateaus~\cite{Brauns.etal2020}.
The profile equation, Eq.~\eqref{eq:profile-equation}, shows that the inflection point (${\partial_x^2\rho|_\mathrm{infl}=0}$) also corresponds to a FBS-NC intersection point.
For the plateaus to be stable, the corresponding homogeneous steady state has to be linearly stable, i.e., in geometrical terms the nullcline slope needs to be positive (see Sec.~\ref{sec:HSS-instability}).
In contrast, the inflection point corresponds to a FBS-NC intersection point with a negative nullcline slope~\cite{Brauns.etal2020}.

Depending on the shape of the nullcline we distinguish two qualitatively different pattern types [see~Fig.~\ref{fig:MC-stat-construction}].\footnote{
We consider only systems for which $\rho$ stays bounded from below as is the case for example if $\rho$ describes a concentration.}
If the nullcline is $\mathsf{N}$-shaped, three FBS-NC intersection points exist [see Fig.~\ref{fig:MC-stat-construction}(a)].
The resulting pattern is \textit{mesa-shaped} with a low- and a high-density plateau connected by an interface [see Fig.~\ref{fig:MC-stat-construction}(b)].
The densities $\rho_+$ and $\rho_-$ of the upper and lower plateaus are given by the outer FBS-NC intersection points, the inflection point density $\rho_\mathrm{infl}$ by the middle FBS-NC intersection point.
The average density $\bar{\rho}$ sets the position of the interface, i.e., the lengths $L_\pm$ of the upper and lower plateaus [see Fig.~\ref{fig:MC-stat-construction}(b)].
To make the mesa size explicit, we define the (surplus) mesa mass $M$ with respect to the background density $\rho_-$ as
\begin{equation}
\label{eq:mesa-peak-mass}
    M 
    = 
    2\int_0^{\frac{\Lambda}{2}}
    \mathrm{d}x\, 
    \big(
    \rhostat(x)-\rho_-
    \big)
    \, .
\end{equation}
To analyze the coarsening process, we will exploit that the mesa mass $M$ can be used instead of the average density $\bar{\rho}$ to parametrize the stationary patterns of different total masses. 

In contrast, if the nullcline is $\mathsf{\Lambda}$-shaped, only two FBS-NC intersection points exist which correspond to a low-density plateau $\rho_-$ and the inflection point $\rho_\mathrm{infl}$ [see Fig.~\ref{fig:MC-stat-construction}(e)].
For this setting, the pattern does not saturate in a high-density mesa but attains a \textit{peak-shaped} profile [see Fig.~\ref{fig:MC-stat-construction}(f)].
We define the peak mass $M$ in the same way as for mesa patterns.
While for mesa patterns only the width depends on the average density $\bar{\rho}$, it determines both the peak height and width for peak patterns.
Again, the peak mass $M$ can be used instead of the average density $\bar{\rho}$ to parametrize the different stationary peak profiles (see supplementary material of Ref.~\cite{Brauns.etal2021}).
Finally, we note that one also observes peak patterns in systems with highly asymmetric $\mathsf{N}$-shaped nullclines if the mass $M$ is low and the maximal pattern density does not saturate in the high-density plateau given by the third FBS-NC intersection point $\rho_+$.

In the context of intracellular pattern formation, one observes membrane-bound protein patterns that exhibit narrow interfaces \cite{Goehring.etal2011,Cezanne.etal2020} or strongly localized high-density clusters (concentration peaks) \cite{Chiou.etal2021}.
Sharp interfaces can form in the concentration profiles of membrane-associated proteins because protein diffusion on the membrane is very slow.
The diffusion constant for membrane-associated proteins, corresponding to $D_u$ in the 2cRD system, is about two to three orders of magnitude slower than diffusion in the cytosol \cite{Meacci.etal2006, Chiou.etal2018}.
Specifically, in the 2cMcRD system, the interface or `half-peak' width $\ell_\mathrm{int}$ [see Fig.~\ref{fig:MC-stat-construction}(b,f)] is set by the diffusion length ${\ell_\text{diff}:=\sqrt{D_u/r}}$, where the reaction rate $r$ describes the reactive timescale at the interface~\cite{Brauns.etal2020} [see also Eq.~\eqref{eq:interface-width-infl-point}].
Thus, slow diffusion allows for steep gradients in the protein density.
Motivated by this characteristic feature of membrane-bound protein patterns, we will consider peak and mesa patterns in the limit\footnote{
This is called the singular limit in the mathematical study of reaction--diffusion systems because gradients at the interface become infinitely steep.}
\begin{equation}
\label{eq:sharp-int-approx}
    \frac{\ell_\mathrm{int}}{\Lambda}
    \ll 
    1
    \, .
\end{equation}
This will allow one to make a \emph{sharp-interface approximation} and apply asymptotic theory to quantitatively describe the profiles and dynamics of fully nonlinear patterns (cf.\ for example \cite{Iron.etal2001,Kolokolnikov.etal2006,Kawasaki.Ohta1982}).

Recall that the stationary mass-redistribution potential $\etastat$, that is, the position of the FBS is fixed by the average density $\bar{\rho}$.
We have also introduced the peak- or mesa-mass $M$, which can alternatively be used as a parameter of the stable elementary patterns (see supplementary material of Ref.~\cite{Brauns.etal2021}).
The key to understanding the coarsening process will be how the stationary mass-redistribution potential $\etastat$ depends on the domain mass $M$.
Therefore, let us work out the relation $\etastat(M)$.

Integration of the profile equation, Eq.~\eqref{eq:profile-equation}, yields under consideration of the no-flux boundary conditions
\begin{equation}\label{eq:reactive-turnover-balance}
    0 
    = 
    \int_0^\frac{\Lambda}{2}\mathrm{d}x\, 
    \tilde{f}(\rhostat,\etastat)
    \, .
\end{equation}
In the (biologically relevant) limit ${D_v\gg D_u}$ this condition for the value $\etastat$ is easy to interpret since we have ${\eta\approx v}$:
Molecules detach from (attach onto) the membrane and increase (decrease) the cytosolic density ${v\approx\eta}$ until attachment and detachment balance, that is, the \emph{total reactive turnover} vanishes. 

To evaluate this balance condition in phase space, we can multiply the profile equation, Eq.~\eqref{eq:profile-equation}, by $\partial_x\rhostat$ before integrating and find the modified condition \cite{Brauns.etal2020}
\begin{equation}\label{eq:ttb}
    0 
    = 
    \int_{\rhostat(0)}^{\rhostat(\frac{\Lambda}{2})}\mathrm{d}\rho \,
    \tilde{f}(\rho,\etastat)
    \, .
\end{equation}
This total turnover balance is qualitatively represented by an (approximate) balance of the areas enclosed between NC and FBS [(red-) shaded areas in Fig.~\ref{fig:MC-stat-construction}(a,e)], similarly to a Maxwell construction (see Ref.~\cite{Brauns.etal2020} and Appendix~\ref{app:stat-state-MC}).
Total turnover balance agrees exactly with the area equality, i.e., the Maxwell construction, if the reaction term has the simple form ${\tilde{f}\sim \eta-\eta^*(\rho)}$ as in the CH system.

Total turnover balance Eq.~\eqref{eq:ttb} determines the relation $\etastat(M)$:
For mesa patterns in the sharp-interface limit, a change of the mesa mass $M$ changes the lengths $L_\pm$ of the upper and lower plateaus [see Fig.~\ref{fig:MC-stat-construction}(b)].
Now, linearization of the profile equation, Eq.~\eqref{eq:profile-equation}, around the plateau densities $\rho_\pm$ shows that the pattern profile approaches these plateau densities exponentially [see Fig.~\ref{fig:MC-stat-construction}(d)].
Since $\etastat$ depends on the maximal and minimal pattern densities $\rhostat(\Lambda/2)$ and $\rhostat(0)$,  $\etastat$ changes exponentially with increasing length of the upper or lower plateau [see Fig.~\ref{fig:MC-stat-construction}(c); detailed analysis in Appendix~\ref{app:stat-state-MC} and supplementary material of Ref.~\cite{Brauns.etal2021}]:
\begin{equation}\label{eq:mesa-eta-M-scaling}
    \partial_{L_\pm}\etastat \propto \mp \exp\left(-\frac{2 L_\pm}{\ell_\pm}\right)
    ,
\end{equation}
where $\ell_\pm$ describes the length scale of the exponential approach of the upper or lower plateau density.
Because a change ${\delta M \approx (\rho_+-\rho_-)\delta L}$ in the mesa mass is proportional to a length change $\delta L$ of the plateaus, $\etastat$ also changes only exponentially slowly with the mesa mass $M$.

For peak patterns the peak amplitude increases with the peak mass $M$ [see Fig.~\ref{fig:MC-stat-construction}(f)].
Thus, the stationary mass-redistribution potential $\etastat$ has to decrease to balance the total turnover [see Fig.~\ref{fig:MC-stat-construction}(e,g) and Ref.~\cite{Brauns.etal2021}], and we obtain
\begin{equation}
\label{eq:peak-eta-M-scaling}
    \eta_\mathrm{stat}(M)
    \sim 
    M^{-\alpha}
    \, ,
\end{equation}
if the reaction term $\tilde{f}$ [and thus the nullcline $\eta^*(\rho)$] is of power-law form at large densities $\rho$ (see the scaling analysis in Appendix~\ref{app:peak-scaling}).

\subsection{The effect of weak source terms}
\label{sec:stat-patterns-nMC}
The profile of the stationary pattern $[\rhostateps(x),\etastateps(x)]$ under the influence of weak source terms (${0 < \varepsilon \ll 1}$) shows only small deviations (of order ${\sim\varepsilon}$) from the stationary profile $[\rhostat(x),\etastat]$ of the mass-conserving system (${\varepsilon=0}$).

Importantly, as for the homogeneous steady state, the average total density $\bar{\rho}$ is no longer a control parameter of the system with source terms as it is not set by the initial condition.
Instead, the average density evolves as [integrate Eq.~\eqref{eq:cont-eq}]
\begin{equation}\label{eq:dens-evol-source-terms}
    \partial_t \bar{\rho} 
    = \varepsilon\frac{2}{\Lambda}\int_0^\frac{\Lambda}{2}\mathrm{d}x\,s_\mathrm{tot}(\rho(x),\eta(x))
    \, .
\end{equation}
Thus, the average density $\bar{\rho}$, that is, the `size' of the stationary peak or mesa adapts until the overall production and degradation of molecules in the system are balanced.
The mass of the stationary pattern is thus fixed by the source balance condition ${\partial_t\bar{\rho} = 0}$.

In summary, we have shown that (nearly) mass-conserving 2cRD systems form peak or mesa patterns.
The type of pattern is determined by the shape of the nullcline $\eta^*(\rho)$.
In the mass-conserving system, the stationary mass-redistribution potential is constant and its value depends on the domain mass $M$.
It decreases exponentially slowly with mass for mesa patterns, and (typically) like a power law for peak patterns.
In both cases, we have ${\partial_M\etastat <0}$.
Weak source terms fix the peak or mesa mass because the stationary pattern size has to balance the overall production and degradation in the system.

\section{Growth rate of the mass-competition instability}
\label{sec:MC-mass-competition}
Having classified the possible patterns, we now study the growth rate $\sigma$ of the mass-competition instability in 2cMcRD systems in terms of the interaction of two neighboring elementary patterns, each approximated by a quasi-steady-state density profile.
This instability underlies the coarsening dynamics (see Sec.~\ref{sec:phenomenology}).
Detailed, systematic derivations of the results presented in the following section can be found in Appendix~\ref{app:MC-mass-competition}.

\subsection{Assumptions and approximations}
\label{sec:approximations}
To determine the growth rates, we build on two assumptions.
First, we employ the sharp-interface approximation (${\ell_\mathrm{int}\ll \Lambda}$) which is justified for well-separated peaks and interfaces (see section~\ref{sec:stat-patterns}).
It allows the use of singular perturbation theory to determine growth rates of perturbations around stationary patterns in the fully nonlinear regime.
Second, we assume that mass competition is slow compared to the relaxation of single peaks and mesas toward their stationary profile.
With this separation of timescales, we account for the observation that---during the coarsening process---the pattern profiles of single peaks or mesas are well approximated by the stationary profiles (fast `regional' relaxation) [see Fig.~\ref{fig:coarsening-phenomenology}(b,d)].
Thus, we use a QSS approximation for the density profile of single peaks and interfaces but not for the value of $\eta$ in the peak or interface region (as used in Ref.~\cite{Brauns.etal2021}).
As a result of the QSS approximation, mass competition is described in terms of a few \emph{collective variables}: peak masses and positions or interface positions.

Quantitatively the timescale separation between local relaxation and mass redistribution can be analyzed using singular perturbation theory.
Specifically, in Appendix~\ref{app:stability}, we derive the relaxation rates of the relaxation modes of the elementary pattern that redistribute mass from the peak or interface region to the plateau regions.\footnote{
Other relaxation modes exist that locally deform the peak or interface profile.
These modes are not accessible via singular perturbation methods where one assumes the peak or interface as infinitely sharp.
Within the sharp-interface approximation, however, we expect these modes to relax rapidly because mass redistribution within the narrow interface or peak regions will be fast.
Also, we find numerically that those modes are relaxed during the coarsening process [see Fig.~\ref{fig:coarsening-phenomenology}(b,d)].}
These growth rates describe how fast a single peak or mesa relaxes toward its steady state profile if mass is added (or removed) from its plateaus.
If this plateau relaxation is fast compared to mass redistribution between different peaks or mesa interfaces (cf.\ Fig.~\ref{fig:coarsening-modes}),
we can assume that both the pattern plateaus and the peak or mesa mass $M$ are fully relaxed during the competition process (timescale separation, see also Appendix~\ref{app:MC-mass-competition}).
The detailed analysis shows that local relaxation is fast compared to the mass-competition process if the following condition is satisfied
\begin{equation} \label{eq:small-plateau-mass}
    \frac{1}{\Lambda \,  \partial_{\etastat}\rho_\pm} 
    \gg 
    \left|\partial_M^{}\etastat\right|
    \, ,
\end{equation}
where $\rho_\pm$ denotes the densities in the high- and low-density plateau.
Heuristically this condition can be understood as a consequence of the fact that the strength of changes in the mass-redistribution potential $\eta$ determines how fast mass redistribution proceeds.
Consider adding a small amount of mass $\delta M$ in a plateau [l.h.s.\ of Eq.~\eqref{eq:small-plateau-mass}] or a peak/mesa [r.h.s.\ of Eq.~\eqref{eq:small-plateau-mass}].
Within the sharp-interface approximation, additional mass in the plateaus (of length~${\approx \Lambda}$) leads to the change ${\delta\etastat \approx  \partial_{\rho_\pm}\eta^* \, \delta M/\Lambda}$ of the (stationary) mass-redistribution potential. 
Using the implicit function theorem this can be written as ${\delta\etastat \approx \delta M /(\Lambda \partial_\etastat\rho_\pm)}$.
Similarly, additional mass in a domain induces a change ${\delta\etastat\approx (\partial_M\etastat) \, \delta M}$.
Thus, the condition Eq.~\eqref{eq:small-plateau-mass} ensures that the $\eta$ gradients induced by redistribution of mass into the pattern plateaus are large compared to the gradients arising during mass competition between neighboring domains. 

Importantly, the properties of stationary patterns in 2cMcRD systems ensure that the condition Eq.~\eqref{eq:small-plateau-mass} is generically fulfilled for sufficiently large peaks and mesas because $\partial_M^{}\etastat$ is strongly suppressed at large peak and mesa masses ${M = \Lambda \left(\bar{\rho}-\rho_-\right)}$ [see Eqs.~\eqref{eq:mesa-eta-M-scaling},~\eqref{eq:peak-eta-M-scaling}].

Conveniently, the condition Eq.~\eqref{eq:small-plateau-mass} not only allows us to neglect the time evolution of the pattern plateaus and assume these as fully relaxed to a (quasi-)steady state, but we can also neglect the total mass stored in the background plateau $\rho_-$ during the mass-competition process.
Specifically, the total mass of an elementary pattern changes with the peak or mesa mass $M$, that is, with the stationary mass-redistribution potential $\etastat$ by [see Eq.~\eqref{eq:mesa-peak-mass}]
\begin{equation}
    \partial_\etastat \int_0^\frac{\Lambda}{2}
    \! \mathrm{d}x\, \rhostat 
    =
    \partial_\etastat
    \big(
    \Lambda  \rho_- 
    \,{+}\, 
    M
    \big)
    \approx\partial_\etastat M
    \, ,
\end{equation}
where the last estimate follows by inverting the above condition Eq.~\eqref{eq:small-plateau-mass}.
Therefore, redistribution of mass into the background plateau $\rho_-$ does not play a significant role in the mass-competition dynamics if condition Eq.~\eqref{eq:small-plateau-mass} is fulfilled.
The same simplification arises in LSW theory for coarsening in phase-separating systems \cite{Lifshitz.Slyozov1961}.

\subsection{Instability growth rates for competition between two peaks/mesas}
\label{sec:mc-growth-rates}
\begin{figure}
    \centering
    \includegraphics{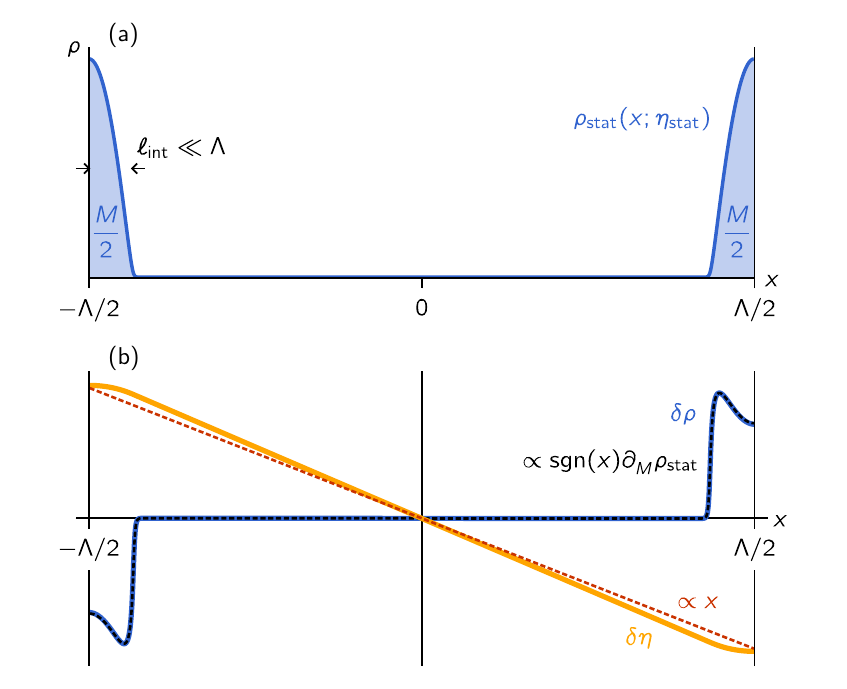}
    \caption{
    The mass-competition instability of two half peaks.
    (a) The stationary pattern $[\rhostat(x),\etastat]$ to analyze peak competition is symmetric and composed of two equally sized half peaks of mass $M/2$ each [cf.\ Fig.~\ref{fig:coarsening-modes}(a) bottom].
    (b) The antisymmetric mass-competition eigenmode [blue (dark gray), orange (light gray)] destabilizes this symmetric stationary state.
    The density profile $\delta\rho$ [blue (dark gray)] is localized to the peak regions.
    In these regions, it is well approximated by the `mass mode' $\partial_M^{}\rhostat(x)$ (black dashed).
    In the sharp interface limit, the mass-redistribution potential $\delta\eta$ [orange (light gray), strongly magnified] is approximately linear (red dashed).
    The linear stability analysis is exemplified in this figure for the peak-forming model ${\tilde{f} = \eta - 10 \rho/(1+\rho^2)}$ with parameters ${D_u = 1}$, ${D_v = 10}$, ${\bar{\rho} = 4}$, and ${\Lambda=200}$ (see Appendix~\ref{app:const-reac-rate-peak-model}).
    }
    \label{fig:peak-MC-modeApprox}
\end{figure}

Applying the approximations discussed in the previous section, we now explain the processes that underlie the mass-competition instability.
We find that the mass-competition process is composed of three substeps: Particle release at one peak or mesa, diffusive mass transport through the cytosol, and particle incorporation into the other domain.
The timescale of mass competition is shown to be the sum of the timescale of reactive conversion (release and incorporation) and mass transport.
Depending on which of the two contributions is rate-limiting, we find, alike to the instability of the HSS (see Sec.~\ref{sec:HSS-instability}), a diffusion- and a reaction-limited regime which correspond to CH and cAC dynamics, respectively.

\textit{The mass-competition process.\;---}
For all coarsening scenarios (see Fig.~\ref{fig:coarsening-modes}), mass competition follows the same principles.
In essence, a gradient $\delta\eta(x)$ induced in the mass-redistribution potential between the two competing domains drives the mass-competition instability:
A mass increase in one peak/mesa reduces the mass-redistribution potential there [cf.\ Eqs.~\eqref{eq:mesa-eta-M-scaling},~\eqref{eq:peak-eta-M-scaling}], and the resulting gradient leads to mass redistribution that enhances the initial perturbation \cite{Brauns.etal2021}.
For concreteness, we discuss the competition between peaks (see Fig.~\ref{fig:peak-MC-modeApprox}).
The mathematical analysis of the eigenmode corresponding to peak competition [see Fig.~\ref{fig:peak-MC-modeApprox}(b)] is given in Appendix~\ref{app:MC-mass-competition}.

\begin{figure}
    \centering
    \includegraphics{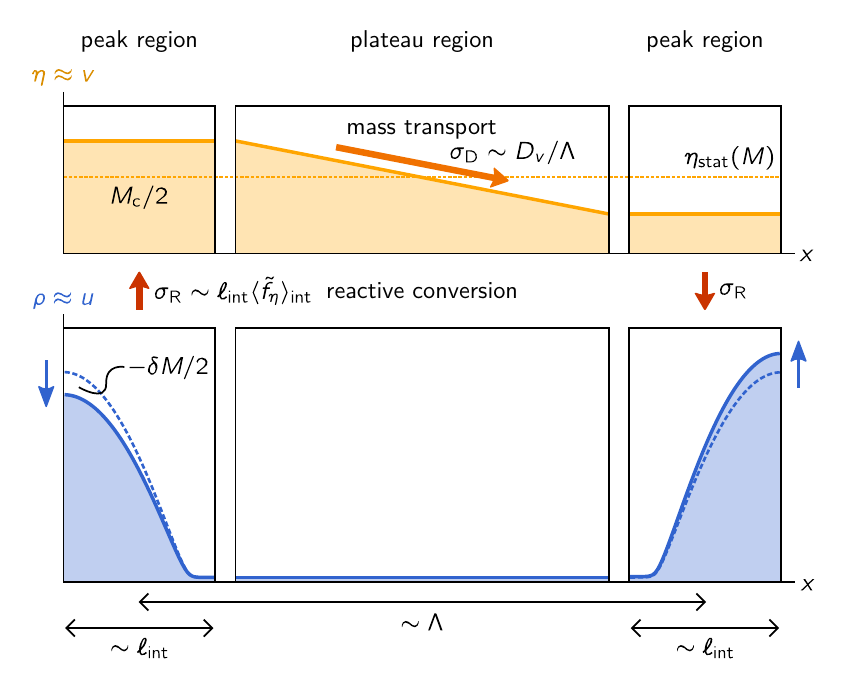}
    \caption{The three steps of mass competition between two peaks (similarly for mesas) in the limit ${D_v\gg D_u}$.
    In this limit, particles accumulate into sharp peaks on the membrane while the cytosolic concentration is nearly uniform at each peak due to the much faster cytosolic diffusion.
    Therefore, most of the peak mass $M$ accumulates in the membrane species $u$, and one has ${\rho\approx u}$.
    Moreover, it holds ${\eta= v + d u\approx v}$.
    During mass competition, chemical reactions convert particles from the slow into the fast species at the smaller peak, where the half-peak mass is reduced by $-\delta M/2$ [particle release; red (dark gray) arrow].
    The resulting change in the cytosolic density yields a gradient that transports these particles toward the other peak [orange (light gray) arrow] where they are reactively re-converted to the slow species and increase the peak mass of the larger peak [particle incorporation; red (bold, dark gray) arrow]. 
    The rate of release and incorporation is given by $\sigma_\mathrm{R}$ while the rate of mass transport between the peaks is given by $\sigma_\mathrm{D}$.
    }
    \label{fig:sigma-conceptual}
\end{figure}

To understand the competition process in detail, we focus on the (biologically relevant) limit ${D_u\ll D_v}$.
In this limiting case, the pattern mainly forms in the slow-diffusing (membrane-bound) species $u$ while, in comparison, the density profile in the species $v$ is smoothed out by fast diffusion.
At the same time, redistribution of mass between the peaks mainly proceeds through the fast-diffusing species $v$; note that  ${\partial_x\eta\approx \partial_x v}$ in the plateau between the peaks.
As a consequence of these different roles of $u$ and $v$ in the dynamics, redistribution of mass from smaller into larger peaks, i.e., the mass competition process driving coarsening, must involve three subsequent steps illustrated in Fig.~\ref{fig:sigma-conceptual}:
First, particles at the smaller peak detach from the membrane into the cytosol (``particle release'', left part of Fig.~\ref{fig:sigma-conceptual}).
Second, these additional cytosolic particles create a gradient ${\delta v(x)\approx\delta\eta(x)}$ in the fast-diffusing species which causes net diffusion of particles toward the other, larger peak (top middle of Fig.~\ref{fig:sigma-conceptual}).
In a third step, chemical reactions revert the particles back into the slow state $u$, thus letting the larger peak grow (``particle incorporation'', the opposite of the release process at the first peak; right part of Fig.~\ref{fig:sigma-conceptual}).
In the next paragraphs, we explain why particles are released from the smaller and incorporated into the larger peak.
This leads to the criterion for coarsening given in Eq.~(4) in Ref.~\cite{Brauns.etal2021}.

Consider the weakly perturbed initial configuration where, starting from a symmetric configuration of equal masses, a small amount of mass $\delta M/2$ is redistributed from the left to the right half peak.
This creates a mass imbalance $\delta M$ between the two half peaks
(see Fig.~\ref{fig:sigma-conceptual}).\footnote{
Note that the peak mass $M$ as defined in Eq.~\eqref{eq:mesa-peak-mass} decreases/increases by $\pm\delta M$ at the left/right peak because we analyze only half peaks.}
This mass imbalance implies that reactive turnover balance, Eq.~\eqref{eq:reactive-turnover-balance}, can no longer be fulfilled for the same cytosolic densities at the two peaks.
At the smaller peak, the stationary cytosolic concentration ${v\approx\etastat(M-\delta M)}$ increases compared to the uniform concentration ${v\approx\etastat(M)}$ of the symmetric steady state [${\partial_M\etastat<0}$, see Eqs.~\eqref{eq:mesa-eta-M-scaling},~\eqref{eq:peak-eta-M-scaling}].
Thus, the local relaxation of the smaller peak toward the shifted stationary state corresponding to its reduced peak mass drives the detachment of particles from the membrane to increase the local cytosolic density toward ${v\approx\etastat(M-\delta M)>\etastat(M)}$.
The same but reverse process occurs at the larger peak: Particles attach onto the membrane and deplete the cytosolic density toward ${v\approx\etastat(M+\delta M)<\etastat(M)}$.
The resulting cytosolic gradient couples both processes by diffusive transport of the released particles from the smaller toward the larger peak, inducing the positive feedback that drives the mass-competition instability.
Taken together, the characteristic timescale $1/\sigma$ of the mass-competition instability is given by the sum of the timescales of reactive conversion $1/\sigma_\mathrm{R}$ at the peaks (particle release/incorporation) and diffusive redistribution $1/\sigma_\mathrm{D}$ between the peaks.
Next, we determine the rates $\sigma_\mathrm{R}$ and $\sigma_\mathrm{D}$.

In the limit ${D_v\to\infty}$, diffusive transport in the cytosol becomes instantaneous and the only dynamic processes are the release and incorporation processes (${1/\sigma_\mathrm{D}\to 0}$).
The dynamics of release and incorporation follows from integration of Eq.~\eqref{eq:2cRD-u} (${\varepsilon=0}$) over, say, the left half peak $[-\Lambda/2,0]$.
One can use that the gradients of the $u$ profile in the plateau are small and that the peak mainly forms on the membrane such that $-\partial_t\delta M/2\approx\partial_t\int_{-\Lambda/2}^{0}\mathrm{d}x\, u$.
The factor 1/2 accounts for the fact that we consider only a half-peak.
It follows that [cf.\ reactive turnover balance, Eq.~\eqref{eq:reactive-turnover-balance}, and Fig.~\ref{fig:sigma-conceptual}]
\begin{equation}
    \partial_t \frac{-\delta M}{2} 
    \approx \int_{-\frac{\Lambda}{2}}^{0}\mathrm{d}x\, \tilde{f}(\rho(x),\etastat(M))
    \, .
\end{equation}
The value of the mass-redistribution potential stays constant at its unperturbed value $\etastat(M)$ of the symmetric steady state because the cytosolic density is uniform in the limit of instantaneous redistribution and any inflow of particles at the smaller peak must be balanced by an outflow of particles at the larger peak (antisymmetry of the eigenmode associated with mass competition, see Fig.~\ref{fig:peak-MC-modeApprox}).
The reactive turnover integral (right-hand side) can be evaluated approximately using that the plateau regions of the pattern rapidly relax toward reactive equilibrium ${\tilde{f}\approx 0}$ (see Secs.~\ref{sec:stat-patterns},~\ref{sec:approximations}). 
Thus, the integral is dominated by contributions from the peak, and one can approximate the integral by the product of half the peak width $\ell_\mathrm{int}$ and the (linearized) average net reactive flux ${\bar{f}_\mathrm{peak}(\delta M)= \langle\tilde{f}_\eta\cdot [\etastat(M)-\etastat(M{-}\delta M)]\rangle_\mathrm{int}}$, introducing the shorthand notation $\tilde{f}_\eta = \partial_\eta\tilde{f}$. Taken together, this yields
\begin{align}
    \partial_t \frac{-\delta M}{2} 
    &\approx \ell_\mathrm{int}\langle\tilde{f}_\eta\rangle_\mathrm{int} \Big[
    \etastat(M)-\etastat(M{-}\delta M)
    \Big]
    \nonumber\\
    &= \ell_\mathrm{int}\langle\tilde{f}_\eta\rangle_\mathrm{int} \, \partial_M^{}
    \etastat \, \delta M
    \, ,
\label{eq:incorp-release-heuristic}
\end{align}
where the second line applies for the linear regime where $\delta M$ is small.
This gives the following estimate for the rate $\sigma_\mathrm{R}$ of the release and incorporation processes
\begin{equation}\label{eq:sigma-R-heuristic}
    \sigma_\mathrm{R} 
    \approx 
    -2 \ell_\mathrm{int}
    \langle\tilde{f}_\eta\rangle_\mathrm{int}
    \partial_M^{} \etastat
    \, .
\end{equation}
The mathematical analysis given in Appendix~\ref{app:MC-mass-competition} yields explicit expressions for the `half-peak' width or interface width $\ell_\mathrm{int}$ as well as the average conversion rate $\langle \tilde{f}_\eta\rangle_\mathrm{int}$ [see Eqs.~\eqref{eq:interface-average},~\eqref{eq:interface-width}].
The resulting expression for $\sigma_\mathrm{R}$ is exact in the sense of a singular-perturbation analysis.

If the diffusive transport between the peaks is not instantaneous, its rate $\sigma_\mathrm{D}$ is determined by the strength of the gradient $\delta\eta(x)$ resulting from the shifts in the cytosolic density at the peaks.
In the limit of fast reactive conversion ${1/\sigma_\mathrm{R}\to 0}$, the only dynamic process is the diffusive mass transport between the peaks, and the cytosolic density at the peaks directly relaxes toward its equilibrium values ${\etastat(M\pm\delta M)}$ (QSS approximation).
Thus, the mass transport between the peaks is approximately given by [integrate the continuity equation, Eq.~\eqref{eq:cont-eq} (${\varepsilon =0}$), over the domain half $[-\Lambda/2,0]$]
\begin{align}
    \partial_t\frac{-\delta M}{2} &\approx D_v \frac{\etastat(M{+}\delta M)-\etastat(M{-}\delta M)}{\Lambda}\nonumber\\
    &\approx \frac{2 D_v}{\Lambda}(\partial_M^{}\etastat) \delta M
    \, ,
\label{eq:mass-transport-heuristic}
\end{align}
where the second approximation again holds in the linear regime for small mass differences $\delta M$.
The rate $\sigma_\mathrm{D}$ of diffusive mass transport between the peaks can then be read off as
\begin{equation}
    \sigma_\mathrm{D} \approx -\frac{4 D_v}{\Lambda} \, \partial_M^{}\etastat
    \, .
\end{equation}
Taken together, the above analysis explains that mass competition proceeds in three subsequent steps.
Its characteristic timescale is the sum of the characteristic timescales of the ``reactive-conversion steps'', $1/\sigma_\mathrm{R}$, (particle release and incorporation) and diffusive mass transport, $1/\sigma_\mathrm{D}$, between the competing domains.

\textit{The rates of the coarsening modes.\;---}
Indeed, the singular-perturbation analysis given in Appendix~\ref{app:MC-mass-competition} shows that the growth rate $\sigma$ of the mass-competition instability in peak and mesa patterns has the structure
\begin{equation}
\label{eq:sigma-MC-massComp}
    \frac{1}{\sigma} 
    = 
    \frac{1}{\sigma_\mathrm{D}}
    +
    \frac{1}{\sigma_\mathrm{R}}
    \, .
\end{equation}
Please note that the mathematical derivation of this relationship and the following expressions for each individual rate does not use the assumption ${D_u\ll D_v}$ we have used above to simplify the interpretation (see Appendix~\ref{app:MC-mass-competition}).

For peak competition (cf.\ Fig.~\ref{fig:coarsening-modes}), the singular perturbation analysis yields the rates expected from the above heuristic arguments (see Appendix~\ref{app:MC-mass-competition})
\begin{subequations}\label{eq:sigma-MC-peak-D-R}
\begin{align}
    \sigma_\mathrm{D}  &= - \frac{4 D_v}{\Lambda} \, \partial_{M}^{} \etastat
    \, , \label{eq:sigma-MC-peak-D}\\
    \sigma_\mathrm{R} &= - \frac{2\ell_\mathrm{int}\langle \tilde{f}_\eta\rangle_\mathrm{int}}{1+d} \, \partial_{M}^{} \etastat
    \, .
\end{align}
\end{subequations}
For the two scenarios of mesa competition and coalescence, we obtain (see Appendix~\ref{app:MC-mass-competition})
\begin{subequations}\label{eq:sigma-MC-mesa-D-R}
\begin{align}
    \sigma_\mathrm{D}^\pm &= - \frac{4 D_v}{\xi_\mp \Lambda} \, \partial_{M}^\pm \etastat
    \, , \label{eq:sigma-MC-mesa-D}\\
    \sigma_\mathrm{R}^\pm 
    &= 
    - \frac{2\ell_\mathrm{int}\langle \tilde{f}_\eta\rangle_\mathrm{int}}{1+d} \, 
    \, \partial_{M}^\pm \etastat
    \, ,
\end{align}
\end{subequations}
where ${\xi_\pm = 2 L_\pm/\Lambda}$ are the relative plateau widths. 
The superscript $+$ indicates the mode driven by competition of the high-density plateaus [mesa competition; Fig.~\ref{fig:coarsening-modes}(a)], while $-$ indicates the mode driven by competition of the low-density plateaus [mesa coalescence; Fig.~\ref{fig:coarsening-modes}(b)].
We introduced the notation ${\partial_M^\pm\etastat = \pm \frac{1}{2(\rho_+-\rho_-)}\partial_{L_\pm}\etastat}$ which describes the change $\delta\etastat$ of the stationary mass-redistribution potential only due to the change in length ${\delta L_\pm = \pm \delta M /(\rho_+-\rho_-)}$ of the high-density (low-density) plateau, respectively.
The rates for mesa competition and coalescence depend on the distinct, modified derivatives $\partial_M^\pm\etastat$ since for the competition (coalescence) mode only the high-density (low-density) plateaus change in length, respectively (see Fig.~\ref{fig:coarsening-modes}).
The other plateau in each scenario merely shifts as a whole and does not change in length during the mass-competition process.
Because peak coalescence also depends on the length change of the lower plateaus, it is analogous to the case of mesa coalescence and is described by the same rate $\sigma^-_\mathrm{R,D}$.\footnote{
For peak coalescence, the rate expressions $\sigma_\mathrm{D,R}^-$ only hold qualitatively because the gradient $\delta\eta(x)$ between the competing (low-density) domains lies within the peak and not within a plateau.
Therefore, it cannot simply be approximated by a linear gradient within the sharp-interface limit (cf.\ Appendix~\ref{app:MC-mass-competition} and Fig.~\ref{fig:mesa-MC-modeApprox}).}

\textit{The condition for uninterrupted coarsening.\;---}
In the above heuristic discussion of the mass-competition process, we have argued that the mass-competition instability is driven by the \emph{decrease} of the stationary mass-redistribution potential $\etastat(M)$ as a function of the domain mass $M$. Indeed, both rates $\sigma_\mathrm{D}$ and $\sigma_\mathrm{R}$ are proportional to the derivative $\partial_M^{}\etastat$. Thus, the singular perturbation analysis recovers the condition that uninterrupted coarsening occurs if ${\partial_M^{}\etastat < 0}$ holds for all stationary peaks that are stable as isolated elementary patterns.
Similarly, mesa patterns undergo uninterrupted coarsening if ${\partial_M^+\etastat < 0}$ or ${\partial_M^-\etastat < 0}$ if fulfilled for all stable stationary mesas.
Then, these stable stationary domains are all destabilized by mass exchange with neighboring domains:
The growth rate $\sigma$ is always positive and any mass difference $\delta M$ between two domains grows exponentially by $\partial_t\delta M =\sigma \delta M$, i.e., one has $\delta M \sim \operatorname{e}^{\sigma t}$.
This stability condition was previously derived mathematically  \cite{Otsuji.etal2007,Ishihara.etal2007}.
In Ref.~\cite{Brauns.etal2021} it is shown that the criterion ${\partial_M^{}\etastat < 0}$ (and ${\partial_M^\pm\etastat < 0}$) is generically fulfilled in strictly mass-conserving two-component reaction--diffusion systems, implying that these systems always exhibit uninterrupted coarsening [see also Eqs.~\eqref{eq:mesa-eta-M-scaling},~\eqref{eq:peak-eta-M-scaling} Sec.~\ref{sec:stat-patterns}].
In mass-conserving systems with more than two components, it is possible for $\partial_M \etastat$ to become positive, resulting in interrupted coarsening.
For example, interrupted coarsening is observed for specific three-component mass-conserving systems where the total density of the three components is conserved \cite{Murray.Sourjik2017,Jacobs.etal2019,Chiou.etal2021,Gai.etal2020}.
For two-component systems, the coarsening processes can be brought to a halt by introducing weak production and degradation terms that break mass conservation.
This latter scenario may serve as a prototype for the analysis of interrupted coarsening in more-component models and will be presented in Sec.~\ref{sec:nMC-mass-competition} below.

\textit{Diffusion-~and~reaction-limited~coarsening.\;---}
The growth rate of the mass-competition instability exhibits two distinct limits, depending on whether the diffusive mass transport or the reactive conversion step is rate-limiting.
In the first case we find ${\sigma\to \sigma_\mathrm{D}}$ while in the second case we have ${\sigma \to \sigma_\mathrm{R}}$.
Therefore, we term $\sigma_\mathrm{D}$ the \emph{diffusion-limited} and $\sigma_\mathrm{R}$ the \emph{reaction-limited} growth rate (cf.\ Ref.~\cite{Brauns.etal2021}).
We have actually applied these two limiting cases in the above heuristic discussion to determine the two growth rates $\sigma_\mathrm{D,R}$.
From the explicit expression for the growth rates $\sigma$, $\sigma_\mathrm{D}$, and $\sigma_\mathrm{R}$, Eqs.~\eqref{eq:sigma-MC-massComp}--\eqref{eq:sigma-MC-mesa-D-R}, one infers that the crossover between these regimes is located at
\begin{equation}\label{eq:diff-reac-crossover}
    \frac{D_v}{\Lambda} 
    \sim 
    \ell_\mathrm{int}\langle \tilde{f}_\eta\rangle_\mathrm{int}
    \, ,
\end{equation}
i.e., where the timescale of diffusive redistribution between the peaks/mesas, $1/\sigma_\mathrm{D}$, is comparable to the timescale of reactive conversion at the peak or mesa interface, $1/\sigma_\mathrm{R}$.

\subsubsection{The diffusion-limited regime}
In the diffusion-limited regime $\sigma\to \sigma_\mathrm{D}$ one recovers the result Eq.~(3) from Ref.~\cite{Brauns.etal2021}.\footnote{
The expressions differ by a factor of $2$ because here we consider the competition of two half peaks or half mesas.}
There, the mass-redistribution potential $\eta$ at each peak (or mesa) was approximated to be in a quasi-steady state (QSS), i.e., ${\eta|_\mathrm{peaks}\approx \etastat(M\mp\delta M)}$.
This assumption holds in the diffusion-limited regime considered here because the reactive conversion at the peaks and the ensuing relaxation of $\eta$ toward its QSS are fast.
In this regime, the growth rate of the mass-competition instability is determined solely by the amplitude of the gradients in $\eta$ that are induced by a mass difference between the peaks/mesas.
During coarsening, this regime will always be reached at late times because the peak or interface separation increases further and further, and mass redistribution becomes slower over larger distances.
Hence, the asymptotic long-time behavior of the coarsening process only depends on $\sigma_\mathrm{D}$.

The diffusion-limited regime manifests itself in the generalized CH equation, in which the mass-redistribution potential $\eta$ does not have a time evolution of its own, but instantly adapts to the density profile $\rho$ (see Eq.~\eqref{eq:gen-CH-eta}).
Consequently, the only dynamic process is mass redistribution, and $\sigma_\mathrm{D}$ gives the total growth rate of the mass competition instability in the generalized CH equation.
The derivation of the growth rate $\sigma=\sigma_\mathrm{D}$ of the mass-competition instability in the generalized CH equation proceeds analogously to Appendix~\ref{app:MC-mass-competition}.

\subsubsection{The reaction-limited regime}
If, in contrast, diffusive mass redistribution between domains is fast compared to the local reactive conversion, we have ${\sigma\to \sigma_\mathrm{R}}$.
Due to fast mass redistribution through the cytosol, the species $v$ acts as a global pool for mass exchange and the coarsening rate is set by the release of mass into the pool by the shrinking peak/mesa and the intake of mass from the pool by the growing peak/mesa.

The reaction-limited regime is reached in the shadow limit ${D_v\to\infty}$ such that $\sigma_\mathrm{R}$ is the growth rate expression found in cAC systems.
In contrast to CH equations, which describe the dynamics based solely on mass redistribution, cAC systems describe only the reactive dynamics of peak growth and shrinking (``interface-controlled kinetics'').

\section{Scaling analysis of the coarsening dynamics}
\label{sec:coarsening-scaling}
Having calculated the growth rate $\sigma$ of the mass-competition instability, we now apply a scaling argument \cite{Langer1971,Glasner.Witelski2003,Brauns.etal2021} to argue how the coarsening law, i.e., the time evolution of the average wavelength $\langle \Lambda \rangle(t)$ of the pattern on a large domain, can be obtained from $\sigma$.
For specificity, we again focus on peak patterns [cf.\ Fig.~\ref{fig:coarsening-phenomenology}(a)].

An extended pattern consisting of many peaks can, in principle, show both peak competition and peak coalescence.
Also, mass can be exchanged not only between nearest neighbor peaks but also between peaks that are further apart.
Out of these coarsening scenarios the fastest competition (or coalescence) process will lead to the fastest collapse of domains and thus drive the coarsening process.
For peak patterns, we can focus on the mass-competition instability between two neighboring peaks.\footnote{
Mass competition is faster over shorter distances.
Moreover, in the shadow limit (cAC system) mass exchange is instantaneously mediated via a ``global pool''.
Thus, the competition rate is (approximately) independent of the spatial arrangement of the peaks.
Refer to Appendix~\ref{app:MC-mass-competition} for a detailed discussion.}
The coalescence mode can be neglected because its rate ${\sigma\sim\partial_M^-\etastat}$ is exponentially suppressed with the plateau length.
The change $\partial_M^{}\etastat$ of the stationary mass-redistribution potential induced by the height change of the peaks---which induces mass transport in the peak-competition scenario---falls off much more slowly [cf.\ Eq.~\eqref{eq:peak-eta-M-scaling}, in agreement with the observed phenomenology [cf.\ Fig.~\ref{fig:coarsening-phenomenology}(a)].

To find the time evolution of the average peak separation $\langle\Lambda\rangle(t)$, we need to determine how fast the total number of peaks $N(t)$ changes.
For this, we first determine the characteristic collapse timescale $t_\mathrm{col}$ for the smaller of two neighboring peaks with masses $M_1$ and $M_2$.
Let us denote the average mass by ${\bar{M} = (M_1 + M_2)/2}$ and the mass difference by ${2\, \delta M = M_1 - M_2}$ (assuming ${\delta M>0}$).
The symmetric state ${M_1=M_2}$ is a stationary state, and we approximate the time evolution of the mass difference $\delta M$ by the linear linearized growth law for small mass differences.
Thus, we have
\begin{equation}
    \partial_t \delta M 
    \approx \sigma(\bar{M},\Lambda) \delta M
    \, ,
\end{equation}
where the growth rate $\sigma$ of the mass-competition instability depends on the average mass $\bar{M}$ and the spatial separation $\Lambda$ of the peaks [see Eqs.~\eqref{eq:sigma-MC-peak-D-R}].
One can then estimate the collapse timescale $t_\mathrm{col}$ as the time until all the mass is transferred from the smaller to the larger peak, i.e., ${0 \approx \bar{M} - \delta M \operatorname{e}^{\sigma(\bar{M},\Lambda)\, t_\mathrm{col}}}$.
This yields
\begin{equation}
    t_\mathrm{col} \approx \frac{\log\left(\bar{M}/\delta M\right)}{\sigma(\bar{M},\Lambda)}
    \, .
\end{equation}
Thus, the collapse timescale depends on on the relative size of the initial mass difference $\delta M/\bar{M}$ and the rate $\sigma$ of mass-competition.
On a large domain showing many simultaneous collapse events, the magnitude of the mass differences $\delta M$ of the different, competing pairs of peaks can be estimated by the standard deviation of the peak-size distribution.
Within the scaling hypothesis common to coarsening~\cite{Lifshitz.Slyozov1961,Wagner1961,Bray2002}, the shape of the peak-size distribution is invariant under the coarsening dynamics and only scales with the average peak mass.
Under this scaling hypothesis, one can thus assume that $\log\left(\bar{M}/\delta M\right)$ is constant on average during the coarsening process.
Moreover, the scaling hypothesis suggests that the initial mass $\bar{M}$ and the peak separation $\Lambda$, which enter the rate of mass competition $\sigma$ as arguments, scale with the average peak mass $\langle M\rangle$ and average peak separation $\langle\Lambda\rangle$, respectively.
Consequently, the average rate of a peak collapse is proportional to the rate of mass competition ${\sim\sigma(\mu\langle M\rangle(t), \nu \langle\Lambda\rangle(t))}$ during the coarsening process, where
$\mu$ and $\nu$ are scaling amplitudes.

Using that the average collapse rate per peak is determined by $\sigma$, the total number of peaks $N(t)$ in a large system of many peaks evolves as
\begin{equation}
    \partial_t N 
    \sim 
    -
    \sigma
    \big(
    \mu \langle M\rangle (t), \nu \langle \Lambda \rangle(t)
    \big) \, 
    N
    \, .
\end{equation}
We are still missing the connection between the total peak number $N(t)$, the average peak mass $\langle M\rangle (t)$, and the average peak separation $\langle\Lambda\rangle (t)$ to write this evolution equation in a closed form.
This relationship is given by the mass-conservation constraint:
The total mass in the system has to be distributed between the peaks and the background plateau $\rho_-$ from which the peaks rise.
This gives the scaling ${\langle M \rangle \approx \left(\bar{\rho}-\rho_-\right)\langle\Lambda\rangle \sim N^{-1}}$.
For simplicity, we set ${\rho_-\approx 0}$.
\footnote{At late times during coarsening, we can neglect any changes of the mass in the low-density plateau from which the peaks rise (see Sec.~\ref{sec:approximations}).
}
With this mass-conservation constraint, we arrive at the closed evolution equation for the typical pattern length scale $\langle\Lambda\rangle(t)$,
\begin{equation}
\label{eq:dyn-length-scale-eq}
    \partial_t \langle \Lambda\rangle \sim \sigma
    \big(\mu\bar{\rho}\langle\Lambda\rangle,\nu\langle\Lambda\rangle\big)\,\langle\Lambda\rangle
    \, .
\end{equation}
This relation shows that the functional form of the growth rate $\sigma$ of the mass-competition instability completely determines the coarsening law.

Finally, one can reduce the relationship between $\sigma$ and the length-scale evolution $\langle\Lambda\rangle(t)$ given in Eq.~\eqref{eq:dyn-length-scale-eq} to an algebraic relation if the collapse rate $\sigma$ decreases sufficiently strongly as the peaks grow during the coarsening process [cf.\ Eqs.~\eqref{eq:mesa-eta-M-scaling},~\eqref{eq:peak-eta-M-scaling}].
If the decrease in the collapse rate is sufficiently strong, the growth of the average length scale is limited by the duration of the last collapse events.
Then, the time $t$ which is necessary to increase the average peak separation to $\langle\Lambda\rangle (t)$, scales with the duration of the peak collapses at size ${\langle M\rangle (t) \approx (\bar{\rho}-\rho_-)\langle\Lambda\rangle (t)}$, and one finds the coarsening law by inversion of the scaling relation
\begin{equation}
    t 
    \sim 
    \sigma
    \big(
    \mu\bar{\rho}\langle\Lambda\rangle (t),
    \nu\langle\Lambda\rangle (t)
    \big)^{-1}
    \, .
\end{equation}
This asymptotic scaling can be verified explicitly by integration of the dynamic equation, Eq.~\eqref{eq:dyn-length-scale-eq}, for power-law (peak patterns) or exponential suppression (mesa patterns) of $\sigma$ with $\langle\Lambda\rangle$.

In summary, we employed a scaling argument \cite{Langer1971,Glasner.Witelski2003,Brauns.etal2021} to show that the elementary motive of coarsening---mass competition between neighboring domains with rate $\sigma$---determines the macroscopic evolution of the average pattern length scale $\langle\Lambda\rangle(t)$.
In particular, for a one-dimensional system, the scaling analysis results in power-law coarsening for peaks and logarithmically slow coarsening in mesa-forming systems, independent of whether these are CH, cAC or 2cMcRD systems.
The scaling relation is tested numerically in Ref.~\cite{Brauns.etal2021}.
This comparison also shows that the crossover from peak to mesa patterns in systems with highly asymmetric $\mathsf{N}$-shaped nullclines is faithfully predicted \cite{Brauns.etal2021}.

\section{Mass competition in the presence of source terms}
\label{sec:nMC-mass-competition}

Building on the understanding of coarsening in 2cMcRD systems, in Ref.~\cite{Brauns.etal2021} a simple criterion was found to determine the length scale $\Lambda_\mathrm{stop}(\varepsilon)$ of interrupted coarsening in 2cRD systems with source terms of strength ${0<\varepsilon\ll 1}$.
Because the mass-competition instability is weak, and coarsening slow for large peak or mesa masses, one expects that only weak source terms are necessary to suppress the instability and interrupt the coarsening process (cf.\ Sec.~\ref{sec:coarsening-phenomenology}).
To give a basis of interrupted coarsening in 2cRD systems beyond the QSS approximation used in Ref.~\cite{Brauns.etal2021}, we discuss in this section the full growth rate of the mass-competition instability under the influence of weak source terms.
The detailed mathematical analysis using singular perturbation theory can be found in Appendix~\ref{app:nMC-mass-competition}.
We restrict ourselves here to stating the expressions thus obtained and focus on heuristic considerations that explain their mathematical structure.
We then discuss when the simple criterion for coarsening arrest based on the QSS approximation for the mass-redistribution potential $\eta$ in the peak or interface regions applies.

\subsection{The growth rate and stability threshold}
\label{sec:growth-rate-nMC}
To derive the growth rate, we employ the same assumptions as in the mass-conserving case, i.e., we use the sharp-interface approximation and assume that mass competition is slow compared to the local relaxation of elementary stationary patterns (see Sec.~\ref{sec:approximations}).
We analyze the contribution from the weak source terms perturbatively to first order in the source strength $\varepsilon$.
To analyze the regime in which the mass-competition instability competes with the stabilizing effect of the source terms, one chooses $\varepsilon$ such that the stabilization rate $\varepsilon\sigma_\mathrm{S}$ (to be specified below) is of the same order as the mass-competition rate $\sigma$.
For peak patterns, one has to additionally assume that the profile of the stationary mass-redistribution potential is approximately constant at the peak (for details see Appendices~\ref{app:stat-state-nMC},~\ref{app:nMC-mass-competition}).
While in the mass-conserving case $\etastat$ is strictly constant, this is no longer the case under the influence of weak source terms [see Eq.~\eqref{eq:eta-stat} and Appendix~\ref{app:stat-state-nMC}].
Importantly, this assumption does not imply that the mass-redistribution potential $\etastateps$ is uniform \emph{on the scale of the wavelength} $\Lambda$, i.e., that diffusion on the scale of the wavelength is fast compared to the reactive timescales.
That is, we do not assume a well-mixed cytosolic reservoir (shadow limit).

The growth rate of the mass-competition instability under the influence of weak source terms for peak and mesa patterns then has the form (see Appendix~\ref{app:nMC-mass-competition})
\begin{equation}\label{eq:sigma-eps}
    \sigma^\varepsilon \approx \frac{\sigma_\mathrm{R}}{\sigma_\mathrm{D} + \sigma_\mathrm{R}} \, 
    \big[
    \sigma^\varepsilon_\mathrm{D} - \varepsilon \sigma_\mathrm{S}^{}
    \big]
    \, .
\end{equation}
For mesa patterns, $\sigma_\mathrm{D}$ and $\sigma_\mathrm{R}$ have to be replaced by $\sigma^\pm_\mathrm{D}$ and $\sigma^\pm_\mathrm{R}$ that describe mesa competition or mesa coalescence, respectively [cf.\ Eqs.~\eqref{eq:sigma-MC-mesa-D-R}].
The rate $\sigma^\varepsilon$ consists of two contributions:
The first is a generalization of the growth rate $\sigma$ found in mass-conserving systems [note that Eq.~\eqref{eq:sigma-MC-massComp} implies ${\sigma = \sigma_\mathrm{D}\sigma_\mathrm{R}/(\sigma_\mathrm{D}+\sigma_\mathrm{R})}$], while the second, negative term is genuinely new and reflects a stabilization process due to the added source terms.
The influence of the stabilizing process is quantified by the ``stabilization rate'' $\sigma_\mathrm{S}$.
For mesa competition (superscript $+$) and mesa coalescence (superscript $-$), it is given by
\begin{equation}\label{eq:sigma-stab-mesas}
    \sigma_\mathrm{S}^\pm 
    = 
    \frac{\left| s_\mathrm{tot}^\pm\right|}{\Delta\rho}
    \, ,
\end{equation}
where ${s_\mathrm{tot}^\pm = s_\mathrm{tot}(\rho_\pm,\eta_\mathrm{stat}^\infty)}$ and ${\Delta\rho = \rho_+ - \rho_-}$.
For peak patterns, we find the stabilization rate as
\begin{equation}\label{eq:sigma-stab-peak}
    \sigma_\mathrm{S} 
    = 
    |\langle \partial_\rho s_\mathrm{tot}\rangle_\mathrm{int}|
    \, .
\end{equation}

The rate $\sigma^\varepsilon_\mathrm{D}$ constitutes a generalization of the diffusion-\-limited mass-competition rate $\sigma_\mathrm{D}$ [cf.\ Eqs.~\eqref{eq:sigma-MC-peak-D},~\eqref{eq:sigma-MC-mesa-D}],
\begin{equation}\label{eq:sigma-D-eps-correction}
    \sigma^\varepsilon_\mathrm{D} 
    = 
    - 
    \frac{4 D_v}{\Lambda} \, 
    \partial_{M}^{} 
    \big(
    \etastat +\varepsilon\, \delta\eta_\mathrm{stat}^\varepsilon
    \big)
    \, ,
\end{equation}
and implicitly depends on the source strength $\varepsilon$.
The additional term $\varepsilon\, \delta\eta_\mathrm{stat}^\varepsilon$ accounts for the shift of the \emph{stationary} mass-redistribution potential at the stationary peak or interface due to the weak source terms.
It is given by (see Appendix~\ref{app:stat-state-nMC})
\begin{equation}\label{eq:nMC-etastat-shift}
    \delta\eta_\mathrm{stat}^\varepsilon = 
    - 
    \frac{\langle s_1 + d\, s_2\rangle_\mathrm{int}}{\langle\tilde{f}_\eta\rangle_\mathrm{int}}
    \, .
\end{equation}
Notably, for mesa patterns, the shift in the stationary mass-redistribution potential is independent of the mass $M$, (${\partial_M^\pm \delta\eta_\mathrm{stat}^\varepsilon =0}$) because the shift, Eq.~\eqref{eq:nMC-etastat-shift}, only depends on the interface profile, which does not change with the mesa $M$ within the sharp-interface approximation (the interface is only shifted, see Sec.~\ref{sec:stat-patterns}). In other words, $\delta\eta_\mathrm{stat}^\varepsilon$ does not contribute to gradients in $\eta$ between mesas with different masses. This implies that the contribution to the growth rate $\sigma^\varepsilon$ that stems from mass competition [first term in Eq.~\eqref{eq:sigma-eps}] is unchanged compared to the mass-conserving case for mesa patterns (${\sigma_\mathrm{D}^\varepsilon = \sigma^\pm_\mathrm{D}}$).

\textit{Diffusion- and reaction-limited regimes.\;---}
As in the mass-conserving case, we distinguish a diffusion- and a reaction-limited regime for the growth rate $\sigma^\varepsilon$.
In the diffusion-limited regime [${D_v\ll \Lambda\ell_\mathrm{int}\langle\tilde{f}_\eta\rangle_\mathrm{int}}$, see Eq.~\eqref{eq:diff-reac-crossover}], the growth rate $\sigma^\varepsilon$ simplifies to (we reason in Appendix~\ref{app:nMC-mass-competition} why $\sigma^\varepsilon_\mathrm{D}$ may be replaced by $\sigma_\mathrm{D}$)
\begin{equation}\label{eq:sigma-eps-diffLim}
    \sigma^\varepsilon \approx \sigma_\mathrm{D}-\varepsilon\sigma_\mathrm{S}
    \, ,
\end{equation}
where one again substitutes ${\sigma_\mathrm{D}=\sigma_\mathrm{D}^\pm}$ in the case of mesa competition and mesa coalescence, respectively.
In this limiting regime, the rates of mass transport and source production or degradation simply add up because single domains and the value of the mass-redistribution potential at the domains can be approximated by a QSS.
The mass of the quasi-stationary domains then slowly changes by mass transport between these and the production or degradation of particles within the domains.

In contrast, in the reaction-limited regime (${D_v\gg \Lambda\ell_\mathrm{int}\langle\tilde{f}_\eta\rangle_\mathrm{int}}$), the shift $\varepsilon\,\delta\eta_\mathrm{stat}^\varepsilon$ becomes relevant and one has to distinguish between peak and mesa patterns (a scaling argument is given at the end of Appendix~\ref{app:nMC-mass-competition}).
The growth rate $\sigma^\varepsilon$ for mesa patterns takes the form
\begin{equation}
\label{eq:sigma-eps-mesa-reacLim}
    \sigma^\varepsilon \approx \sigma_\mathrm{R}^\pm -\varepsilon \frac{\sigma_\mathrm{R}^\pm}{\sigma_\mathrm{D}^\pm}\sigma_\mathrm{S}
    \, .
\end{equation}
This corresponds to the result in the diffusion-limited regime but scaled by the ratio of the reaction- and diffusion-limited rates of mass competition.
The scaling of the magnitude of the rate accounts for the reactive limitation of the rate for the mass change of single domains while the shift $\varepsilon\,\delta\eta_\mathrm{stat}^\varepsilon$ does not contribute for mesa patterns (${\partial_M^{}\delta\etastateps \approx 0}$).
At first order in $\varepsilon$, these growth rates for mesa patterns in the reaction-limited regime agree with the growth rates derived by McKay and Kolokolnikov~\cite{McKay.Kolokolnikov2012} when evaluating their expressions for nearest-neighbor competition.\footnote{
For comparison, we identify their reaction terms $f$ and $g$ with our terms ${f+\varepsilon s_1}$ and ${-f+\varepsilon s_2}$, the diffusion constants ${\varepsilon^2=D_u}$, ${D = D_v}$, and set ${\tau=1}$.
Then one obtains ${g_\pm = \varepsilon s_\mathrm{tot}^\pm}$ and ${\alpha_\pm = -\Delta\rho\, \partial_M^\pm \etastat [1 + \order{\varepsilon}]}$.}

For peak patterns in the reaction-limited regime, we find instead
\begin{equation}\label{eq:sigma-eps-peak-reacLim}
    \sigma^\varepsilon \approx \sigma_\mathrm{R} - 2 \ell_\mathrm{int} \langle\tilde{f}_\eta\rangle_\mathrm{int} \,\varepsilon\, \partial_M^{}\delta\etastateps-\varepsilon \frac{\sigma_\mathrm{R}}{\sigma_\mathrm{D}}\sigma_\mathrm{S}
    \, ,
\end{equation}
where the middle term is due to the shift of the stationary mass-redistribution potential at the peak.
The last term is negligible in comparison to the second if the source terms fulfill ${s_1 \gtrsim s_2}$ [see Eq.~\eqref{eq:nMC-etastat-shift}].
We will explain the structure of the growth rates in the following section and also discuss what causes the shift $\varepsilon\,\delta\eta_\mathrm{stat}^\varepsilon$.
Before, the stability threshold is analyzed.

\textit{Stability threshold.\;---}
For a sufficiently large source strength or large wavelength, $\sigma^\varepsilon$ may become negative as the stabilization rate $\varepsilon\sigma_\mathrm{S}$ increases or the rate of destabilizing mass transport $\sigma_\mathrm{D}^\varepsilon$ decreases, respectively.
A negative growth rate $\sigma^\mathrm{\varepsilon}$ indicates that the mass-competition instability is suppressed, and a mass difference $\delta M$ between the domains decreases exponentially in time, ${\delta M\sim \operatorname{e}^{\sigma^\varepsilon t}}$.
Because the coarsening process increases the pattern wavelength $\Lambda$ due to the collapse or coalescence of peaks or mesas, it is expected to interrupt at the stability threshold ${\sigma^\varepsilon(\Lambda,\varepsilon) = 0}$.
This behavior of patterns with several peaks or mesas is analyzed in Sec.~\ref{sec:interrupted-coarsening}.
Using Eqs.~\eqref{eq:sigma-eps}, \eqref{eq:sigma-stab-mesas}, the criterion ${\sigma^\varepsilon(\Lambda,\varepsilon) = 0}$ yields  for mesa patterns the critical source strength
\begin{equation}\label{eq:eps-threshold-mesas}
    \varepsilon_\mathrm{stop}(\Lambda) = \frac{\sigma^\pm_\mathrm{D}(\Lambda)}{\left| s_\rho^\pm\right| / \Delta\rho}
    \, .
\end{equation}
For peak patterns, the stability threshold follows from Eqs.~\eqref{eq:sigma-eps}, \eqref{eq:sigma-D-eps-correction}, \eqref{eq:sigma-stab-peak} as
\begin{equation}\label{eq:eps-threshold-peaks}
    \varepsilon_\mathrm{stop}(\Lambda) = \frac{\sigma_\mathrm{D}(\Lambda)}{-\langle \partial_\rho s_\rho\rangle_\mathrm{int} + \frac{4 D_v}{\Lambda}\partial_M\delta\eta_\mathrm{stat}^\varepsilon}
    \, ,
\end{equation}
where the dependence on $\Lambda$ appears implicitly in the various terms.
The critical wavelength $\Lambda_\mathrm{stop}(\varepsilon)$ is found by inverting the above relations.
In general, it is not possible to perform this inversion analytically.

The growth rate Eq.~\eqref{eq:sigma-eps-diffLim} in the diffusion-limited regime as well as the threshold of interrupted coarsening for mesa patterns, Eq.~\eqref{eq:eps-threshold-mesas}, recover the expressions found based on the QSS approximation in Ref.~\cite{Brauns.etal2021}.
However, the mathematical structure of the general growth rate Eq.~\eqref{eq:sigma-eps} is quite intriguing and demands further analysis.

\subsection{The mathematical structure of the growth rates}
\label{sec:growth-rate-nMC-discussion}
In the following, we will analyze how the different terms of the growth rate $\sigma^\varepsilon$ arise [Eq.~\eqref{eq:sigma-eps}]. 
We will find that the rate $\sigma_\mathrm{S}$ accounts for net production and degradation at the peaks or mesas.
We then show that the source terms determine, together with mass transport in the cytosol, whether a peak or mesa accumulates or loses mass.
The rate $\sigma_\mathrm{R}$ of reactive conversion (particle release and incorporation) will be shown to only affect how fast the accumulated mass can be incorporated into a peak or mesa (or how fast mass is released from the peak/mesa). 
Thus, the rate $\sigma_\mathrm{R}$ only affects the magnitude of the growth rate $\sigma^\varepsilon$, and enters $\sigma^\varepsilon$ only as a prefactor.
To begin, we first consider the effect of the source terms alone and explain the stabilization rate $-\varepsilon\sigma_\mathrm{S}$.

\textit{Stabilization rate due to weak source terms.\;---}
The relaxation rate $-\varepsilon\sigma_\mathrm{S}$ describes the direct stabilizing effect of the source terms.
In the case of two competing half-mesas (cf.\ Fig.~\ref{fig:coarsening-modes}, mesa competition/coalescence and also peak coalescence), a shift in the interface positions (the peak) by $\delta\ell$ due to mass transfer from one mesa to the other only changes the length of the outer plateaus, i.e., the high-density plateaus for mesa competition and the low-density plateaus for mesa/peak coalescence.
For specificity, we now analyze mesa competition.
In the high-density plateaus, the source terms lead to net degradation, i.e., ${s_\mathrm{tot}(\rho_+,\etastat)=s_\mathrm{tot}^+<0}$.
\footnote{As shown in Appendix~\ref{app:stat-state-nMC}, single (half-)mesas are stable only if ${s_\mathrm{tot}(\rho_+,\etastat) < 0 < s_\mathrm{tot}(\rho_-,\etastat)}$.
Similarly, in peak patterns, stability mandates that degradation increases for larger peaks, i.e., ${\langle \partial_\rho s_\mathrm{tot}\rangle_\mathrm{int} < 0}$.
}
Thus, an increase of mass ${\delta M = \Delta\rho\, \delta\ell}$ in one (half-)mesa there leads to additional degradation ${\varepsilon s_\mathrm{tot}^+ \delta\ell = \partial_t^\mathrm{source} \delta M < 0}$, and to additional production ${-\partial_t^\mathrm{source} \delta M >0}$ in the other (half-)mesa.
Consequently, production and degradation together result in a relaxation of the initial perturbation and drive the interfaces back to their symmetric rest position with the rate ${-\varepsilon\sigma_\mathrm{S} = -\varepsilon |s_\mathrm{tot}^+|/\Delta\rho}$ (respectively, ${-\varepsilon\sigma_\mathrm{S} = -\varepsilon |s_\mathrm{tot}^-|/\Delta\rho}$ for mesa/peak coalescence).
Analogously, in the case of peak competition, additional mass $\delta M$ increases the peak size and leads to increased degradation
\begin{align}\label{eq:source-derivation}
    \partial_t^\mathrm{source}\delta M &\approx \varepsilon\int_0^\frac{\Lambda}{2}\mathrm{d}x\, s_\mathrm{tot}(\rhostat+2\delta M \partial_M^{}\rhostat,\nonumber\\
    &\hspace{2.6cm}\etastat + 2\delta M \partial_M^{}\etastat)
    \, .
\end{align}
Using that $\partial_M^{}\etastat$ becomes negligible for sufficiently large peaks [cf.\ Eqs.~\eqref{eq:mesa-eta-M-scaling},~\eqref{eq:peak-eta-M-scaling} and Appendix~\ref{app:stat-state-nMC}] one arrives at the simplified mass evolution
\begin{align}
    \partial_t^\mathrm{source}\delta M &\approx 2\delta M \varepsilon \int_0^\frac{\Lambda}{2}\mathrm{d}x\, \left(\partial_\rho s_\mathrm{tot}\right)\, 2\partial_M^{}\rhostat\nonumber\\
    &= \varepsilon\langle\partial_\rho s_\mathrm{tot}\rangle_\mathrm{int} \delta M
    \, ,\label{eq:source-prod-heuristic}
\end{align}
where the last equality follows from the definition of the interface average $\langle\cdot\rangle_\mathrm{int}$ [see Eq.~\eqref{eq:interface-average}].
Because ${\langle\partial_\rho s_\mathrm{tot}\rangle_\mathrm{int} < 0}$ is mandated by the stability of a single peak (see Appendix~\ref{app:stability}), particle production and degradation at the peaks stabilize the symmetric peak configuration.
The stabilization rate is read off as ${-\varepsilon\sigma_\mathrm{S} = \varepsilon \langle\partial_\rho s_\mathrm{tot}\rangle_\mathrm{int} < 0}$.
With this we have found expressions for the stabilization rate.
While the mass of a single peak (mesa) is arbitrary in the mass-conserving system, a fixed mass is selected by the source terms (see Sec.~\ref{sec:stat-patterns-nMC}).
For this domain mass to be stable, production and degradation have to degrade mass if the domain mass is larger than the mass of the stationary peak, while the source terms have to produce mass if the domain mass lies below its stationary value.
The same process stabilizes the symmetric stationary state of two domains: The masses of both peaks (mesas) are driven back toward their stationary value.
With this understanding of the stabilizing source effect, we discuss the effect of the relaxation rate $-\varepsilon\sigma_\mathrm{S}$ on the mass-competition instability.

\textit{Competition between mass transport and the source terms.\;---}
The presence of a mass-competition instability is determined by the sign of the growth rate $\sigma^\varepsilon$, Eq.~\eqref{eq:sigma-eps}.
Hence, the stability properties of a given pattern are dictated by the competition between the stabilizing production and degradation processes (described by $-\varepsilon\sigma_\mathrm{S}$) and the (modified) diffusion-limited rate of mass exchange $\sigma_\mathrm{D}^\varepsilon$.
Reaction limitation of the mass exchange between the domains only enters through an overall factor [the prefactor in Eq.~\eqref{eq:sigma-eps}].
It is remarkable that $\sigma_\mathrm{R}$ does not enter explicitly in the stability threshold ${\sigma^\varepsilon = 0}$ although the strength of the destabilizing mass-exchange process depends on both $\sigma_\mathrm{D}$ and $\sigma_\mathrm{R}$ [cf.\ Eq.~\eqref{eq:sigma-MC-massComp}].

\begin{figure}
    \centering
    \includegraphics{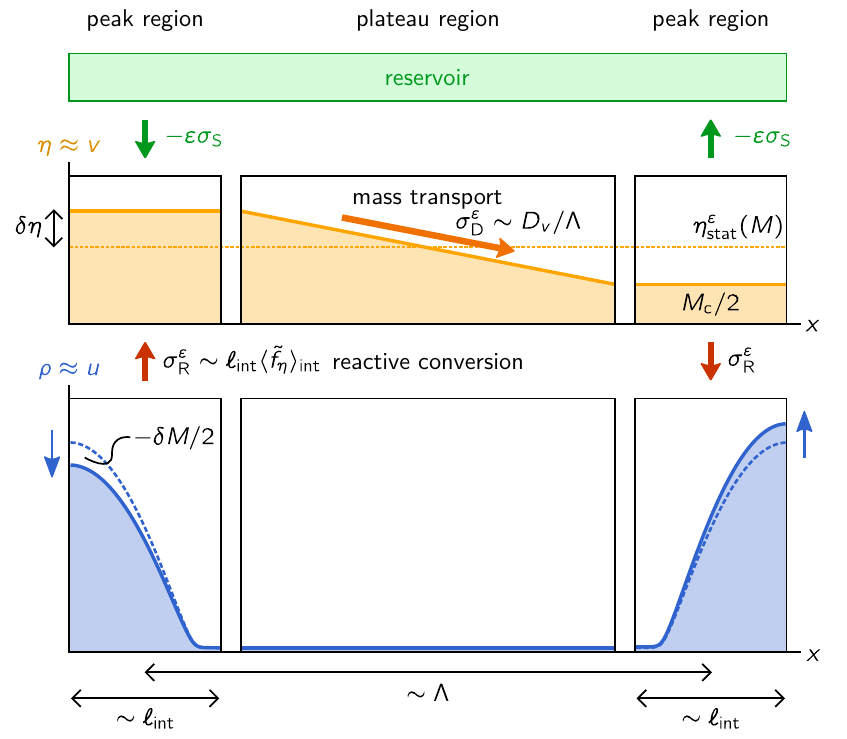}
    \caption{
    The effect of source terms on the mass-competition instability in the limit $D_v\gg D_u$.
    The source terms lead to an interaction of the mass-conserving 2cRD system with an implied reservoir [green (top) rectangle].
    The source terms (with rate $\varepsilon\sigma_\mathrm{S}$) together with the mass exchange between the two peaks (rate $\sigma_\mathrm{D}^\varepsilon$) determine the evolution of the cytosolic pool at each peak [orange (light gray) and green (top-most) arrows; total mass $M_\mathrm{c}/2$ of the cytosolic pool at right half-peak].
    The changed cytosolic density then induces peak growth or shrinking by reactive conversion of particles between the $u$ and $v$ species [$\rho\approx u$; blue (dark gray) profile and red (bold, dark gray) arrows].
    }
    \label{fig:sigma-eps-conceptual}
\end{figure}

To resolve this puzzling finding, we employ the freedom we have in defining the reaction terms.
Given a 2cRD system, one has some freedom in distributing the reaction terms between the conversion term $\tilde{f}$ and the source terms $s_{1,2}$ [cf.\ Eq.~\eqref{eq:2cRD-u}].
If one defines
\begin{equation}\label{eq:new-f}
    \tilde{f}' = \tilde{f} + \varepsilon (s_1 + d s_2)
    \, ,
\end{equation}
the source terms exactly cancel in the stationary profile equation, Eq.~\eqref{eq:profile-equation}, and the shift $\delta\etastateps$ induced by the remaining source terms vanishes [see Eq.~\eqref{eq:nMC-etastat-shift} and Appendix~\ref{app:stat-state-nMC}].
At the same time, the 2cRD system, Eqs.~\eqref{eq:2cRD}, becomes for ${D_v \gg D_u}$
\begin{subequations}
\label{eq:2cRD-mod-source-term}
\begin{align}
    \partial_t u 
    &\approx D_u \partial_x^2 u + f'
    \, ,\\
    \partial_t v 
    &\approx D_v \partial_x^2 v - f' + \varepsilon s_\mathrm{tot}
    \, ,
\end{align}
\end{subequations}
with ${\tilde{f}' = (1-d) f'\approx f'}$.
The redefinition underlines that the shift $\delta\etastateps$ results from source terms in the membrane species $u$, i.e, from the additional reactive conversion from the $v$ into the $u$ species, which is necessary in steady state to balance degradation in the $u$ species at the peak by replenishment with particles from the plateau region where production prevails [cf.\ the reaction-limited growth rate $\sigma_\mathrm{R}$, Eq.~\eqref{eq:sigma-R-heuristic}, and see Appendix~\ref{app:stat-state-nMC}].
Equations~\eqref{eq:2cRD-mod-source-term} shows that source terms in the slowly diffusing species $u$ can also be considered as source terms in the fast-diffusing species if we account for a change of the mass-conserving chemical conversion reactions $\tilde{f}$, i.e., if we account for a deformation of the nullcline $\eta^*(\rho)\to {\eta^*}'(\rho)$ and the ensuing deformation of the stationary pattern (cf.\ Sec.~\ref{sec:stat-patterns}).

We now use this argument to analyze the mass change of two competing peaks in the limit ${D_v\gg D_u}$ (see Fig.~\ref{fig:sigma-eps-conceptual}).
The same approach as used below also works for mesa competition and for mesa/peak coalescence.
With all source terms moved to the cytosol (and neglecting slow membrane diffusion), the mass in a peak region changes only through the cytosol, by diffusive transport and production/degradation.
As in the mass-conserving case, mass is  transported  from the cytosolic pool at the smaller toward the larger peak with rate $\sigma_\mathrm{D}^\varepsilon$ [orange (light gray, diagonal) arrow in Fig.~\ref{fig:sigma-eps-conceptual}].
The rate $\sigma_\mathrm{D}^\varepsilon$ accounts, to first order in $\varepsilon$, for the changed stationary mass-redistribution potential due to the modified reaction term $\tilde{f}'$ [see Appendix~\ref{app:stat-state-nMC}, Eq.~\eqref{eq:MC-stat-state-deviations}].
The mass transport is counteracted by the source terms that deplete the cytosol at the larger peak and increase the pool at the smaller peak, described by the stabilization rate $-\varepsilon\sigma_\mathrm{S}$ [green (top-most) arrows in Fig.~\ref{fig:sigma-eps-conceptual}].
Hence, the sign of the total rate ${\sigma_\mathrm{D}^\varepsilon-\varepsilon\sigma_\mathrm{S}}$ determines whether the larger peak accumulates additional mass in its cytosolic pool (positive sign), or loses mass (negative sign).
In addition, we have already found in the mass-conserving case that an increase in the cytosolic pool at a peak of mass $M$ above its stationary density $v\approx\etastat(M)$ also induces growth of the density peak formed on the membrane [red (dark gray) arrows in Fig.~\ref{fig:sigma-eps-conceptual}].
Together, this explains why the sign of the rate ${\sigma_\mathrm{D}^\varepsilon-\varepsilon\sigma_\mathrm{S}}$ alone decides between self-enhancing peak growth or stabilizing shrinking of the larger peak, solving the puzzle that the reaction-limited rate $\sigma_\mathrm{R}$ only affects the magnitude of the growth rate $\sigma^\varepsilon$.
In the following, we explain the prefactor ${\sigma_\mathrm{R}/(\sigma_\mathrm{D}+\sigma_\mathrm{R})}$ by showing that the timescale $1/\sigma^\varepsilon$ is composed of the timescale for the change of the cytosolic pool $1/(\sigma_\mathrm{D}^\varepsilon-\varepsilon\sigma_\mathrm{S})$ and a timescale describing reactive conversion at the peak, that is, the incorporation and release processes at the peaks. This is fully analogous to the reasoning behind  $1/\sigma$ in the mass-conserving case, that is, Eq.~\eqref{eq:sigma-MC-massComp}.

\textit{The effect of local reactive conversion.\;---}
In the mass-conserving system, the rate of reactive conversion is given by $\sigma_\mathrm{R}$ (see Sec.~\ref{sec:mc-growth-rates}).
We now determine this rate under the influence of weak source terms.
To this end, assume that the cytosolic pool is depleted or enlarged by mass transport or source terms as discussed in the previous paragraph. Then, the reactive conversion rate, which describes how fast particles are exchanged between the cytosolic pool at a peak and the density peak formed on the membrane, can be determined by analyzing the time evolution of the cytosolic mass. For specificity consider the cytosolic mass ${M_\mathrm{c}/2}$ at the right peak in Fig.~\ref{fig:sigma-eps-conceptual}, which changes via three processes: mass transfer between the peaks, net production or degradation, and the reactive-conversion flux into or from the membrane species $u$ [see Fig.~\ref{fig:sigma-eps-conceptual}; orange, green, and red (bold) arrows].
Mass transport is, as used in Eq.~\eqref{eq:mass-transport-heuristic}, determined by the cytosolic gradient between the peaks that is induced by the change in the cytosolic density $\delta v\approx -\delta\eta$ (at the right peak; see Fig.~\ref{fig:sigma-eps-conceptual}).
This gives ${\partial_t^\mathrm{transport}M_\mathrm{c}/2 \approx 2D_v\delta\eta/\Lambda}$.
Furthermore, the production or degradation process is described by ${\partial_t^\mathrm{source}M_\mathrm{c}/2 \approx-\varepsilon\sigma_\mathrm{S}\delta M/2}$ where $\delta M$ denotes the mass difference between the left and right peak [see Eq.~\eqref{eq:source-prod-heuristic}].
Finally, mass release from the density peak on the membrane or incorporation into this peak is driven by the deviation ${\Delta v\approx \Delta\eta^\varepsilon}$ of the cytosolic density ${v\approx\etastateps(M)-\delta\eta}$ from its stationary value ${v_\mathrm{stat}\approx\etastateps(M+\delta M)\approx\etastateps(M)+\delta M \partial_M^{}\etastateps}$ (at the right peak).
It is thus described by ${\partial_t^\mathrm{conversion}M_\mathrm{c}/2 \approx \big(\delta M\partial_M^{}\etastateps+\delta\eta\big) \ell_\mathrm{int}\langle\tilde{f}_\eta\rangle_\mathrm{int}}$ [see Eq.~\eqref{eq:incorp-release-heuristic}].
Taken together, one has\footnote{
One can use the interface width $\ell_\mathrm{int}$ and average conversion rate $\langle\tilde{f}_\eta\rangle_\mathrm{int}$ of the corresponding mass-conserving system (setting ${\varepsilon=0}$) because the change of the stationary peak profile due to weak source terms only yields higher order corrections to these quantities.
The term $\varepsilon\partial_M^{}\delta\etastateps$ is not negligible because $\partial_M^{}\etastat\sim \partial_M^{}\varepsilon\, \delta\etastateps$ by assumption (see Appendix~\ref{app:MC-mass-competition}).
}
\begin{align}
    \partial_t \frac{M_\mathrm{c}}{2} &\approx \frac{2 D_v}{\Lambda}\delta\eta -\varepsilon\sigma_\mathrm{S}\frac{\delta M}{2}\nonumber\\
    &\hspace{4mm} + \left(\delta M\partial_M^{}\etastateps+\delta\eta\right) \ell_\mathrm{int}\langle\tilde{f}_\eta\rangle_\mathrm{int}
    \, .\label{eq:time-evol-cytosolic-pool-sigma-eps}
\end{align}

Because the peaks mainly form on the membrane, the change in the cytosolic mass is negligible, and we set ${\partial_t M_\mathrm{c} \approx 0}$.
It then follows from Eq.~\eqref{eq:time-evol-cytosolic-pool-sigma-eps}
\begin{equation}
    \Delta\eta^\varepsilon \approx -\frac{\sigma_\mathrm{D}^\varepsilon-\varepsilon \sigma_\mathrm{S}}{\sigma_\mathrm{D}} \frac{\sigma_\mathrm{D}}{\sigma_\mathrm{D} + \sigma_\mathrm{R}} (\partial_M^{}\etastat)\delta M
    \, .
\end{equation}
Consequently, the offset ${\Delta\eta^\varepsilon = -\left(\delta M\partial_M^{}\etastateps+\delta\eta\right)}$ is scaled compared to the corresponding offset ${\Delta\eta = -\left(\delta M\partial_M^{}\etastat+\delta\eta\right)}$ in the mass-conserving system (setting ${\varepsilon=0}$) by ${\Delta\eta^\varepsilon = \frac{\sigma_\mathrm{D}^\varepsilon-\varepsilon\sigma_\mathrm{S}}{\sigma_\mathrm{D}}\Delta\eta}$.
Because reactive conversion is driven by the offset $\Delta\eta^\varepsilon$ and changes the peak mass by ${\partial_t\delta M/2 \approx \ell_\mathrm{int}\langle\tilde{f}_\eta\rangle_\mathrm{int}\Delta\eta^\varepsilon}$ [see Eq.~\eqref{eq:incorp-release-heuristic}], the reactive conversion rate is changed as well by
\begin{equation}
    \sigma_\mathrm{R}\to \frac{\sigma_\mathrm{D}^\varepsilon-\varepsilon\sigma_\mathrm{S}}{\sigma_\mathrm{D}}\sigma_\mathrm{R}
    \, .
\end{equation}
Combining this finding with the rate of mass in- or outflow $1/(\sigma_\mathrm{D}^\varepsilon-\varepsilon \sigma_\mathrm{S})$ into or from the peak region, the total growth rate of mass competition under the influence of weak source terms follows as [cf.\ Eq.~\eqref{eq:sigma-MC-massComp}]
\begin{equation}
    \frac{1}{\sigma^\varepsilon}=\frac{1}{\sigma_\mathrm{D}^\varepsilon-\varepsilon\sigma_\mathrm{S}} + \frac{1}{\sigma_\mathrm{R}\frac{\sigma_\mathrm{D}^\varepsilon-\varepsilon\sigma_\mathrm{S}}{\sigma_\mathrm{D}}}
    \, ,
\end{equation}
which agrees with Eq.~\eqref{eq:sigma-eps}.
This calculation shows that the reactive limitation indeed only gives rise to the prefactor ${\sigma_\mathrm{R}/(\sigma_\mathrm{D}+\sigma_\mathrm{R})}$ in the rate $\sigma^\varepsilon$.
The stability properties, that is, the sign of $\sigma^\varepsilon$ is determined by those processes alone which change the total mass in one peak region.
The timescale of reactive conversion only limits how fast the positive or negative feedback in the mass change of the cytosolic pools at the peaks can be translated into actual peak growth and shrinking.
It does not affect whether additional mass is accumulating at the smaller or larger peak, i.e., whether mass competition will result in stabilization or destabilization of the symmetric steady state.

\begin{figure*}
\centering
\includegraphics{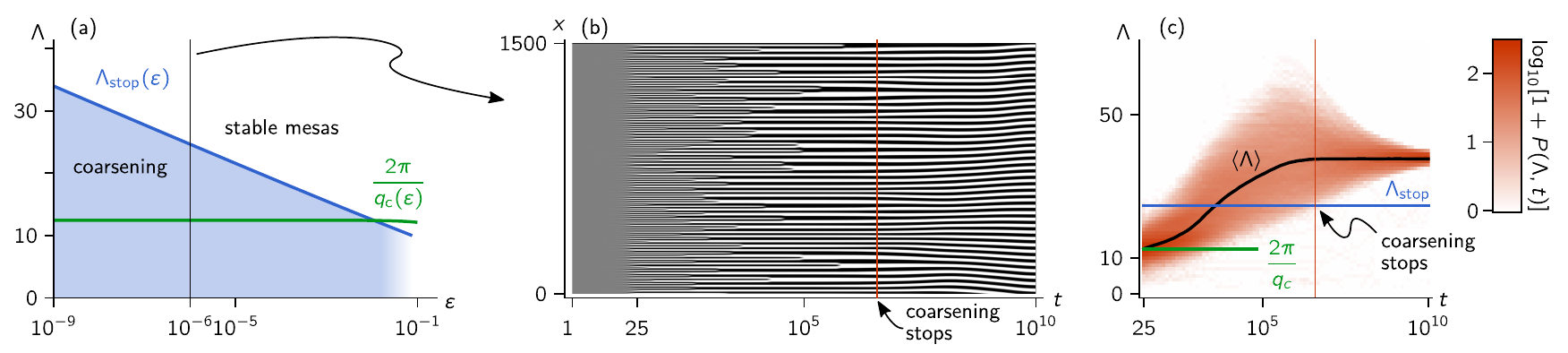}
\caption{
Weak source terms suppress the mass-competition instability at large wavelengths and thereby interrupt the coarsening process (compare with conceptual Fig.~\ref{fig:interrupted-coarsening-phenomenology}).
(a) Patterns grow initially out of the homogeneous steady state with the length scale ${\Lambda = 2\pi/q_\mathrm{c}}$ [green (approximately horizontal) line] set by the fastest-growing mode $q_\mathrm{c}$ in the dispersion relation. 
A source strength $\varepsilon$ is sufficiently strong to stabilize patterns with a length scale ${\Lambda>\Lambda_\mathrm{stop}(\varepsilon)}$ [blue line (top edge of the shaded region)].
(b) The kymograph obtained from the numerical simulations (snippet from a larger domain) shows the formation of mesa patterns (grayscale, $-1$ to $1$) of which some subsequently collapse such that the average domain size increases.
At later times [red (vertical) line], no further collapses occur and domains merely rearrange into the periodic stationary state.
(c) The histogram $P(\Lambda,t)$ of the wavelengths of the single elementary patterns on the domain [(red) density plot; averaged over 7 independent runs starting from different random initial fluctuations around the HSS] clearly shows that coarsening stops and the average length scale $\langle\Lambda\rangle$ (black) becomes constant when the smallest domains cross the stability threshold $\Lambda_\mathrm{stop}$ [blue (dark gray) line] approximately at the red (vertical) line.
The histogram is normalized by $\int_0^\infty\mathrm{d}\Lambda\,P(\Lambda,t) = 2 N(t)$ where $2 N(t)$ is the number of elementary patterns (single mesa interfaces or half peaks) contained in the full pattern.
The cubic model ${\tilde{f}=\eta -\rho^3+\rho}$ (see Appendix~\ref{app:cubic-model}) with source terms ${(s_1,s_2) = (0,p-\rho)}$ was simulated on a domain of length ${L=20 \, 000}$ with periodic boundary conditions using Comsol Multiphysics \cite{Comsol}.
The parameters are ${D_u = 1}$, ${D_v=10}$, ${p=0}$ and ${\varepsilon = 10^{-6}}$.
}
\label{fig:interrupted-coarsening}
\end{figure*}

\subsection{Suppression of the mass-competition instability determines interrupted coarsening}
\label{sec:interrupted-coarsening}
Let us now analyze how the suppression of the mass-competition instability at the threshold ${\sigma^\varepsilon=0}$ translates into the interruption of the coarsening process in a large system containing many peaks or mesas.
To this end, we now revisit the introductory example of interrupted coarsening shown in Fig.~\ref{fig:interrupted-coarsening-phenomenology}.

A more detailed view on the simulation is given in Fig.~\ref{fig:interrupted-coarsening}.
First, because the mass-competition instability of mesa-forming 2cMcRD systems is exponentially weak as a function of the pattern wavelength, i.e., the interface or peak separation, an exponentially small (as a function of the wavelength) source term is sufficient to interrupt coarsening at some pattern wavelength $\Lambda_\mathrm{stop}$ [see Eqs.~\eqref{eq:mesa-eta-M-scaling},~\eqref{eq:eps-threshold-mesas} and Fig.~\ref{fig:interrupted-coarsening}(a)].
As in a large system containing many peaks or mesas the coarsening process can proceed by competition and coalescence, the wavelength $\Lambda_\mathrm{stop}$ corresponds to the threshold ${\sigma^\varepsilon=0}$ for the most unstable coarsening scenario.

In numerical simulations of large systems containing many peaks [see Fig.~\ref{fig:interrupted-coarsening}(b)], the behavior of the characteristic pattern length scale(s) is best read off from the distribution $P(\Lambda,t)$ of wavelengths $\Lambda$ for the single elementary patterns, each of which comprises half a period of the full pattern [see Fig.~\ref{fig:interrupted-coarsening}(c)].
Initially, the pattern develops at the length scale $2\pi/q_\mathrm{c}$ set by the fastest growing mode $q_\mathrm{c}$ in the dispersion relation (see Sec.~\ref{sec:HSS-instability}).
Then the average pattern length scale $\langle\Lambda\rangle(t)$ grows due to the mass-competition instability.
As soon as the lower edge of the distribution $P(\Lambda,t)$ passes the threshold $\Lambda_\mathrm{stop}$ of interrupted coarsening, the mass-competition instability becomes stabilized for all pattern domains.
Thus, coarsening stops and the average length scale as well as the total number of mesas become constant.
At later times, the interfaces slowly rearrange toward the periodic stationary state which leads to a narrowing of the length-scale distribution $P(\Lambda,t)$.
This rearrangement is driven by the (now stabilized) coarsening scenarios that lead to an equalization of the masses in the different mesas and troughs.

In conclusion, the suppression of the mass-competition instability indeed arrests the coarsening process.
The threshold $\Lambda_\mathrm{stop}$ determines the wavelength selected.
Depending on the width of the length-scale distribution $P(\Lambda, t)$, the threshold $\Lambda_\mathrm{stop}$ directly gives a rough estimate of the final wavelength.
Especially in higher spatial dimensions, we expect that the distribution becomes narrower as each pattern domain interacts with more different neighbors.
The threshold $\Lambda_\mathrm{stop}$ then becomes a better estimate for the selected length scale.
It is an interesting open question whether the domain-size distributions determined for coarsening processes \cite{Lifshitz.Slyozov1961,Wagner1961,Kawasaki.Nagai1983} can be used to translate the threshold size $\Lambda_\mathrm{stop}$ into the average length scale selected by interrupted coarsening.

\subsection{Comparison to numerical examples}
\label{sec:numerics}
At last, we test the derived analytic expressions for the growth rate as well as the stability threshold and compare these with the numerical linear stability analysis for different example systems.
The numerical linear stability analysis of two competing half-peaks or mesas was performed by spatially discretizing the system using a finite-differences approach implemented in Mathematica v12.2 (code is available under \url{https://github.com/henrikweyer/2cRD-wavelength-selection}).

To examine the results of the analytic treatment, we compare the numerically obtained leading eigenvalue of the linearized 2cRD dynamics, Eqs.~\eqref{eq:2cRD}, with the analytic expression Eq.~\eqref{eq:sigma-eps} for the growth rate $\sigma^\varepsilon$ in $(D_v,\varepsilon)$-parameter space.
Tuning the diffusion constant $D_v$ drives the transition between the diffusion- and reaction-limited regimes of the growth rate, and
increasing the source strength $\varepsilon$ leads through the stability threshold $\varepsilon_\mathrm{stop}(\Lambda)$.

To facilitate the overview of the different phenomena, we consider only linear source terms.
However, the growth rate $\sigma^\varepsilon$ given in Eq.~\eqref{eq:sigma-eps} also applies to nonlinear source terms. 
Other effects apart from interrupted coarsening may occur for nonlinear source terms as, for example, oscillations driven by a cycle of overall production and degradation \cite{Li.Cates2020,Kuwamura.Izuhara2017}.
These effects are excluded in our analysis because we assume that single peaks or mesas are stable.
Starting from these stable domains, we are interested in domain destabilization due to the interaction of several domains which then changes the characteristic pattern length scale.

\begin{figure}
    \centering
    \includegraphics{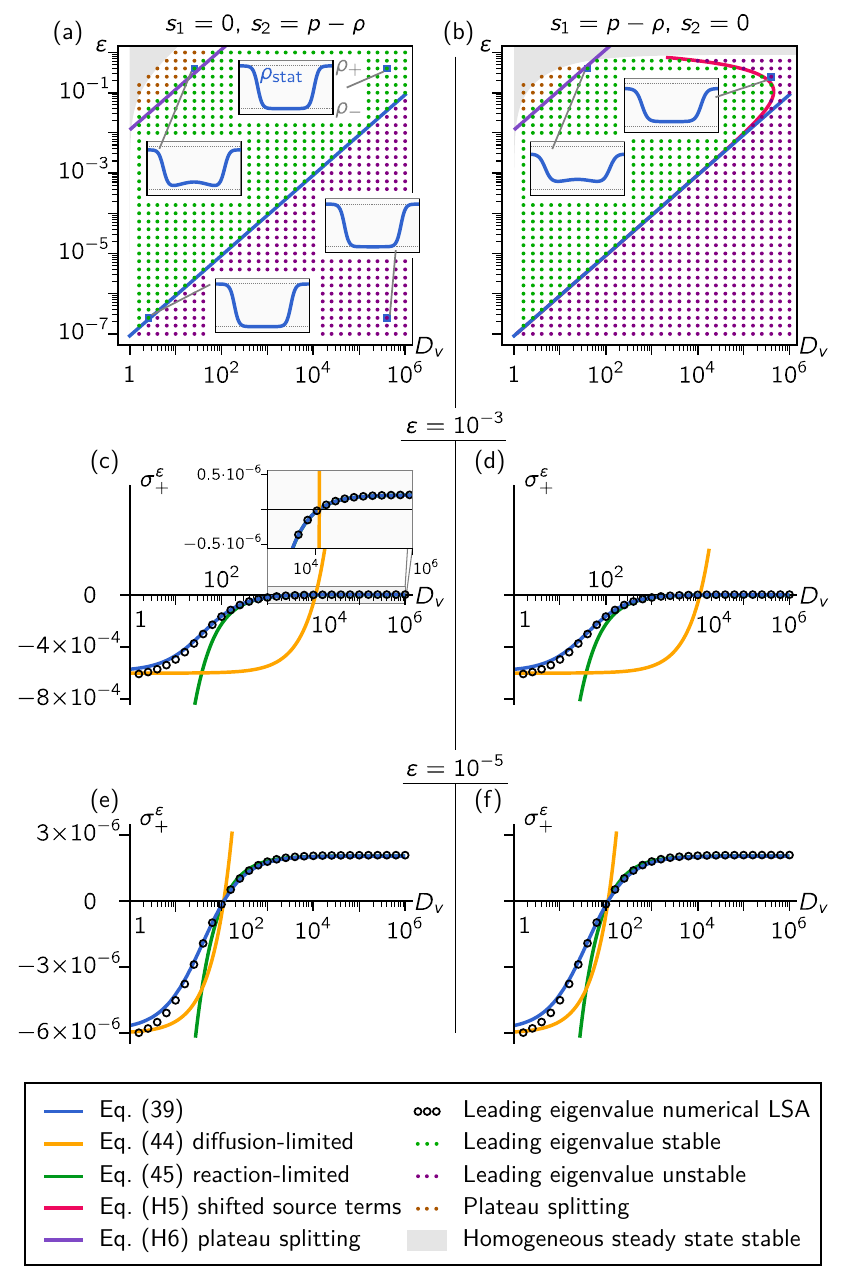}
    \caption{Stability threshold $\varepsilon_\mathrm{stop}$ and growth rate $\sigma^\varepsilon$ in the mesa-forming cubic model ${\tilde{f}(\rho,\eta) = \eta - \rho^3 + \rho}$ with a linear source term ${s=p-\rho}$ in the fast- (a,c,e) and slowly diffusing (b,d,f) species.
    (a,b) We compare the numerically determined stability regions for two half mesas (mesa competition) with the threshold of interrupted coarsening, Eq.~\eqref{eq:eps-threshold-mesas} [blue (lower diagonal) line].
    The improved approximation obtained by shifting the source terms into the $v$ species [cf.\ Eq.~\eqref{eq:new-f}] is shown in (b) as well [red (curved) line].
    The insets show the stationary profile $\rhostat(x)$ on the domain $[-\Lambda/2,\Lambda/2]$ whose stability is analyzed.
    Numerical results are shown where an elementary stationary pattern of length $\Lambda/2$ exists.
    The regime of mesa splitting [orange dots, (gray dots in the top-left corner)] can be estimated by the criterion put forward in Ref.~\cite{Brauns.etal2021} [see Eq.~\eqref{eq:cubicModel-splitting}; purple (top-most, diagonal) line].
    In the grey-shaded parameter region, the homogeneous steady state is linearly stable against arbitrarily-large-wavelength perturbations.
    (c-f) The value of the numerically determined leading eigenvalue (circles) is compared with the growth rate approximation, Eq.~\eqref{eq:sigma-eps} [blue (dark gray) line].
    The transition from the diffusion- [orange (light gray) line; Eq.~\eqref{eq:sigma-eps-diffLim}] into the reaction-limited [green (concave) line; Eq.~\eqref{eq:sigma-eps-mesa-reacLim}] regime occurs as the diffusion constant $D_v$ increases.
    The parameters of the model are chosen as ${\Lambda = 30}$, ${D_u = 1}$ and ${p = -0.2}$.
    }
    \label{fig:numerics-eigenvalueMapping-cubicModel}
\end{figure}

\subsubsection{The cubic model}
\label{sec:cubic-model}
The cubic model is constructed to closely resemble classical models of phase separation.
The reaction term is defined by (cf.\ \cite{McKay.Kolokolnikov2012} and Appendix~\ref{app:cubic-model})
\begin{equation}\label{eq:cubic-model}
    \tilde{f}(\rho,\eta) 
    = 
    \eta - \rho^3 + \rho
    \, .
\end{equation}
Thus, the nullcline reproduces the cubic nonlinearity used in the classical Cahn--Hilliard and Allen--Cahn equations (cf.\ Sec.~\ref{sec:stat-patterns}).

Figure~\ref{fig:numerics-eigenvalueMapping-cubicModel} shows the numerical and analytic results for the mesa-competition scenario in the cubic model [see Fig.~\ref{fig:coarsening-modes}(a), top].
The source terms are chosen as ${(s_1,s_2) = (0,p-\rho)}$ [Fig.~\ref{fig:numerics-eigenvalueMapping-cubicModel}(a,c,e)] and ${(s_1,s_2) = (p-\rho,0)}$ [Fig.~\ref{fig:numerics-eigenvalueMapping-cubicModel}(b,d,f)].
In both cases, this yields ${\sigma_\mathrm{S} = (1-p)/2}$.
Moreover, the functional form of the interface pattern on the infinite line is known for this cubic model such that the growth rate $\sigma^\varepsilon$ can be determined analytically (see Appendix~\ref{app:cubic-model}).
This yields a stability threshold $\varepsilon_\mathrm{stop}$ linear in $D_v$ because the stationary pattern $(\rhostat(x),\etastat)$ is independent of $D_v$, and the only dependence on $D_v$ is the explicit linear dependence of $\sigma_\mathrm{D}^+$ [see Eq.~\eqref{eq:sigma-MC-mesa-D}].
The threshold reproduces the zero crossing of the leading eigenvalue determined in the numerical stability analysis very well [blue (dark gray) line in Fig.~\ref{fig:numerics-eigenvalueMapping-cubicModel}(a,b)].

Analyzing the effect of a strong source term (large $\varepsilon$) in the slowly diffusing species $u$ on mass competition, we observe that the numerically determined stability threshold deviates from the analytic result.
We attribute the breakdown of the analytical approach to the fact that in the presence of strong sources, the stationary state becomes strongly deformed and can no longer be approximated by the stationary pattern of the mass-conserving system [see the top right corner in Fig.~\ref{fig:numerics-eigenvalueMapping-cubicModel}(b) and the inset].
The deviations are less pronounced when the source term is added in the fast-diffusing species [see Fig.~\ref{fig:numerics-eigenvalueMapping-cubicModel}(a)].  
Intuitively this is due to the fast diffusive mixing which averages out the effect of cytosolic source terms.
This effect can even be explicitly read off from Eqs.~\eqref{eq:gen-stat-pattern}, which determines the stationary pattern.
There, the source term $s_2$ only enters with strength $\varepsilon/D_v$, implying that the approximation of the stationary pattern $[\rhostateps(x),\etastateps(x)]$ of the full system by the solution $[\rhostat(x),\etastat]$ of the mass-conserving system remains accurate even at large source strengths $\varepsilon$ if $D_v$ is sufficiently large.

This observation can be used to verify that the deviations in Fig.~\ref{fig:numerics-eigenvalueMapping-cubicModel}(b) result from the deformation of the stationary profile by shifting the source term $s_1$ into the $v$ species via the replacement ${\tilde{f}\to\tilde{f}'=\eta+\varepsilon p - \rho^3+(1-\varepsilon)\rho}$ [cf.\ Eq.~\eqref{eq:new-f}].
By this, the profile equation, Eq.~\eqref{eq:profile-equation}, is free of any direct dependence on the source terms.
These only enter through Eq.~\eqref{eq:eta-stat} for $\etastateps$ and again only contribute with strength ${\sim \varepsilon/D_v}$.
Because the modified source term $\tilde{f}'$ is again a cubic polynomial in the density $\rho$, again the interface profile can be determined analytically.
However, due to the implicit dependence of $\tilde{f}'$ on the source strength $\varepsilon$, the condition Eq.~\eqref{eq:eps-threshold-mesas} has to be solved numerically for $\varepsilon_\mathrm{stop}$.
This modified approximation describes the stability threshold of the leading eigenvalue to high accuracy even at very large source strengths [red (curved) line in Fig.~\ref{fig:numerics-eigenvalueMapping-cubicModel}(b)].

Beyond the threshold $\varepsilon_\mathrm{stop}$, the analytic expression Eq.~\eqref{eq:sigma-eps} also predicts the magnitude of the growth rate $\sigma^\varepsilon$ of the mass-competition process. 
The behavior of the growth rates away from the stability threshold is analyzed in Figs.~\ref{fig:numerics-eigenvalueMapping-cubicModel}(c-f).
The full growth rate $\sigma^\varepsilon$ [blue (dark gray) lines] clearly shows a crossover between the diffusion- and the reaction-limited regime.
For small values of $D_v$, the growth rate is described by the diffusion-limited expression, Eq.~\eqref{eq:sigma-eps-diffLim} [orange (light gray) lines].
For large values of $D_v$, the growth rate follows the reaction-limited expression, Eq.~\eqref{eq:sigma-eps-mesa-reacLim} [green (left-most) lines].
The deviations between the analytically and the numerically determined growth rate at small values of $D_v$ can be attributed to the sharp-interface approximation:
As $D_v$ is decreased, the gradient in $\eta$ between the interfaces becomes steeper [see Fig.~\ref{fig:mesa-MC-modeApprox} and Eq.~\eqref{eq:mesa-int-cont-eq}], and it is less well justified to approximate $\eta$ as linear (cf.\ Fig.~\ref{fig:peak-MC-modeApprox}).
Overall, the behavior of the growth rates is described excellently by Eq.~\eqref{eq:sigma-eps}.

\begin{figure}
    \centering
    \includegraphics{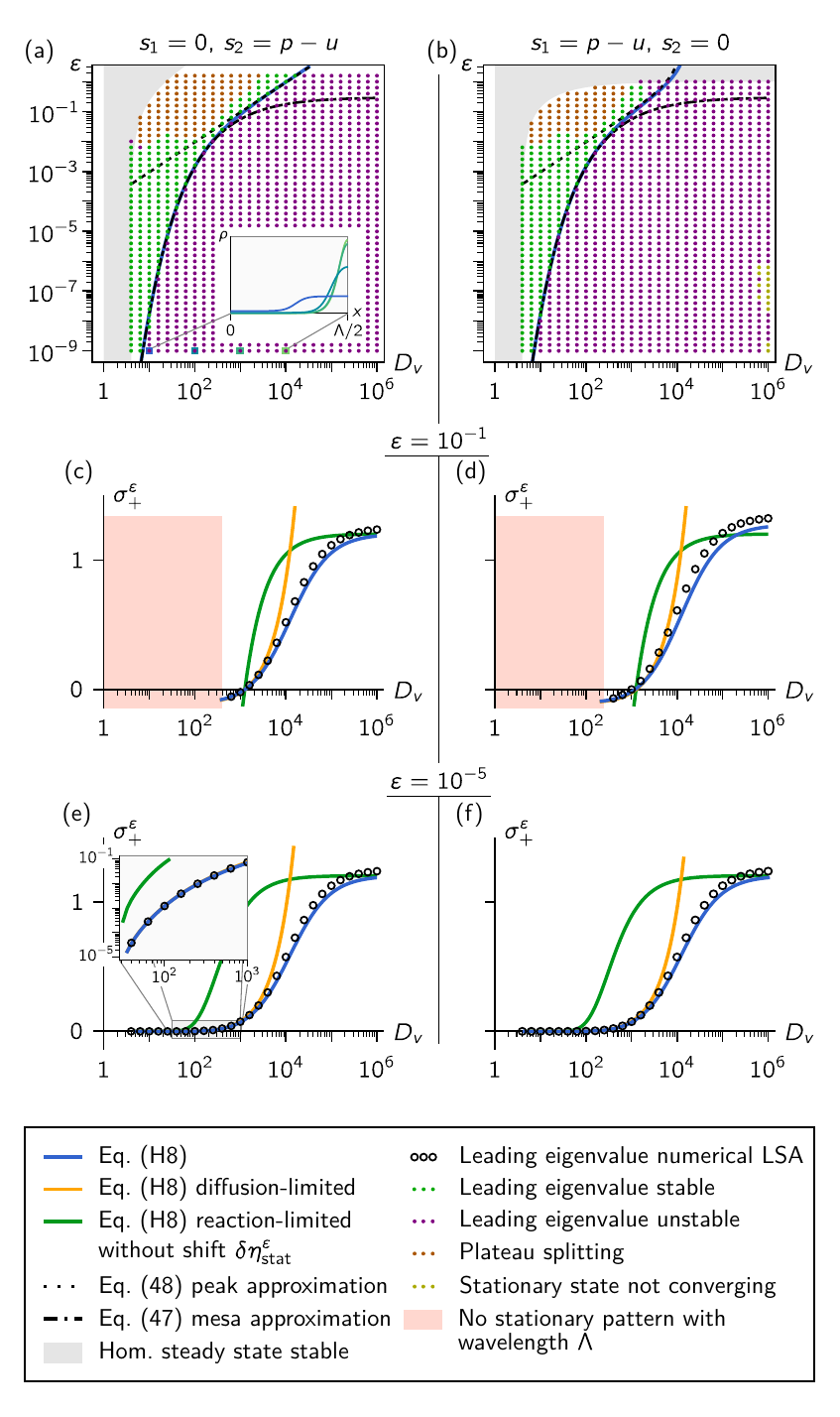}
    \caption{
    Stability threshold $\varepsilon_\mathrm{stop}$ and growth rate $\sigma^\varepsilon$ in the Brusselator model ${f(u,v) = u^2 v - u}$ with a linear source term $s=p-u$ in the fast- (a,c,e) and slowly diffusing (b,d,f) species (coloring as in the previous figure).
    In panels (a,b), the numerically determined stability regions for two half mesas or peaks (mesa/peak competition) are compared with the threshold of interrupted coarsening [see Eq.~\eqref{eq:sigma-eps-crossover}].
    A transition from mesa to peak patterns occurs with increasing diffusion constant $D_v$ [inset in (a)].
    Analytic approximations for the threshold $\varepsilon_\mathrm{stop}$ are found in the mesa-forming (black dash-dotted line) and the peak-forming (black dashed line) regimes (see Appendix~\ref{app:Brusselator}).
    (c-f) The value of the leading eigenvalue determined by numerical linear stability analysis (circles) is compared with the growth rate approximation, Eq.~\eqref{eq:sigma-eps} [see Eq.~\eqref{eq:sigma-eps-crossover}; blue (dark gray)].
    Here, the shown expression in the reaction-limited regime [green (left-most) line] does not include the effect of the shifted stationary mass-redistribution potential [Eq.~\eqref{eq:sigma-eps-mesa-reacLim}].
    In (d), this additional effect is visible as an increase in the full growth rate [blue (dark gray) line] at large $D_v$.
    The parameters of the model are ${\Lambda = 40}$, ${D_u = 1}$ and ${p = 2}$.
    }
    \label{fig:numerics-eigenvalueMapping-Brusselator}
\end{figure}

\subsubsection{The Brusselator model}
\label{sec:Brusselator}
A classical 2cRD system to study pattern formation is the Brusselator model \cite{Prigogine.Lefever1968}, a two-species chemical system originally introduced as the well-mixed system.
Accounting for the diffusion of both species, its dynamic equations read
\begin{subequations}
\begin{align}
    \partial_t u &= D_u\nabla^2 u + u^2 v - u + \varepsilon p - \varepsilon u
    \, ,\\
    \partial_t v &= D_v\nabla^2 v -(u^2 v - u)
    \, ,
\end{align}
\end{subequations}
where $\varepsilon p$ is a production term and $\varepsilon$ a degradation rate.
We analyze this 2cRD system in the limit of weak source terms by defining the mass-conserving `core' system ${f(u,v) = u^2 v-u}$ and adding the source terms ${(s_1,s_2) = (0,p-u)}$ [Fig.~\ref{fig:numerics-eigenvalueMapping-Brusselator}(a,c,e)] or ${(s_1,s_2) = (p-u,0)}$ [Fig.~\ref{fig:numerics-eigenvalueMapping-Brusselator}(b,d,f)].\footnote{
The Hopf bifurcation of the local reaction kinetics is avoided for a sufficiently low source strength $\varepsilon$ and ${p > 1}$.
For ${p<1}$ the homogeneous steady state lies on the unstable branch of the nullcline (the Brusselator core shows local bistability of the reaction kinetics).
For the construction of the final (large-amplitude) stationary patterns in the nearly mass-conserving regime and the slow, long-time dynamics close to the stationary states, the position of the homogeneous steady state and the bistability of the reaction kinetics is irrelevant (cf.\ discussion of bistability in Ref.~\cite{Brauns.etal2020}).}

In $\rho,\eta$-coordinates the reaction term,
\begin{equation}
\label{eq:Brusselator-reaction-term}
    \tilde{f}(\rho,\eta) 
    = 
    \frac{1}{1 -d}
    \left(
    \frac{\rho-\eta}{1-d}\right)^2 
    \bigg[
    \eta - 
    \bigg(
    \frac{\left(1-d\right)^2}{\rho-\eta} + d \rho
    \bigg)
    \bigg]
    \, ,
\end{equation}
depends on the relative diffusivity ${d = D_u/D_v}$.
In particular, the nullcline is $\mathsf{N}$-shaped and becomes strongly asymmetric for ${d \ll 1}$ because the (lower stable branch of the) nullcline approaches ${\eta^*(\rho) \approx \frac{1}{\rho} + d \rho}$ for large densities ${\rho\gg 1}$ and small relative diffusivity ${d\ll 1}$.
Consequently, the density $\rho_+$ of the high-density plateau of stationary patterns shifts to higher values as the relative diffusivity is increased.
If the mass of an elementary stationary pattern is kept constant, the half-mesa will become narrower and narrower as $\rho_+$ increases until the mesa pattern transitions into a peak pattern when the half-mesa width becomes of the order of the interface width $\ell_\mathrm{int}$.
The same occurs in the system with weakly broken mass conservation if the production term $p$ is kept constant [inset in Fig.~\ref{fig:numerics-eigenvalueMapping-Brusselator}(a)] because this approximately fixes the total peak/mesa mass (see Sec.~\ref{sec:stat-patterns-nMC}).
This change in the stationary profile results in a different behavior of the threshold ${\sigma^\varepsilon=0}$ which we analyze now.
In particular, in the peak-shaped regime, the shift of the stationary mass-redistribution potential ${\etastat\to\etastateps}$ may become relevant [cf.\ \eqref{eq:sigma-D-eps-correction}].

If the source terms act in the fast-diffusing species $v$ [see Fig.~\ref{fig:numerics-eigenvalueMapping-Brusselator}(a)], the shift $\varepsilon\,\delta\eta_\mathrm{stop}^\varepsilon$ in the stationary mass-redistribution potential is approximately zero for ${D_v\gg D_u = 1}$ [see Eq.~\eqref{eq:nMC-etastat-shift}].
At lower values of $D_v$, the pattern is of mesa shape in which case the shift ${\varepsilon\,\delta\eta_\mathrm{stop}^\varepsilon}$ does not affect mass competition (see Sec.~\ref{sec:growth-rate-nMC}).
Thus, in the analyzed scenario of mesa/peak competition the threshold $\varepsilon_\mathrm{stop}$ is determined by ${\varepsilon_\mathrm{stop} = \sigma_\mathrm{D}^+/\sigma_\mathrm{S}}$ [see Eq.~\eqref{eq:eps-threshold-mesas}].
This expression can be calculated by numerically determining $\partial_M^+\etastat$ and $\sigma_\mathrm{S}$ throughout the crossover from mesa to peak patterns [blue (dark gray) line].
In Appendix~\ref{app:Brusselator} the details of the analysis of the transition from mesa to peak patterns are discussed.
Moreover, in the limit of mesa and peak patterns, that is, $D_v\to 1$ and $D_v\to\infty$, analytic approximations for the stability threshold $\sigma^\varepsilon=0$ can be derived [black dashed and dash-dotted lines in Fig.~\ref{fig:numerics-eigenvalueMapping-Brusselator}(a,b); see Appendix~\ref{app:Brusselator}].
In both limits as well as in the crossover region the stability threshold describes the zero crossing of the numerically determined leading eigenvalue very well [see Fig.~\ref{fig:numerics-eigenvalueMapping-Brusselator}(a)].

In comparison to the threshold observed for cytosolic source terms, source terms in the slow-diffusing (membrane) species $u$ in the reaction-limited regime of large $D_v$ lead to an increase of the stability threshold $\varepsilon_\mathrm{crit}$ [see Fig.~\ref{fig:numerics-eigenvalueMapping-Brusselator}(b)].
This increase is due to the shift ${\varepsilon\,\delta\eta_\mathrm{stat}^\varepsilon}$ of the stationary mass-redistribution potential which fulfills ${\partial_M^{}\etastateps < 0}$ and becomes relevant in the regime ${D_v\gg D_u}$ for source terms in the slow-diffusing species [see Eq.~\eqref{eq:nMC-etastat-shift}].
In particular, patterns are unstable for any source strength $\varepsilon$ if ${D_v \gtrsim 10^4}$.
Also for source terms in the slowly diffusing species $u$ the stability threshold is well described.
As for the cubic model, deviations in the stability threshold occur at large $\varepsilon$, which is consistent with our derivation that assumes a small source strength.

In both cases with source terms in the $v$ or $u$ species, also the growth rates [see Fig.~\ref{fig:numerics-eigenvalueMapping-Brusselator}(c-f)] away from threshold are well described by the numerically calculated, full expression for $\sigma^\varepsilon$ [blue (dark gray) line; see Appendix~\ref{app:Brusselator}].
The diffusion-limited regime at low values of $D_v$ is accurately captured by Eq.~\eqref{eq:sigma-eps-diffLim} [orange (light gray) line].
We also show the expression derived in the reaction limit for mesa patterns [see Eq.~\eqref{eq:sigma-eps-mesa-reacLim}; green (left-most) line].
Thus, the additional effect of the shift $\varepsilon\,\delta\eta_\mathrm{stat}^\varepsilon$ is not included here.
Consequently, the increase in the growth rate in the reaction-limited regime induced by this shift is seen by comparison of the blue (dark gray) and green (left-most) lines in Fig.~\ref{fig:numerics-eigenvalueMapping-Brusselator}(d).
This already indicates that the shift of the stationary mass-redistribution potential $\varepsilon\,\delta\eta_\mathrm{stat}^\varepsilon$ is indeed relevant in the reaction-limited regime.
This becomes clearer in the third example analyzed below.

Comparing the growth rate predictions with the numerically determined values, we find that the growth rates are well approximated.
Deviations arise in the reaction limit, which are explained by the large [order $\mathcal{O}(1)$] reaction-limited growth rate of the peak patterns.
Such fast mass competition violates the assumption that the competition process is slow compared to the relaxation rates of the single peaks.
The slowest relaxation modes can be numerically determined to be of order $\mathcal{O}(1)$ as well.
In fact, the approximation is accurate in a surprisingly large parameter range.

\begin{figure}
    \centering
    \includegraphics{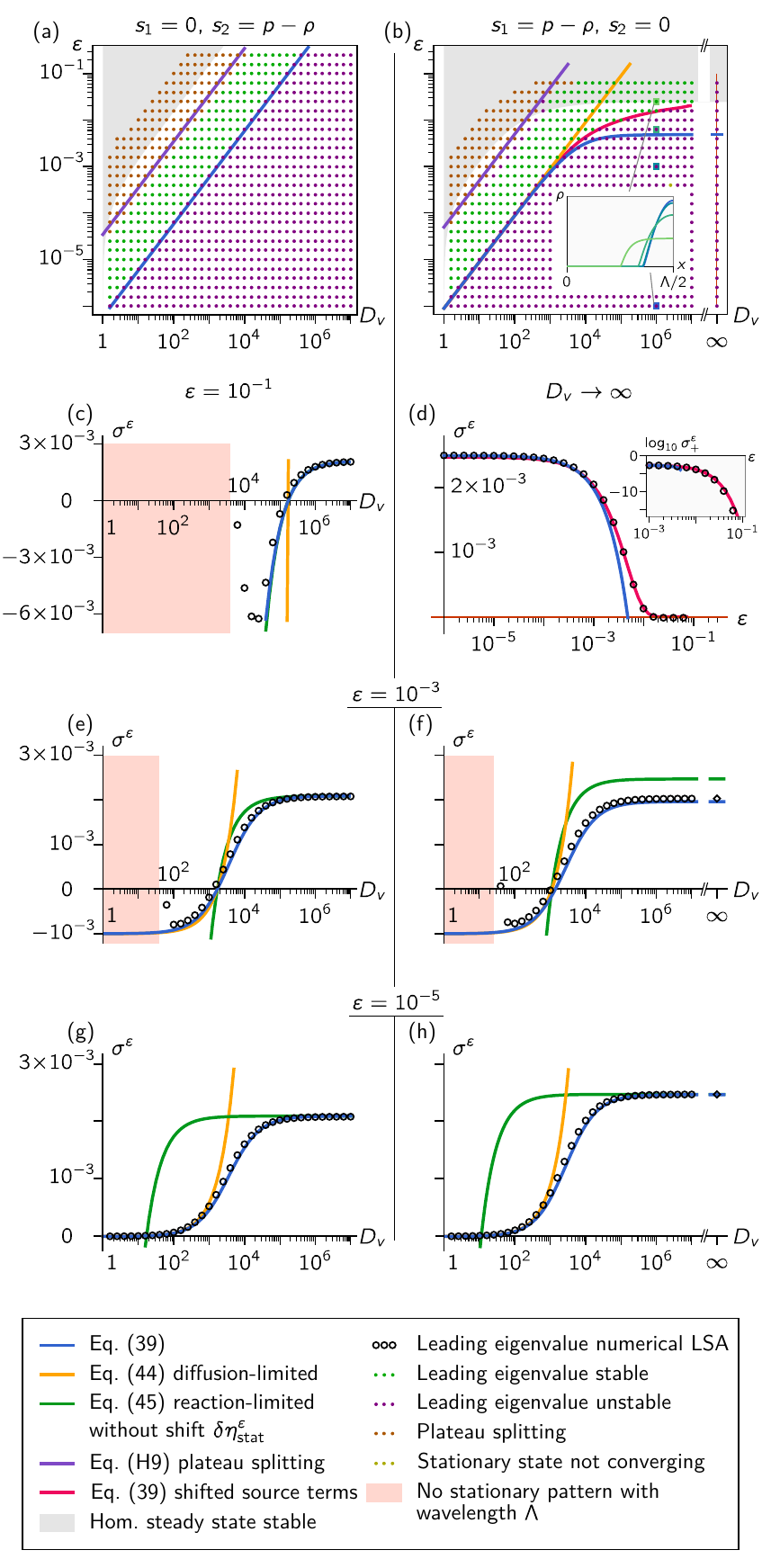}
    \caption{
    Stability threshold $\varepsilon_\mathrm{stop}$ and growth rate $\sigma^\varepsilon$ in the peak-forming model ${\tilde{f}(\rho,\eta) = \eta - a \rho / \left(1+\rho^2\right)}$ with source term ${s=p-\rho}$ in the $v$ (a,c,e,g) and $u$ (b,d,f,h) species (coloring as in the previous figures).
    The stationary pattern transitions from a peak to a mesa shape with increasing source strength $\varepsilon$ for source terms in the slowly diffusing species $u$ [inset in (b)].
    The red (light gray, curved) line denotes the threshold $\varepsilon_\mathrm{crit}$ derived by shifting all source terms into the fast species $v$ [cf.\ Eq.~\eqref{eq:new-f}].
    The purple (top-most, diagonal) line depicts the threshold of plateau splitting given in Ref.~\cite{Brauns.etal2021} [see Eq.~\eqref{eq:constReacRateModel-splitting}].
    Moreover, in (b,d,f,h) results for the cAC (shadow) limit are included (${D_v\to\infty}$).
    (d) In the cAC (shadow) limit at high source strength $\varepsilon$, the growth rate is well described by the modified approximation with all source terms acting in the $v$ species.
    The parameters are ${\Lambda = 200}$, ${D_u = 1}$, ${a = 10}$ and ${p = 20}$.
    }
    \label{fig:numerics-eigenvalueMapping-constReacRateModel}
\end{figure}

\subsubsection{A prototypical model for peaks patterns}
\label{sec:const-reac-rate-peak-model}
So far we have analyzed the cubic model as a simple example of a mesa-forming system and the Brusselator model as a system showing a transition from mesa to peak patterns.
As a final example, let us analyze a model that is prototypical for peak-forming systems.
Such a model is obtained by constructing a reaction term with a $\mathsf{\Lambda}$-shaped nullcline.
For simplicity, we choose a conceptual model with the reaction term
\begin{equation}\label{eq:constReacRateModel}
    \tilde{f}(\rho,\eta) = \eta -a \frac{\rho}{1+\rho^2}
    \, ,
\end{equation}
which has the $\mathsf{\Lambda}$-shaped nullcline ${\eta^*(\rho)=a\rho/(1+\rho^2)}$ with parameter $a$.
Because $\tilde{f}$ is independent of the diffusion constants, the corresponding 2cMcRD system yields stationary patterns that are also independent of the diffusion constants [see Eqs.~\eqref{eq:gen-stat-pattern}].

We are particularly interested in how the shift of the stationary mass redistribution potential ${\etastat\to\etastateps}$ affects the mass-competition rate $\sigma^\varepsilon$, since the shift is only relevant for peak patterns [see Sec.~\ref{sec:growth-rate-nMC}].
To examine the significance of the shift for source terms in the $u$ or $v$ species [cf.\ Eq.~\eqref{eq:sigma-eps-peak-reacLim}], we again compare the source terms ${(s_1, s_2) = (0, p-\rho)}$
[Fig.~\ref{fig:numerics-eigenvalueMapping-constReacRateModel}(a,c,e,g)] with the reverse assignment ${(s_1, s_2) = (p-\rho, 0)}$ [Fig.~\ref{fig:numerics-eigenvalueMapping-constReacRateModel}(b,d,f,h)].
For the case that the source terms affect the membrane species $u$, simulation results in the shadow limit are provided, which show the effect of the source terms in the limiting cAC system [results for ``${D_v\to\infty}$'' in Fig.~\ref{fig:numerics-eigenvalueMapping-constReacRateModel}(b,d,f,h)].

Because the stationary profile for the chosen source term $\tilde{f}$ is not known analytically, we have determined $\partial_M^{}\etastat$ and the various other terms in $\sigma^\varepsilon$ as well as the stability threshold $\varepsilon_\mathrm{stop}$ numerically [see Eqs.~\eqref{eq:sigma-eps},~\eqref{eq:eps-threshold-peaks}]. 
For the stability threshold $\varepsilon_\mathrm{stop}$ we find a linear scaling ${\varepsilon_\mathrm{stop}\propto D_v}$ with the larger diffusion constant $D_v$ in the diffusion-limited regime and, for the case where the source terms only affect the $v$ species, also in the reaction-limited regime ${D_v\gg 1}$ [see Fig.~\ref{fig:numerics-eigenvalueMapping-constReacRateModel}(a) and orange (light gray, straight) line in Fig.~\ref{fig:numerics-eigenvalueMapping-constReacRateModel}(b)].
In these cases the threshold is independent of the shift $\varepsilon\, \delta\etastateps$ [cf.\ Eqs.~\eqref{eq:nMC-etastat-shift},~\eqref{eq:sigma-eps-diffLim}].
In addition, the stationary patterns of the mass-conserving model, Eq.~\eqref{eq:constReacRateModel}, are independent of $D_v$.
Therefore, as for the cubic model, we find a linear threshold here.

If the source terms are introduced in the slow-diffusing species $u$ [see Fig.~\ref{fig:numerics-eigenvalueMapping-constReacRateModel}(b)], the shift $\varepsilon\,\delta\eta_\mathrm{stat}^\varepsilon$ becomes relevant in the reaction-limited regime and affects the stability threshold $\varepsilon_\mathrm{stop}$.
Opposite to the Brusselator model in Sec.~\ref{sec:Brusselator}, here we find an increase ${\varepsilon\,\partial_M^{}\delta\eta_\mathrm{stat}^\varepsilon >0}$ of the shift in the stationary mass-redistribution potential for larger peaks.
The reason is that the reaction rate ${\tilde{f}_\eta=\partial_\eta\tilde{f}}$ is constant here while it increases strongly with increasing peak size in the Brusselator model [${\tilde{f}_\eta\sim \rho^2}$, see Eq.~\eqref{eq:Brusselator-reaction-term}], and the shift $\delta\etastateps$ becomes smaller with increasing conversion rate $\tilde{f}_\eta$ [see Eq.~\eqref{eq:nMC-etastat-shift}].
The increase in the stationary mass-redistribution potential caused by the increase ${\varepsilon\,\partial_M^{}\delta\eta_\mathrm{stat}^\varepsilon >0}$ counteracts the mass-competition instability driven by the \emph{decrease} of the stationary mass-redistribution potential for larger peaks.
Consequently, the threshold $\varepsilon_\mathrm{stop}$ of interrupted coarsening is strongly decreased in the reaction-limited regime compared to the diffusion-limited bound [compare the blue (dark gray) and orange (light gray) lines in Fig.~\ref{fig:numerics-eigenvalueMapping-constReacRateModel}(b)].

Although the full expression for the stability threshold $\varepsilon_\mathrm{stop}$, Eq.~\eqref{eq:eps-threshold-peaks}, qualitatively captures that the stability threshold is decreased, this formula does not quantitatively describe the zero crossing of the numerically determined leading eigenvalue.
Again, the deviations from the zero crossing of the leading eigenvalue are caused by the deformation of the stationary peak profile by the source terms in the slow-diffusing species $u$ [see inset in Fig.~\ref{fig:numerics-eigenvalueMapping-constReacRateModel}(b)].
Because of strong degradation at large densities $\rho$, the peak amplitude is reduced and a transition toward a mesa-shaped pattern occurs as the source strength $\varepsilon$ is increased.
Again, we can shift the source term $s_1$ into the $v$ species by the modification ${\tilde{f}\to\tilde{f}'=\tilde{f}+\varepsilon s_1}$ to account for this deformation caused by source terms in the slowly diffusing species $u$.
The resulting modified nullcline is $\mathsf{N}$-shaped, explaining the transition from peak to mesa patterns.
The implicit dependence of $\tilde{f}'$ on the source strength $\varepsilon$ requires that the stationary pattern be calculated numerically for each source strength $\varepsilon$.
Based on these stationary profiles
the stability threshold $\varepsilon_\mathrm{stop}$ can be recalculated.
Since the modification ${\tilde{f}\to\tilde{f}'}$ ensures that the shift $\delta\etastateps$ vanishes (see Sec.~\ref{sec:growth-rate-nMC-discussion}), the threshold is determined by $\varepsilon_\mathrm{stop}\sigma_\mathrm{S}=\sigma_\mathrm{D}$ [cf.\ Eq.~\eqref{eq:eps-threshold-peaks}], where all terms are evaluated using the modified reaction term $\tilde{f}'$.
This threshold is highly accurate throughout the whole tested parameter regime [red (light gray, curved) line in Fig.~\ref{fig:numerics-eigenvalueMapping-constReacRateModel}(b)].

For mesa-forming systems, it has been shown mathematically that in the shadow limit the mass-competition instability always destabilizes the patterns and that coarsening is always uninterrupted~\cite{McKay.Kolokolnikov2012,Ni.etal2001}.
Due to the transition from peak to mesa patterns at large source strengths $\varepsilon$, we find the same here [see Fig.~\ref{fig:numerics-eigenvalueMapping-constReacRateModel}(b)].
It is an interesting question for future work whether a transition to mesa patterns generally occurs near the threshold to interrupted coarsening for large cytosolic diffusion constants $D_v$, such that it is indeed impossible to observe interrupted coarsening in the shadow limit also for peak-forming systems.

Let us now turn to the magnitude of the growth rate $\sigma^\varepsilon$. Above, we argued that the increase of the shift $\delta\etastateps$ with peak mass reduces the strength of destabilizing mass transport between peaks and thus decreases the stability threshold $\varepsilon_\mathrm{stop}$.
Hence, the magnitude of the growth rate $\sigma^\varepsilon$ must also be reduced by the shift.
Let us, therefore, analyze the growth rate away from the stability threshold.
First, one notes that the growth rates themselves are well described by the full expression $\sigma^\varepsilon$ as long as the source strength $\varepsilon$ is sufficiently small [see Fig.~\ref{fig:numerics-eigenvalueMapping-constReacRateModel}(c-h)].
The diffusion-limited expression, Eq.~\eqref{eq:sigma-eps-diffLim}, approximates the growth rate at low values of $D_v$, the reaction-limited expression [Eq.~\eqref{eq:sigma-eps-peak-reacLim}, again without considering the shift $\varepsilon\,\delta\eta_\mathrm{stat}^\varepsilon$ by setting ${\partial_M^{}\delta\etastateps=0}$] describes the growth rate at large values of $D_v$.
When the source terms affect the slow species $u$, the comparison of the full growth rate $\sigma^\varepsilon$ with the reaction-limited growth rate ${\sigma_\mathrm{R}( 1 -\varepsilon\sigma_\mathrm{S}/\sigma_\mathrm{D})}$ determined without the shift $\delta\etastateps$ [set ${\delta\etastateps=0}$ in Eq.~\eqref{eq:sigma-eps-peak-reacLim}] shows the decrease of the growth rate due to the behavior of the shift $\varepsilon\,\delta\eta_\mathrm{stat}^\varepsilon$ [see Fig.~\ref{fig:numerics-eigenvalueMapping-constReacRateModel}(f)].
Also in the cAC limit [see Fig.~\ref{fig:numerics-eigenvalueMapping-constReacRateModel}(d)], the growth rate is well described by the reaction limit of $\sigma^\varepsilon$, i.e., Eq.~\eqref{eq:sigma-eps-peak-reacLim} at sufficiently small values of the source strength $\varepsilon$.
The modified approximation obtained by replacing $\tilde{f}$ with $\tilde{f}'$ describes the growth rate also well at large values of the source strength $\varepsilon$.

Thus, the overall behavior of the growth rates is well captured by $\sigma^\varepsilon$, Eq.~\eqref{eq:sigma-eps}.
Note at last that large deviations in the growth rate appear close to the regime of plateau splitting because the stationary profile is strongly deformed.
The threshold for plateau splitting can be estimated analytically by the criterion derived in Ref.~\cite{Brauns.etal2021} [purple (top-most, diagonal) lines in Fig.~\ref{fig:numerics-eigenvalueMapping-constReacRateModel}(a,b)].
Therefore, one can predict in which regime the approximation will fail.
\par

In summary, the analysis of the above three exemplary models shows that the expression Eq.~\eqref{eq:sigma-eps} for the growth rate $\sigma^\varepsilon$ offers insights into a wide variety of phenomena.
It explains the suppression of the mass-competition instability at large source strengths $\varepsilon$ and
the different effects source terms, which affect either the slow or the fast diffusing species, have on the stability threshold.
Importantly, we have shown that the stability threshold is well approximated by the simple condition ${\varepsilon_\mathrm{stop}\sigma_\mathrm{S}=\sigma_\mathrm{D}}$ if all source terms are chosen to affect only the cytosolic species [cf.\ Eqs.~\eqref{eq:eps-threshold-mesas},~\eqref{eq:eps-threshold-peaks},~\eqref{eq:nMC-etastat-shift}].
This is the criterion which was obtained in Ref.~\cite{Brauns.etal2021} based on a QSS approximation for the mass-redistribution potential at individual pattern domains.

\section{Discussion}
\label{sec:discussion}

Motivated by complex biochemical protein systems, there is a general interest in studying the dynamics of many-component systems composed of different particle species interacting on a spatially extended domain.
Intriguingly, these systems can form spatially heterogeneous patterns, for example, via a coupling of reactions and particle diffusion, or by phase separation.
Both processes are known to play an important role in the spatial self-organization of the cell \cite{Halatek.etal2018,Burkart.etal2022a,Hyman.etal2014,Shin.Brangwynne2017}.
Phase separation, and biomolecular condensation, as well as intracellular protein-pattern formation are processes that (approximately) conserve the total number of molecules of each species.

\textit{Mass conservation.\;---}
In classical multi-component fluid systems, conservation of mass results in a continuity equation for the densities dictating the dynamics of each component.
For example, in binary mixtures with two particle species A and B, phase-separation dynamics must respect that both densities $\rho_\mathrm{A,B}(\mathbf{x},t)$ follow a continuity equation.
Under an incompressibility constraint for the whole fluid system, the well-known Cahn--Hilliard (Model B) dynamics follows. 

In contrast, for protein pattern formation by a reaction--diffusion mechanism, it is important to realize that the diffusible proteins take on different conformations, with (chemical) reactions describing the transition between these different states.
The concentrations of the individual states are not conserved; that is, they do not obey continuity equations.
Nevertheless, the total density of all the different conformations for each protein is conserved and follows a continuity equation if pattern formation is fast compared to protein turnover by gene expression and protein degradation.
Thus, the \emph{total} densities are control parameters of the (local) dynamics, and the redistribution of the total densities is crucial to the system dynamics on long scales \cite{Halatek.Frey2018,Brauns.etal2020,Denk.Frey2020,Wurthner.etal2022}.
Both in Cahn--Hilliard and reaction--diffusion systems, broken mass conservation can be accounted for by source terms in the continuity equation(s).

Building on the common feature of mass conservation, we demonstrated in Ref.~\cite{Brauns.etal2021} that the concept of coarsening dynamics, usually employed in the context of phase separation, is useful to understand wavelength selection in reaction--diffusion systems governed by an (approximate) conservation law.
Here, we have elaborated this approach for (nearly) mass-conserving two-component reaction--diffusion (2cRD) systems, which are paradigmatic models for intracellular protein pattern formation. 
In particular, we have substantiated our findings by a thorough, model-independent mathematical analysis based on singular perturbation theory, which enabled us to find explicit expressions for the growth rates determining the pattern dynamics.
Furthermore, we have developed a systematic link to classical phase separation as described by the Cahn--Hilliard and the conserved Allen--Cahn model.
We anticipate that our mathematical analyses and the correspondences between different types of models will provide a starting point and conceptual basis for the future analysis of systems with a larger number of components.

\textit{The mass-competition process.\;---}
In strictly mass-conserving 2cRD systems, one observes coarsening much like the Cahn--Hilliard or conserved Allen--Cahn dynamics describing phase separation.
In the presence of weak source terms of strength $\varepsilon$, the coarsening process is interrupted at a characteristic pattern length scale determined by  $\varepsilon$.
The same behavior is found when non-equilibrium reaction terms are introduced in equilibrium phase-separating systems \cite{Glotzer.etal1995,Weber.etal2019,Li.Cates2020}.
Our mathematical analysis shows that this length-scale selection is caused by processes that change the mass of single pattern domains (peaks or mesas):
On the one hand, mass is exchanged diffusively between different domains.
Net transport is determined by the gradients in the mass-redistribution potential $\eta$, which acts analogously to the chemical potential in equilibrium phase separation.
If larger domains have a lower mass-redistribution potential $\eta$, the resulting gradients lead to mass transport from smaller to larger domains, inducing further growth of the already larger domain.
Then, mass exchange destabilizes two interacting domains, a process we term \emph{mass-competition instability}.
On the other hand, source terms induce local net production and degradation which restore the mass of each domain to its stationary value and therefore counteract the instability.

\textit{How pattern wavelength determines stability.\;---}
How can one determine the stability of a pattern with a characteristic length scale (wavelength) $\Lambda$?
For this, we need to understand whether mass exchange is destabilizing and whether the destabilizing mass exchange or the stabilizing production or degradation processes are stronger.
Our analysis demonstrates that these questions are answered by analyzing the corresponding mass-conserving system without source terms.
The strength and stability of the mass-competition process are determined by the expression $\partial_M^{}\etastat$ of the corresponding mass-conserving system, i.e., by answering how the stationary mass redistribution potential changes with the mass $M$ of a single pattern domain (peak or mesa).
At this point, the phase-space construction for patterns of 2cMcRD systems gives rise to several model-independent insights.
First, 2cMcRD systems always fulfill the coarsening condition ${\partial_M\etastat<0}$ along the complete branch of stable stationary elementary patterns \cite{Brauns.etal2021}, implying that mass redistribution through gradients in $\eta$ always destabilizes a pattern toward larger length scales and domain sizes [${M\approx (\bar{\rho}-\rho_-)\Lambda}$] via the collapse of small domains.
Second, we find that $\partial_M^{}\etastat$ typically decreases rapidly as a function of the domain size (at least ${\sim M^{-\alpha-1}}$ with ${\alpha>0}$).

For 2cRD systems with source terms, it follows that patterns at sufficiently small length scales ${\Lambda<\Lambda_\mathrm{stop}}$ are destabilized by mass competition between domains.
Above the stability threshold ${\Lambda_\mathrm{stop}}$ however, mass competition is weaker than the stabilizing source terms, and the mass-competition instability is suppressed at length scales $\Lambda>\Lambda_\mathrm{stop}$.

This stability threshold extends the classical Eckhaus stability boundary which is derived from amplitude equations close to the supercritical onset of pattern formation \cite{Cross.Hohenberg1993}.
In our formulation of the 2cRD system, this supercritical onset is reached at large source strengths where the band of unstable modes $q_\mathrm{c}\pm \sqrt{\mu}$ vanishes [type-$\mathrm{I_s}$ instability following Ref.~\cite{Cross.Hohenberg1993}; cf.\ Fig.~\ref{fig:dispersion-relations}(b)].
Here, the parameter $\mu$ parametrizes the distance from the onset.
The Eckhaus instability destabilizes all patterns with wavelengths larger or smaller than ${q_\mathrm{E}^\pm = q_\mathrm{c}\pm \sqrt{\mu/3}}$, respectively.
Starting from a short-wavelength pattern, ${q>q_\mathrm{E}^+}$, this long-wavelength instability leads to an increase of the pattern wavelength into the Eckhaus-stable regime similar to coarsening.
While these results are based on the universal form of the amplitude equations, the form of the stability boundary further away from onset is strongly system-dependent and may be influenced by several different instabilities \cite{Cross.Hohenberg1993,Veerman.etal2021}.

What can we say about the stability threshold further away from the onset of pattern formation in general?
While in the Eckhaus regime the shortest stable wavelength $2\pi/q_\mathrm{E}^+$ decreases with the distance from onset, our analysis implies that---approaching the mass-conserving limit of 2cRD systems and generalized phase-separating systems---this stability boundary turns around and starts increasing, i.e., ${\Lambda_\mathrm{stop} \to \infty}$ as the stabilizing source effects become weaker and weaker.
Thus, the dominant wavelength $2\pi/q_\mathrm{c}$ of the initial instability is crossed at a low source strength, destabilizing patterns at the initial wavelength and inducing the coarsening process [cf.\ Fig.~\ref{fig:interrupted-coarsening}(a)].
The coarsening process stops once it has driven the pattern length scale $\Lambda$ above the stability boundary $\Lambda_\mathrm{stop}$ [cf.\ Fig.~\ref{fig:interrupted-coarsening}(c)].
Consequently, $\Lambda_\mathrm{stop}$ predicts the final pattern length scale in this case of interrupted coarsening.
On the contrary, the wavelength $2\pi/q_\mathrm{c}$ determines the length scale of the final pattern if the threshold $\Lambda_\mathrm{stop}$ lies at lower wavelengths, that is, in the regime of strong source terms.
The fact that mass competition is weak due to the typically strong decrease of $\partial_M^{}\etastat$ as a function of $\Lambda$ (in particular in mesa-forming systems) explains the large parameter regimes observed in general 2cRD systems where the length scale $2\pi/q_\mathrm{c}$ is informative even for highly nonlinear patterns far from onset.
Interrupted coarsening is only observed for very weak source strengths (near the mass-conserving limit).
Although weak mass competition is necessary to obtain patterns with a final wavelength ${\Lambda\approx2\pi/q_\mathrm{c}}$, we note that broken mass-conservation may give rise to other instabilities such as oscillatory instabilities \cite{Kolokolnikov.etal2006} or domain splitting \cite{Kolokolnikov.etal2007,Brauns.etal2021}. 
Taken together, the comparison of $2\pi/q_\mathrm{c}$ and $\Lambda_\mathrm{stop}$ answers the important question of how the final pattern wavelength is determined given that all patterns with wavelengths ${\Lambda>\Lambda_\mathrm{stop}}$ are stable against mass competition (cf.\ Ref.~\cite{Politi.Torcini2015,Veerman.etal2021}).

\textit{Pattern formation and phase separation.\;---}
In the reaction--diffusion dynamics, the mass-redistribution potential $\eta$ plays a role analogous to the chemical potential $\mu$ in close-to-equilibrium systems describing phase separation.
This allows us to find a systematic link between (nearly) mass-conserving 2cRD systems and (generalized) CH as well as cAC systems, two standard models for phase separation.
Gradients in $\eta$ induce mass transport in the same way as gradients in the chemical potential $\mu$ act in CH (Model B) dynamics.
However, the mass-redistribution potential $\eta$ is not the derivative of a free-energy functional.
It does not adjust instantaneously to a given (total) density profile [cf.\ Eq.~\eqref{eq:Model-B-mu}] but by reactive conversion between the chemical species $u$ and $v$.
In the biologically relevant limit ${D_v\gg D_u}$, ${\eta\approx v}$ is the cytosolic concentration.
Consequently, the timescale of the mass-competition process is the sum of a reactive timescale---describing the particle release from a domain (mainly formed in $u$) into the cytosol and incorporation at a different domain---and a diffusive timescale---accounting for diffusive mass transport in the cytosol by gradients in $\eta$.
In phase separation, the former process is modeled by the cAC equation.
Thus, it approximates the reaction--diffusion dynamics in the \emph{reaction-limited} regime in which the reactive timescale is rate-limiting.
In contrast, the diffusive mass transport is captured by the CH equation (Model B) which describes the \emph{diffusion-limited} regime of 2cRD dynamics.
In the context of precipitates, it was already noted by Wagner that these two regimes of the coarsening dynamics occur \cite{Wagner1961}.

\textit{Outlook.\;---}
Remarkably, reaction limitation, that is, the dynamics of the mass-redistribution potential $\eta$ is irrelevant for the threshold of interrupted coarsening and the wavelength thereby selected.
The functional dependence of the stationary mass-redistribution potential $\etastat(M)$ on the domain mass $M$---and the gradients in $\eta$ thereby created---alone determine the wavelength where coarsening stops.
Therefore, we expect our analysis to apply quite generally to systems in which a (total) density is governed by a modified continuity equation [cf.\ Eq.~\ref{eq:cont-eq}].
As similar techniques have been applied to explain wavelength selection in models of active phase separation \cite{Tjhung.etal2018}, it seems promising to apply these to further active extensions of CH systems, for example, models for non-reciprocally interacting active matter \cite{Saha.etal2020, You.etal2020, Frohoff-Hulsmann.etal2021}.
Our findings may also help to understand the structuring of the cell interior by intracellular condensates interacting with reaction--diffusion systems \cite{John.Bar2005a,Zwicker2022}.

Moreover, this work builds the basis to analyze wavelength selection in reaction--diffusion systems with more than two components.
It is an important open question how wavelength selection proceeds in such systems  \cite{Raskin.deBoer1999,Halatek.etal2018,Chiou.etal2021,Ren.etal2023a}.
Understanding the connection between model parameters and the typical pattern wavelength will enable further insights into biological pattern-forming protein systems based on the phenomenology these show.
It is interesting to analyze whether the mechanisms may be reduced to the elementary motive of interrupted coarsening we elaborated here, and which mechanistically different pathways occur.
This question is particularly interesting for systems showing oscillating pattern domains \cite{Brauns.etal2021a}, traveling waves \cite{Beta.etal2020,Bement.etal2015}, or spatiotemporal chaos \cite{Halatek.Frey2018}.

We focused here on one-dimensional systems to isolate the effects underlying the mass-competition instability but a new effect arises in higher dimensions.
The interfaces of the protein domains can be curved which leads to a shift of the stationary mass-redistribution potential proportional to their curvature (see supplementary material of Ref.~\cite{Brauns.etal2021} and Ref.~\cite{Tateno.Ishihara2021}).
The effect of curved interfaces has been discussed in depth in phase separation and bistable media \cite{Pismen.Pomeau2004,Mikhailov1990,Nepomnyashchy2010}.
For mesa patterns in 2cMcRD systems, one recovers the standard coarsening law ${\langle\Lambda\rangle\sim t^{1/3}}$ in the diffusion-limited regime in two or three dimensions while the exponent for peak patterns remains dependent on the reaction kinetics \cite{Brauns.etal2021,Tateno.Ishihara2021}.
Moreover, droplet splitting occurs in higher dimensions for phase-separating systems with additional reactions \cite{Zwicker.etal2017} as well as in 2cRD systems \cite{Pearson1993}.
In one-dimensional systems, the analog is the splitting of peaks or mesas.
This process can be analyzed geometrically for 2cRD systems and determines the wavelength selected for patterns that develop on a growing domain \cite{Crampin1999,Crampin.etal2002,Brauns.etal2021}.
Because of the strong similarities we uncovered between 2cRD, CH and cAC systems, we believe it is interesting to analyze similarities in the shape instabilities as well.
We expect that such an analysis is another important step to explain the diverse pattern types emerging from protein interactions on two-dimensional membranes \cite{Vecchiarelli.etal2016,Denk.etal2018,Glock.etal2019,Brauns.etal2021b}.

Another aspect of domain geometry is bulk-surface coupling.
Owing to the cycling of proteins between membrane-bound and cytosolic states, this is a fundamental property of many protein-based, pattern-forming systems, and has a profound impact on the patterns that emerge \cite{Levine.Rappel2005,Halatek.Frey2012,Gessele.etal2020,Brauns.etal2021b,Wurthner.etal2022,Burkart.etal2022a}.
However, the impact of bulk-surface coupling on wavelength selection remains largely unexplored and is an important open problem for future research.

Lastly, we have focused exclusively on deterministic mass competition here, but noise could be important because deterministic mass competition quickly becomes weak with increasing domain size.
It is an interesting open question whether stochastic coarsening laws as determined for the (mesa-forming) noisy CH equation \cite{Kawakatsu.Munakata1985} may be used to determine the wavelength selected in stochastic systems (cf.\ Ref.~\cite{Li.Cates2020}).

Relating back to the biological context, pattern formation does not (always) start out from a homogeneous steady state but for example via nucleation, boundary effects, or initial, spatial templates such that the pattern is part of a whole pattern cascade \cite{Cates.etal2010,Wigbers.etal2020,Wigbers.etal2021,Burkart.etal2022a}.
Allowing for a transient coarsening process which is interrupted at a scale separated from the scale of the initial instability may give a robust mechanism to select a pattern.
The length scale thereby selected is independent of the initial process which triggers the pattern formation process.
It is an interesting future task to search for imprints of interrupted coarsening in biochemical protein systems.

\begin{acknowledgments}
This work was funded by the Deutsche Forschungsgemeinschaft (DFG, German Research Foundation) through the Collaborative Research Center (SFB) 1032---Project-ID No.\ 201269156---and the Excellence Cluster ORIGINS under Germany’s Excellence Strategy---Grant EXC 2094---390783311.
This research was conducted within the Max Planck School Matter to Life supported by the German Federal Ministry of Education and Research (BMBF) in collaboration with the Max Planck Society.
\end{acknowledgments}

\appendix

\section{Linear stability analysis of the homogeneous steady states}
\label{app:dispRel}
In this section, we provide the mathematical analysis for the dispersion relation of the (nearly) mass-conserving 2cRD system and discuss the limits of small and large wavenumbers.
Further details on the diffusion- and reaction-limited regimes of the mass-redistribution instability of the HSS, its geometric representation in phase space, and bifurcation diagrams are derived in Ref.~\cite{Brauns.etal2020}.

In the nearly mass-conserving 2cRD system, the dispersion relation describing the stability properties of a HSS $(\rho_\mathrm{HSS},\eta_\mathrm{HSS})$ is found as follows.
The growth rates $\sigma_\mathrm{HSS}(q)$ of Fourier modes $(\delta\rho_\mathrm{q},\delta\eta_\mathrm{q})$ with wave vector $q$ are the eigenvalues of the linearized dynamics [cf.\ Eqs.~\eqref{eq:cont-eq},~\eqref{eq:eta-evol}].
Thus, they fulfill
\begin{equation*}
    \sigma_\mathrm{HSS}(q) \begin{pmatrix}
    \delta\rho_\mathrm{q}\\
    \delta\eta_\mathrm{q}
    \end{pmatrix}
    =
    \left(\mathbf{L}+\varepsilon\mathbf{S}\right)
    \begin{pmatrix}
    \delta\rho_\mathrm{q}\\
    \delta\eta_\mathrm{q}
    \end{pmatrix} 
    ,
\end{equation*}
with the Jacobian matrices
\begin{align*}
    \mathbf{L} &= \begin{pmatrix}
    0 & -D_v q^2 \\
    D_u q^2 -\tilde{f}_\rho  & -\left(D_v+D_u\right) q^2-\tilde{f}_\eta
    \end{pmatrix}
    ,\\
    \mathbf{S} &= \begin{pmatrix}
    \partial_\rho s_\mathrm{tot} & \partial_\eta s_\mathrm{tot}\\
    \partial_\rho \left(s_2 + d s_1\right) & \partial_\eta \left(s_2 + d s_1\right)
    \end{pmatrix}
    ,
\end{align*}
where ${\tilde{f}_{\rho,\eta} := \partial_{\rho,\eta}\tilde{f}}$, and all terms are evaluated at the steady-state densities $(\rho_\mathrm{HSS},\eta_\mathrm{HSS})$.
This yields the two branches of the dispersion relation of which one is stable for all wavenumbers $q$ while the other one can show a band of unstable modes.
For small wavenumbers $q\to 0$ and small source strength ${\varepsilon\ll 1}$, the unstable branch approaches
\begin{equation*}
    \sigma_\mathrm{HSS}(q) \approx
    -D_v \left(\partial_\mathrm{\rho}\eta^*\right) q^2 + \varepsilon\, \partial_\rho s_\mathrm{tot}(\rho,\eta^*(\rho))
    \, .
\end{equation*}
This agrees with the dispersion relation of the generalized CH system, Eqs.~\eqref{eq:gen-Cahn-Hilliard}, with  ${\varepsilon\ll 1}$ in the limit ${q\to 0}$.
Choosing ${\kappa = D_u/\tilde{f}_\eta}$, the largest unstable wavenumber $q_\mathrm{max}$ for the CH system agrees with the 2cRD system in the mass-conserving case ${\varepsilon=0}$ (see Fig.~\ref{fig:dispersion-relations}).

At large wavenumbers, the unstable branch $\sigma_\mathrm{HSS}(q)$ can be approximated by [expanding the full expression in terms of $1/q^2$]
\begin{equation*}
    \sigma_\mathrm{HSS}(q) \approx
    -D_u q^2
    \, ,
\end{equation*}
agreeing with the dispersion relation, Eq.~\eqref{eq:disp-rel-cAC}, for the shadow (cAC) system at large wavelengths $q\to\infty$.

On the basis of the dynamic equations, Eqs.~\eqref{eq:cont-eq},~\eqref{eq:eta-evol}, one can understand this close connection between the (mass-conserving) 2cMcRD, CH, and cAC systems by the following considerations.
In the diffusion-limited regime, mass redistribution limits the time evolution of the 2cMcRD system, and the pattern dynamics is slow compared to the local reaction dynamics [cf.\ Sec.~\ref{sec:HSS-instability}].
Thus, we can estimate ${\partial_t\rho, \partial_t\eta \ll \tilde{f}}$.
Since ${\partial_t\rho\sim \nabla^2\eta}$, Eq.~\eqref{eq:eta-evol} for the dynamics of $\eta$ in this limiting case reduces to
\begin{equation*}
    0\approx -D_u \nabla^2\rho - \tilde{f}(\rho,\eta)
    \, .
\end{equation*}
Thus, the value of the mass-redistribution potential $\eta$ becomes an (implicit) functional of the density profile $\rho(x,t)$ just as in the CH equation, Eqs.~\eqref{eq:gen-Cahn-Hilliard}.
In the linear regime around a HSS, the reaction term can be linearized around the nullcline.
This gives ${\tilde{f}\approx \tilde{f}_\eta\, [\eta-\eta^*(\rho)]}$, and explains the above choice ${\kappa=D_u/\tilde{f}_\eta}$ to match the corresponding CH system with the 2cMcRD system.
In contrast, in the reaction-limited regime, mass redistribution is fast compared to the local reaction dynamics.
Thus, the dynamics in this regime can be approximated by assuming instantaneous mass redistribution, that is, by performing the limit ${D_v\to\infty}$.
This limit retrieves the shadow system given by Eqs.~\eqref{eq:cAC-reservoir},~\eqref{eq:cAC-rho}. 
Employing the same linearization ${\tilde{f}\approx \tilde{f}_\eta\, [\eta-\eta^*(\rho)]}$ around the nullcline as for the diffusion-limited regime, the shadow system, Eqs.~\eqref{eq:cAC-reservoir},~\eqref{eq:cAC-rho}, gives
\begin{equation*}
    \partial_t\rho = D_u \nabla^2\rho - \tilde{f}_\eta\, \eta^*(\rho) + \frac{1}{|\Omega|}\int_\Omega\mathrm{d}x\,\tilde{f}_\eta\,\eta^*(\rho)
    \, ,
\end{equation*}
taking the form of the classical cAC equation [cf.\ Eq.~\eqref{eq:cAC-classical}].

\section{Stationary patterns of the mass-conserving systems}
\label{app:stat-state-MC}
\emph{Local equilibria theory}, described in Ref.~\cite{Brauns.etal2020}, allows the construction of stationary patterns of mass-conserving 2cRD, CH, and cAC systems in phase space regardless of the specific mathematical form of the reaction term $\tilde{f}$ (see Sec.~\ref{sec:stat-patterns}).
Here, we first highlight the differences in the construction of stationary patterns when $\rho$ and $\eta$ are used as phase-space coordinates instead of $u$ and $v$ as in Ref.~\cite{Brauns.etal2020}.
We then present detailed properties of stationary mesa and peak patterns that characterize the mass-competition instability.
How the stationary patterns change under the influence of source terms is presented later in Appendix~\ref{app:stat-state-nMC}.

\subsection{Phase-space construction}
The flux-balance subspace $\eta(x) = \etastat$ (FBS) is horizontal in $(\rho,\eta)$ coordinates while it has the finite negative slope $-d$ in $(u,v)$ coordinates.
Therefore, reaction kinetics with $\mathsf{\Lambda}$-shaped nullclines (NC) in $(\rho,\eta)$ coordinates always give rise to peak patterns [only two FBS-NC intersection points; see Fig.~\ref{fig:MC-stat-construction}(e,f)].
In contrast, a $\mathsf{\Lambda}$-shaped nullcline in $(u,v)$ coordinates always shows a third FBS-NC intersection point due to the finite negative slope ${-d<0}$ of the FBS.
In general, for a physically consistent 2cMcRD system describing concentrations $u,v > 0$, the nullcline has to eventually cross the FBS a third time at high densities (`effective' $\mathsf{N}$-shape).

Such $\mathsf{\Lambda}$-shaped nullclines in $(u,v)$ coordinates yield highly asymmetric $\mathsf{N}$-shaped nullclines in $(\rho,\eta)$ coordinates, with the third FBS-NC intersection point at large densities $\rho$ when ${d \ll 1}$.
These systems give rise to peak patterns only if the average density $\bar{\rho}$ is low and the pattern does not saturate in a high-density plateau.
The $\mathsf{\Lambda}$-shape in $(\rho,\eta)$ phase space is a mathematical idealization of this situation.
In the (non-idealized) systems with highly asymmetric $\mathsf{N}$-shaped nullclines there is a crossover from peak to mesa patterns~\cite{Gratton.Witelski2008,Gai.etal2020,Brauns.etal2021}.

Let us now give details on the stationary mesa and peak patterns, such as the interface width, the interface position, the pattern tails in the plateaus, and the scaling of the peak profile with the mass $M$.
Ultimately, these allow us to determine the change $\partial_M^{}\etastat$ which drives mass competition.

\subsection{Stationary mesa patterns}
On the infinite line, the stationary state $\rhostat^\infty(x)$ with a single interface, say at ${x=0}$, asymptotically approaches the plateau densities $\rho_\pm$ given by the two outer FBS-NC intersection points because ${\partial_x\rhostat^\infty\to 0}$ as ${x\to\pm\infty}$ [see Eq.~\eqref{eq:profile-equation}; green profile (dark gray, thin line) in Fig.~\ref{fig:MC-stat-construction}(d)].
Thus, the mass-redistribution potential $\eta_\mathrm{stat}^\infty$ of the infinitely large mesa pattern fulfills total turnover balance [see Eq.~\eqref{eq:ttb}]
\begin{equation}\label{eq:etastat-infty}
    0 
    = 
    \int_{\rho_-(\eta_\mathrm{stat}^\infty)}^{\rho_+(\eta_\mathrm{stat}^\infty)}
    \mathrm{d}\rho\,
    \tilde{f}(\rho,\eta_\mathrm{stat}^\infty)
    \, .
\end{equation}
Linearization of the profile equation, Eq.~\eqref{eq:profile-equation}, around the plateau densities ${\rho_\pm = \rho_\pm(\eta_\mathrm{stat}^\infty)}$ shows that the profiles approach these plateau densities exponentially~\cite{Brauns.etal2021}.
Correspondingly, on a finite domain $[0,\Lambda/2]$ with no-flux boundary conditions we find [see Fig.~\ref{fig:MC-stat-construction}(d)]: 
\begin{subequations}\label{eq:mesa-exp-tails}
\begin{align}
    \rhostat(x)&\approx \rho_+ - \delta\rho_+ \cosh\left[\frac{1}{\ell_+}\left(\frac{\Lambda}{2}-x\right)\right]
    ,\\
    \rhostat(x)&\approx \rho_- + \delta\rho_- \cosh\left(\frac{x}{\ell_-}\right)
    ,
\end{align}
\end{subequations}
in the upper and lower plateau, respectively.
The diffusion length scales ${\ell_\pm  = \big[-D_u/\partial_\rho \tilde{f}(\rho_\pm,\eta_\mathrm{stat}^\infty)\big]^{1/2}}$ describe the exponential approach toward the plateau densities $\rho_\pm$.
Within the sharp-interface approximation, the deviations ${\delta\rho_\pm = 2 a_\pm \exp\left(-L_\pm/ \ell_\pm\right)}$ from the plateau densities follow from asymptotic matching of the tail profile to the interface solution on the infinite line \cite{Ward1996, Kolokolnikov.etal2006, Brauns.etal2021}.
The coefficients $a_\pm$ are constants specific to the reaction term $\tilde{f}$; see Ref.~\cite{Ward1996}, Eq.~(2.3c).

In addition to these asymptotic properties of the tails of the stationary profile, one can also estimate the interface width $\ell_\mathrm{int}$, i.e., the spatial extent of the transition region between the two plateaus [cf.\ Fig.~\ref{fig:MC-stat-construction}(a,b)]. 
Following the reasoning in Ref.~\cite{Brauns.etal2020}, this width can be approximated by ${\ell_\mathrm{int} \approx \pi/q_\mathrm{max}}$, where $q_\mathrm{max}$ is the wavevector of the fastest growing mode (in the dispersion relation) for the total density $\rho_\mathrm{infl}$ determined by the intermediate FBS-NC intersection point.
This yields \cite{Brauns.etal2020}
\begin{equation}\label{eq:interface-width-infl-point}
    \ell_\mathrm{int}\sim \sqrt{\frac{D_u}{\partial_\rho\tilde{f}(\rho_\mathrm{infl},\eta_\mathrm{stat}^\infty)}}
    \, ,
\end{equation}
an expression similar to the diffusion lengths $\ell_\pm$ above.
From this relationship, it can be seen that the sharp-interface limit is approached for $D_u\to 0$.

The position of the interface is determined by the mass in the system.
Given the average density $\bar{\rho}$, the relative plateau lengths $\xi_\pm$ are fixed within the sharp-interface limit by \cite{Kolokolnikov.etal2006, Brauns.etal2020}
\begin{equation*}
    \xi_+ = \frac{2 L_+}{\Lambda} 
    = 
    \frac{\bar{\rho}-\rho_-}{\Delta\rho}
    \, ,
    \quad 
    \xi_- 
    = 
    \frac{2 L_-}{\Lambda} = 1-\xi_+
    \, ,
\end{equation*}
with ${\Delta\rho = \rho_+-\rho_-}$ [cf.\ Fig.~\ref{fig:MC-stat-construction}(b)].

Finally, of central importance for the strength of the mass-competition instability is the dependence of the stationary mass-redistribution potential $\etastat$ on the mesa mass $M$ [see Eq.~\eqref{eq:mesa-peak-mass}].
In the supplemental material of Ref.~\cite{Brauns.etal2021}, we have shown that total turnover balance, Eq.~\eqref{eq:ttb}, within the sharp-interface approximation implies that
\begin{subequations}\label{eq:mesa-delMeta-tot}
\begin{align}\label{eq:mesa-delMeta}
    \partial_M^{}\etastat 
    &= 
    \partial_M^-\etastat + \partial_M^+\etastat
    \, , \\
    \partial_M^\pm\etastat 
    &= 
    \pm\frac{1}{2\Delta\rho} \, 
    \partial_{L_\pm}\etastat
    \nonumber \\
    &= 
    \mp \frac{\partial_\rho\tilde{f}(\rho_\pm,\eta_\mathrm{stat}^\infty)}{F_\eta}\frac{\partial_M^{}\delta\rho_\pm^2}{2} 
    \, ,
\label{eq:mesa-delMeta-formula}
\end{align}
\end{subequations}
with ${F_\eta := \int_{\rho_-}^{\rho_+}\mathrm{d}\rho\,\partial_\eta\tilde{f}(\rho,\eta_\mathrm{stat}^\infty)}$. 
These equations can be rationalized as follows.
Within the sharp-interface approximation, changes in the mass of a mesa pattern only affect the width of the plateaus: Adding some mass $\delta M$ to the mesa mass $M$ results in the changes $\delta L_\pm =\pm \delta M/(2\Delta\rho)$ of the half-lengths of the high- and low-density plateaus, respectively (cf.\ Fig.~\ref{fig:MC-stat-construction}).
Due to the exponential pattern tails in the plateau regions, the plateau lengths affect the plateau height as they determine the offsets $\delta\rho_\pm$ from the plateau densities $\rho_\pm$.
These changed plateau heights then determine the change $\partial_M^{}\etastat$ of the stationary mass-redistribution potential because the changed heights change total turnover balance [cf.\ Eq.~\eqref{eq:ttb}].
The two derivatives  $\partial_M^\pm\etastat$ describe the effects due to the change of either the upper or the lower plateau, respectively.
The expressions for $\partial_M^\pm\etastat$, Eq.~\eqref{eq:mesa-delMeta-formula}, can then be understood graphically [cf.\ the pale-blue/red (pale gray) construction in Fig.~\ref{fig:MC-stat-construction}(c)]:
A change in the plateau height leads to a change in the area of the (white) triangular areas between FBS and NC.
Due to their triangular shape, total turnover balance, represented by the (red-) shaded areas enclosed between FBS and NC, changes proportionally to $\partial_M^{}\delta\rho_\pm^2/2$.
The change $\partial_M^\pm\etastat$ then corresponds to the shift of the FBS necessary to balance the changed (red-) shaded areas above and below the FBS.

\subsection{Peak patterns}
\begin{figure}
    \centering
    \includegraphics{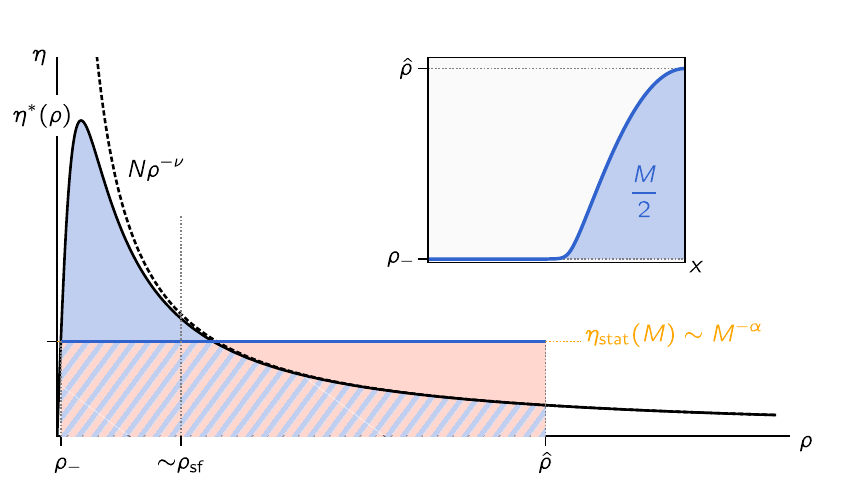}
    \caption{
    Asymptotic scaling of the peak profile.
    For large peak masses $M$ (see inset), the maximal peak density $\hat{\rho}$ lies in the asymptotically scale-free regime ${\rho\gg\rho_\mathrm{sf}}$ of the reaction term $f$.
    Above the density scale $\rho_\mathrm{sf}$ the reaction term can be approximated by a simple polynomial form, Eq.~\eqref{eq:f-scale-free}.
    The nullcline $\eta^*(\rho)$ (solid line) can then be approximated by a power law $N \rho^{-\nu}$ (dashed line).
    The scaling of the stationary mass-redistribution potential ${\etastat(M)\sim M^{-\alpha}}$ follows from total turnover balance, Eq.~\eqref{eq:ttb}, by analyzing the scaling of the blue- (dark gray) and red-shaded (rectangular) regions (the striped region counts to both).
    }
    \label{fig:peak-scaling}
\end{figure}

As with mesa patterns, the low-density plateau of peak patterns is approached by exponential tails emanating from the peak.
Thus, up to exponentially small corrections the pattern density $\rhostat(0)$ in Fig.~\ref{fig:MC-stat-construction}(f) (and in the inset in Fig.~\ref{fig:peak-scaling}) agrees with $\rho_-(\etastat)$.
Again, as with mesa patterns, we are interested in the relation $\etastat(M)$.
In the following, we give a scaling argument for its functional form.

\subsubsection{Scaling analysis of the stationary peak profiles}
\label{app:peak-scaling}
For large densities $\rho$, we assume that the reaction term can be approximated by the simple polynomial form 
\begin{equation}
\label{eq:f-scale-free}
    \tilde{f}
    \sim 
    \rho^\mu \left( \eta - N \rho^{-\nu}\right)
    ,
\end{equation}
where $N \rho^{-\nu}$ approximates the high-density tail of the $\mathsf{\Lambda}$-shaped nullcline $\eta^*(\rho)$ (see Fig.~\ref{fig:peak-scaling}).
This form is suited to describe the asymptotic form of a reaction term composed of an attachment term $a(u) v$ minus a detachment term $b(u) u$ because,
for ${D_v \gg D_u}$, we have ${\eta \approx v}$ and the pattern forms mainly in the slowly diffusing species $u$, i.e., ${\rho \approx u}$.
Moreover, the polynomial form is appropriate because, on the one hand, mass-action kinetics are in general of polynomial form.
On the other hand, composite reaction terms like Hill-type kinetics approach a power law at large densities, e.g., enzymatic detachment might give ${b(u)\sim 1/(K+u^n)\approx 1/(K+\rho^n)\to\rho^{-n}}$.

With $\tilde{f}$ being asymptotically scale-free (being of polynomial form for densities $\rho\gtrsim \rho_\mathrm{sf}$; see Fig.~\ref{fig:peak-scaling}), one expects that the profile equation~\eqref{eq:profile-equation} yields approximately scale-free solutions for the family of stationary peak profiles with different mass $M$---if the maximum density $\hat{\rho}$ lies far in the scale-free regime (in Fig.~\ref{fig:peak-scaling}, one needs ${\hat{\rho}\gg\rho_\mathrm{sf}}$).
As parameterization of the family of solutions, we set
\begin{align*}
    \rhostat(x) 
    &= 
    \hat{\rho}\, \Sigma\left(\frac{x}{\hat{\rho}^\delta}\right)
    , \\
    \etastat 
    &= 
    \hat{\rho}^{-\tau}\, \Theta
    \, ,
\end{align*}
where $\Sigma$ and $\Theta$ are the (approximate) scaling solutions for the density profile and the constant stationary mass-redistribution potential, respectively.
The exponent $\delta$ describes the scaling of the peak width, and the exponent $\tau$ gives the scaling of the stationary mass-redistribution potential with the peak height $\hat{\rho}$.

Total turnover balance, Eq.~\eqref{eq:ttb}, then reads
\begin{align}
    0 
    &\sim 
    \int_0^{\hat{\rho}}\mathrm{d}\rho\, \rho^\mu 
    \left( \eta - N \rho^{-\nu}\right)
    \nonumber\\
    &\sim 
    \hat{\rho}^{1+\mu-\tau}\Theta\int_0^1\mathrm{d}\Sigma\, \Sigma^\mu 
    \nonumber \\
    &\quad - 
    \bigg[
    C +N \hat{\rho}^{1+\mu-\nu}
    \int_{\rho_\mathrm{sf}/{\hat{\rho}}}^1
    \mathrm{d}\Sigma\, \Sigma^{\mu-\nu}
    \bigg]
    . 
\label{eq:peak-scaling-ttb}
\end{align}
Here we used that the density of the lower plateau (approximately given by $\rho_-$) fulfills ${\hat{\rho}\gg \rho_- \approx 0}$ and set the lower integration boundary to zero.
The first term on the right-hand side is qualitatively represented by the red-shaded (rectangular) area below $\etastat$ in Fig.~\ref{fig:peak-scaling} (scaled by the reaction rate ${\sim\rho^\mu}$).
The second term corresponds to the blue-shaded (dark gray-shaded) area below the nullcline, again scaled by the reaction rate ${\sim\rho^\mu}$.
To approximate this second term, we choose an intermediate density $\rho_\mathrm{sf}$ describing the onset of the scale-free regime of the nullcline.
The integral is then dissected into the constant (we set $\rho_\mathrm{sf}$ to a fixed value) integral $C$ over the low-density region up to $\rho_\mathrm{sf}$, and the integration of the asymptotic tail.

To find the (approximate) scaling solution we require that the dominant terms for large peak amplitudes $\hat{\rho}$, i.e., the terms growing fastest with $\hat{\rho}$, have to cancel in Eq.~\eqref{eq:peak-scaling-ttb}.
Hence, these terms must scale with the same exponent in $\hat{\rho}$.
Comparison of the exponents in Eq.~\eqref{eq:peak-scaling-ttb} thus yields the (approximate) scaling exponent of the stationary mass-redistribution potential 
\begin{equation}
\label{eq:tau-scaling}
    \tau = \min(\nu, 1+\mu)
    \, .
\end{equation}
In addition, the profile equation determines the second exponent $\delta$.
Inserting the scaling solution into Eq.~\eqref{eq:profile-equation}, one finds
\begin{equation*}
    0 \sim \hat{\rho}^{1-2\delta}\partial_y^2\Sigma(y) + \hat{\rho}^{\mu-\tau} \Sigma^\mu\Theta - \hat{\rho}^{\mu-\nu}\Sigma^{\mu-\nu}
    \, ,
\end{equation*}
where ${y = x/\hat{\rho}^\delta}$.
Again, the terms growing fastest for ${\hat{\rho}\to\infty}$ have to scale with the same exponent in $\hat{\rho}$.
Because the exponent of the third term is always smaller than the exponent of the second, this gives the exponent identity
\begin{equation}
\label{eq:delta-scaling}
    \delta 
    = 
    \frac{1-\mu +\min(\nu,1+\mu)}{2}
    \, .
\end{equation}
With the peak mass scaling as ${M\sim \hat{\rho}^{1+\delta}}$, the stationary mass-redistribution potential scales as
\begin{equation}
\label{eq:alpha-scaling}
    \etastat\sim M^{-\alpha}
    \sim 
    M^{-\frac{\tau}{1+\delta}}
    \, ,
\end{equation}
which defines the exponent $\alpha$ introduced in the main text (see Sec.~\ref{sec:stat-patterns}).
Thus, the exponent $\alpha$ depends on the reaction kinetics described by $\tilde{f}$.
Because $\alpha$ also determines the coarsening exponent (see Sec.~\ref{sec:coarsening-scaling}), the above relations, Eqs.~\eqref{eq:tau-scaling},~\eqref{eq:delta-scaling},~\eqref{eq:alpha-scaling}, relate the coarsening law to the reaction kinetics via their asymptotic scaling exponents.
The result is power-law coarsening for peaks in 1D with a system-dependent exponent. 
The simple scaling argument put forward in Ref.~\cite{Brauns.etal2021} to determine the coarsening exponent holds if ${\nu \leq 1+\mu}$.
In particular, this simpler scaling argument is valid for systems where the detachment term ${\sim \rho^{\mu-\nu}}$ is non-decreasing with the density $\rho$, i.e., ${\mu > \nu}$.

Similar arguments as above can be used to derive the classical $t^{1/3}$-coarsening law for mesa-forming 2cMcRD systems in two dimensions (see supplementary material of Ref.~\cite{Brauns.etal2021}).

On a technical note, the above analysis shows that our assumption of slow mass competition, Eq.~\eqref{eq:small-plateau-mass}, will always be fulfilled sufficiently late during the coarsening process when peaks have grown sufficiently large.
Because ${\alpha > 0}$, the derivative $\partial_M^{}\etastat$ decreases faster with the domain mass than $\sim 1/M$, and mass conservation ensures ${\langle M\rangle\sim\langle \Lambda\rangle}$ (see Sec.~\ref{sec:coarsening-scaling}).

\section{Linearized dynamics of stationary patterns}
\label{app:lin-dynamics}
The mass-competition instability describes the linearized dynamics around a symmetric stationary pattern.
The analysis of the linear stability of a pattern is analogous to the linear stability analysis of the HSS (see Sec.~\ref{sec:HSS-instability} and Appendix~\ref{app:dispRel}).
The only difference is that the linear operator (the Jacobian) describing the linearized dynamics is space-dependent because the stationary state one linearizes around is not uniform.
Therefore, the Fourier modes are not the eigenmodes of this operator, and one has to determine not only the eigenvalues but must also construct the eigenmodes.
To start out with this analysis, we derive the linearized dynamics in this subsection.

\subsection{Linearized mass-conserving dynamics}
We start by linearizing the (one-dimensional) 2cMcRD system, Eqs.~\eqref{eq:cont-eq},~\eqref{eq:eta-evol} (${\varepsilon = 0}$), around a fully nonlinear stationary pattern $[\rhostat(x),\etastat]$.
To this end, we set ${\rho = \rhostat(x) + \operatorname{e}^{\sigma t} \delta\rho(x)}$ and ${\eta = \etastat + \operatorname{e}^{\sigma t} \delta\eta(x)}$, where we already anticipated that the eigenmode $(\delta\rho,\delta\eta)$ we are looking for grows exponentially with growth rate $\sigma$.
We find---to linear order in the perturbation $(\delta\rho,\delta\eta)$---the Sturm--Liouville eigenvalue problem
\begin{equation}\label{eq:eigenvalue-problem-MC}
    \sigma \begin{pmatrix}\delta\rho\\\delta\eta\end{pmatrix} =
    \begin{pmatrix}
    0 & D_v\,\partial_x^2\\
    -D_u\,\partial_x^2-\tilde{f}_\rho & \left(D_v+D_u\right)\,\partial_x^2 -\tilde{f}_\eta
    \end{pmatrix}
    \begin{pmatrix}
    \delta\rho\\\delta\eta
    \end{pmatrix},
\end{equation}
with the coefficients ${\tilde{f}_\rho = \partial_\rho\tilde{f}(\rhostat(x),\etastat)}$ and ${\tilde{f}_\eta = \partial_\eta\tilde{f}(\rhostat(x),\etastat)}$, which are space-dependent through the spatial profile $\rhostat(x)$.
The eigenmode $(\delta\rho,\delta\eta)$ is defined on the same domain $\Omega$ as the original dynamics and has to fulfill the same  boundary conditions (here always no-flux BCs).
As the coefficients $\tilde{f}_{\rho,\eta}$ depend on the spatial coordinate $x$, the eigenmodes are not simply Fourier modes (as in the linear stability analysis of the homogeneous steady state) and have to be determined together with the values $\sigma$ for which solutions to Eq.~\eqref{eq:eigenvalue-problem-MC} exist.
There is no general method to solve such a Sturm--Liouville eigenvalue problem.
Here, we build on the singular limit ${\ell_\mathrm{int}/\Lambda\to 0}$ [sharp-interface approximation, see Eq.~\eqref{eq:sharp-int-approx}] to make analytic progress and to find approximate solutions.

It is useful to define the linear operator
\begin{equation}\label{eq:lin-op-MC}
    \mathcal{L} = -D_u\,\partial_x^2-\tilde{f}_\rho
    \, ,
\end{equation}
which depends on the stationary pattern $[\rhostat(x),\etastat]$ through $\tilde{f}_\rho$.
Differentiating the stationary profile equation~\eqref{eq:profile-equation} with respect to $x$ and $\etastat$, we find the relations
\begin{subequations}\label{eq:zero-mode-eqs}
\begin{align}
    0 &= \mathcal{L}\,\partial_x\rhostat
    \, ,\\
    \tilde{f}_\eta &= \mathcal{L}\,\partial_{\etastat}^{}\rhostat
    \, .\label{eq:mass-mode-eq}
\end{align}
\end{subequations}
Inserting the second relation into Eq.~\eqref{eq:eigenvalue-problem-MC} shows that ${[\partial_{\etastat}^{}\rhostat(x;\etastat), 1]}$ is an exact zero mode (eigenmode with eigenvalue $0$).
This ``mass mode'' ${\partial_{\etastat}^{}[\rhostat(x;\etastat), \etastat]}$ is a zero mode due to the conservation of the total mass, which can be seen as follows:
Because of this conservation law, the total mass is a control parameter and a continuous family of stationary patterns with different masses (and different $\etastat$, see Sec.~\ref{sec:stat-patterns}) exists.
The mode ${\partial_{\etastat}^{}[\rhostat(x;\etastat), \etastat]}$ describes how the stationary peak/mesa profile changes upon changing the average mass in the system.
As it leads from one stationary pattern to another, it has to be a zero mode of the linearized dynamics.
It is important to note that this mode breaks mass conservation (as it leads from a stationary pattern of mass $M$ to one with a higher or lower mass) and therefore is irrelevant to the dynamics of a closed system.
An in- or outflow of mass is necessary to excite this mode.

In addition, the mode ${[\partial_x\rhostat, 0]=\partial_x[\rhostat(x),\etastat]}$ solves the linearized dynamics, Eq.~\eqref{eq:eigenvalue-problem-MC}, with ${\sigma=0}$.
This mode ${\partial_x[\rhostat(x),\etastat]}$ is the translation mode of the pattern.
In the infinite system, it is a Goldstone mode due to the translational invariance of the system (i.e., an exact zero mode).
In a finite system with no-flux boundary conditions, the boundaries break the translational invariance such that the translation mode is only an approximate zero mode.
It does not fulfill the no-flux boundary conditions because the derivative of the translation mode, $\partial_x(\partial_x\rhostat)$, does not vanish at the boundaries.

\subsection{Linearized dynamics including source terms}
Weak source terms modify the stationary pattern.
Thus, the 2cRD dynamics, Eqs.~\eqref{eq:cont-eq},~\eqref{eq:eta-evol} (for ${\varepsilon > 0}$), have to be linearized around the stationary state $[\rhostateps(x),\etastateps(x)]$ of the non-mass-conserving system, which is distinct from the stationary state $[\rhostat(x),\etastat]$ of the mass-conserving system.
The linearized dynamics can then be written as
\begin{subequations}\label{eq:eigenvalue-problem-nMC}
\begin{align}
    &\left(\sigma^\varepsilon -\varepsilon \partial_\rho s_\mathrm{tot}^\varepsilon\right)\delta\rho = \left(D_v\partial_y^2 + \varepsilon \partial_\eta s_\mathrm{tot}^\varepsilon\right)\delta\eta
    \, ,\label{eq:cont-eq-non-mc}\\
    &\left[\tilde{f}_\eta^\varepsilon + \sigma^\varepsilon + \varepsilon \partial_\eta \left(s_1^\varepsilon + d\, s_2^\varepsilon\right) \right] \delta\eta \nonumber\\
    &\qquad = \left[\mathcal{L}^\varepsilon + (1+d)\sigma^\varepsilon - \varepsilon\partial_\rho \left(s_1^\varepsilon + d\, s_2^\varepsilon\right)  \right] \delta\rho
    \, . \label{eq:lin-eta-eq-non-mc}
\end{align}
\end{subequations}
The superscript $()^\varepsilon$ signifies the evaluation at the stationary state $(\rhostateps,\etastateps)$, e.g., ${\tilde{f}_\eta^\varepsilon = \partial_\eta\tilde{f}[\rhostateps(x),\etastateps(x)]}$. 

Analogously to the mass-conserving case [see Eq.~\eqref{eq:lin-op-MC}], we define the linear operator
\begin{equation*}
    \mathcal{L}^\varepsilon = -D_u\,\partial_x^2-\tilde{f}_\rho^\varepsilon
    \, .
\end{equation*}
Because the conservation law is broken, only the translation mode $\partial_x(\rhostateps,\etastateps)$ is left as (approximate) zero mode. 
The peak mass $M$ is no longer a control parameter.
Rather, it is set by source balance (cf.\ Sec.~\ref{sec:stat-patterns-nMC} and Eq.~\eqref{eq:source-balance}).
Therefore, the system with source terms has only a single (or several) stationary patterns and no continuous family of stationary solutions parametrized by $M$.
Thus, no mass mode ``$\partial_M^{}[\rhostateps,\etastateps]$'' exists here.

\subsection{Linearized dynamics in the pattern plateaus}
\label{app:lin-dynamics-plateaus}
In the next sections, we will repeatedly use that the slow dynamics around the plateaus of stationary peak or mesa patterns can be approximated as purely diffusive.
This is due to the fact that all gradients in the plateaus are shallow.
In addition, weak source terms introduce only a linear production or degradation term with a constant rate.

To show this, we note that the stationary pattern $(\rhostat,\etastat)$ is basically constant in the plateaus, except for the exponential approach toward the plateau densities $\rho_\pm$ (see Appendix~\ref{app:stat-state-MC} and Sec.~\ref{sec:stat-patterns}).
These exponential tails become small in the sharp-interface approximation, or, equivalently, far from the interfaces.
Furthermore, the corrections of the stationary pattern $(\rhostateps,\etastateps)$ due to weak source terms are of order $\varepsilon$.
Consequently, up to the exponential corrections and corrections of order $\varepsilon$, in the pattern plateaus, we can set
\begin{align*}
    \tilde{f}_{\rho,\eta}^{\varepsilon}
    &\approx \partial_{\rho,\eta}\tilde{f}(\rho_\pm,\etastat) 
    = 
    \tilde{f}_{\rho,\eta}^\pm
    \, , \\
    s_\mathrm{tot}^{\varepsilon}
    &\approx s_\mathrm{tot}(\rho_\pm,\etastat) 
    = s_\mathrm{tot}^\pm
    \, , \\
    \left(s_1^\varepsilon + d\, s_2^\varepsilon\right)
    &\approx 
    \left(s_1^\pm + d\, s_2^\pm\right) 
    ,
\end{align*}
which are all spatially constant.
As the mass-competition dynamics is (by assumption, see Sec.~\ref{sec:approximations}) slow compared to the local relaxation of the stationary pattern (${\tilde{f}_{\rho,\eta}^\pm\gg \sigma}$), the linearized dynamics, Eq.~\eqref{eq:eigenvalue-problem-nMC}, simplifies to
\begin{subequations}
\begin{align}
    \left(\sigma -\varepsilon \partial_\rho s_\mathrm{tot}^\pm\right)\delta\rho &\approx 
    \left(D_v\partial_x^2 + \varepsilon \partial_\eta s_\mathrm{tot}^\pm\right)\delta\eta
    \, , \label{eq:plateau-linear-dynamics}\\
    \left[\tilde{f}_\eta^\pm + \order{\sigma,\varepsilon}\right]\delta\eta &\approx -\left[D_u\partial_x^2 +\tilde{f}_\rho^\pm + \order{\sigma,\varepsilon}\right] \delta\rho
    \, .
\end{align}
\end{subequations}
Large gradients in the pattern (${\tilde{f}_\rho\sim D_u\partial_x^2}$) only appear in the narrow interface regions (see Sec.~\ref{sec:stat-patterns}).
Thus, let us assume we can neglect the term $D_u\partial_x^2\delta\rho$ in the plateaus.
One then finds from the second equation that the $\rho$- and $\eta$-profiles fulfill the local equilibrium assumption ${\eta\approx\eta^*(\rho)}$ because (cf.\ discussion of mesa splitting in Ref.~\cite{Brauns.etal2021})
\begin{equation}\label{eq:plateau-local-equilibrium}
    \delta\eta\approx-\frac{\tilde{f}_\rho^\pm}{\tilde{f}_\eta^\pm}\delta \rho = \left(\partial_{\rho_\pm}\eta^*\right) \delta \rho
    \, .
\end{equation}
Inserting this into Eq.~\eqref{eq:plateau-linear-dynamics}, one finds
\begin{equation}\label{eq:eigenvalue-problem-plateaus}
    \sigma \delta\rho \approx D_v\partial_{\rho}\eta^*(\rho_\pm)\, \partial_x^2\delta\rho + \varepsilon\partial_\rho s_\mathrm{tot}^*(\rho_\pm)\,\delta\rho
    \, ,
\end{equation}
which describes diffusive dynamics with an effective diffusion constant ${D_\pm = D_v\partial_{\rho}\eta^*(\rho_\pm)}$.
Importantly, lateral stability of the pattern plateaus ensures a positive nullcline slope at the plateau densities ${\partial_{\rho}\eta^*(\rho_\pm) >0}$ such that the effective diffusion constant is positive (see Sec.~\ref{sec:HSS-instability}).
Additionally, the effects of production and degradation are captured by evaluating the source terms along the nullcline, ${s_\mathrm{tot}^*(\rho) = s_\mathrm{tot}(\rho,\eta^*(\rho))}$, and linearizing for small deviations $\delta\rho$.
Using the final expression, Eq.~\eqref{eq:eigenvalue-problem-plateaus}, we see that the gradient term neglected is indeed small ${\sim \frac{d}{\partial_{\rho}\eta^*(\rho_\pm)} \order{\sigma,\varepsilon}}$.

\begin{figure}
    \centering
    \includegraphics{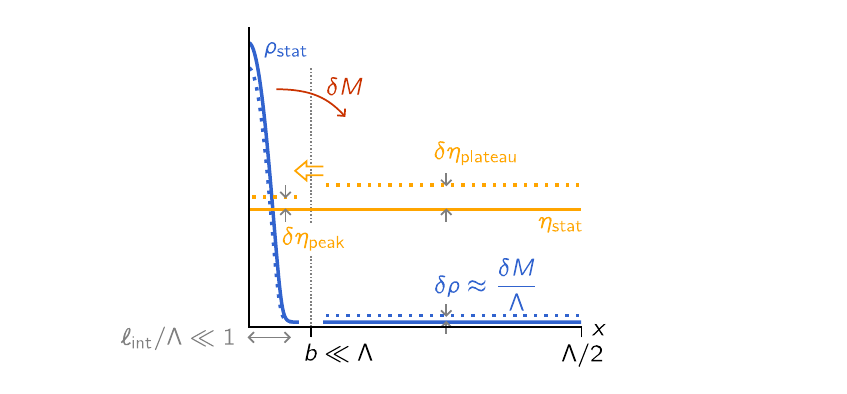}
    \caption{A stable stationary peak has to be stable against redistribution of a small amount of mass $\delta M$ from the peak into the plateau.
    This perturbation reduces the peak size and increases the plateau density (dotted line, shift $\delta\rho$ in the plateau).
    The induced shift $\delta\eta_\mathrm{plateau}$ in the mass-redistribution potential in the plateau has to be larger than the shift $\delta\eta_\mathrm{peak}$ at the peak.
    Then, the resulting gradient moves the redistributed mass [yellow (double) arrow] back into the peak region $[0,b]$.}
    \label{fig:peak-stability}
\end{figure}

\section{Stability of elementary patterns}
\label{app:stability}
This subsection serves a two-fold purpose.
First, we show systematically that the stability of the stationary peak profile in 2cMcRD systems demands (see Fig.~\ref{fig:peak-stability})
\begin{equation}
\label{eq:peak-stability-criterion}
    \Lambda \, \partial_{\etastat}\rho_- 
    < 
    |\partial_{\etastat} M|
    \, .
\end{equation}
This condition was derived heuristically in the supplementary material of Ref.~\cite{Brauns.etal2021}.
It states that the peak is stable if the mass-redistribution potential $\eta$ increases more strongly in the plateau ($\delta\eta_\mathrm{plateau}$ in Fig.~\ref{fig:peak-stability}) than at the peak (shift $\delta\eta_\mathrm{peak}$ in Fig.~\ref{fig:peak-stability}) when a small amount of mass $\delta M$ is redistributed from the peak into the plateau.
This ensures that the ensuing gradient in $\eta$ redistributes the mass back into the peak.

Second, the analysis of the corresponding relaxation rates shows that relaxation due to mass redistribution between the peak and plateau regions is indeed fast if the condition Eq.~\eqref{eq:small-plateau-mass} is fulfilled.
This is a basic assumption underlying our mathematical analysis of the mass-competition instability (see Sec.~\ref{sec:approximations}).

Here we use the sharp-interface approximation and asymptotic matching to derive the relaxation rates of the redistribution modes.
If mass redistribution is the limiting process, these redistribution modes between the peak and plateau will be the slowest relaxation modes because redistribution within the narrow peak or interface region will be fast in comparison.
However, one cannot quantify the relaxation modes and rates of a peak or interface itself based on the sharp-interface approximation.
Therefore, beyond the heuristic argument of rapid mass redistribution within the narrow peak and interface regions, we rely on the observation that the pattern profile is well approximated by a quasi-steady state in the numerical simulations (see Fig.~\ref{fig:coarsening-phenomenology}) to neglect the relaxation modes of the peak or interface profile itself.

For concreteness, we now focus on peak patterns.
The calculations proceed along the same lines for mesa patterns and yield the same condition.
Let us, therefore, consider the stationary elementary peak pattern (see Fig.~\ref{fig:peak-stability}) of a 2cMcRD system.
To determine the eigenmode $(\delta\rho,\delta\eta)$ and relaxation rate $\sigma_\mathrm{relax}$ describing the relaxation of mass redistributed from the peak into the plateau, asymptotic matching uses the separation of scales between the narrow peak and the long plateau (singular perturbation theory).
First, the eigenvalue problem is solved in the plateaus assuming the stationary pattern profile as spatially uniform there.
Afterward, one approximates the eigenmode at the peak under the assumption that the variation $\delta\eta$ of the mass-redistribution potential---which is constant in the stationary state---only varies on long scales and is constant within the narrow peak region.
Both results are then matched at an intermediate scale between the peak and plateau regions (scale $b$ in Fig.~\ref{fig:peak-stability}).

In the plateau, the eigenmode fulfills Eq.~\eqref{eq:eigenvalue-problem-plateaus} (with ${\varepsilon = 0}$), which is solved by
\begin{subequations}\label{eq:plateau-profile-relax-mode}
\begin{align}
    \delta\rho_\mathrm{plateau}(x) &\propto \cos\left[\sqrt{\frac{-\sigma_\mathrm{relax}}{D_-}} \left(\frac{\Lambda}{2}-x\right)\right]
    , \label{eq:plateau-profile-relax-mode-rho}\\
    \delta\eta_\mathrm{plateau}(x) &= \partial_\rho\eta^*(\rho_-)
    \, \delta\rho_\mathrm{plateau}(x)
    \, ,
\end{align}
\end{subequations}
where we anticipated that the mode should be stable, that is, ${\sigma_\mathrm{relax} < 0}$, and used the effective diffusion constant ${D_- = D_v\partial_{\rho}\eta^*(\rho_-)}$.
Thus, we find at the peak location ${x = 0}$ and the matching scale $b$ [choosing ${\ell_\mathrm{int}\ll b\ll\Lambda}$ within the `sharp-peak approximation']
\begin{equation}
\label{eq:peak-relax-mode-plateau}
    \frac{\partial_x\delta\eta_\mathrm{plateau}{\big |}_{x = b}}{\delta\eta_\mathrm{plateau}{\big |}_{x = b}} 
    \approx
    \frac{\partial_x\delta\eta_\mathrm{plateau}{\big |}_{x = 0}}{\delta\eta_\mathrm{plateau}{\big |}_{x = 0}} 
    = 
    \frac{2\chi}{\Lambda}
    \tan(\chi)
    \, ,
\end{equation}
with $\chi = \frac{\Lambda}{2}\sqrt{|\sigma_\mathrm{relax}|/D_-}$.
This (relative) gradient determines the mass in- or outflow at the peak and thereby how fast the peak mass changes.\footnote{
Only the relative gradient is important because the amplitude of the plateau perturbation [proportionality constant in Eq.~\eqref{eq:plateau-profile-relax-mode-rho}] is arbitrary.}

Assuming that the relaxation within the peak is fast compared to mass transport between the plateau and the peak, the peak profile adiabatically follows the changing peak mass.
Thus, we approximate the mode at the peak by the mass mode: ${(\delta\rho\approx \delta M \partial_M^{} \rhostat}$,  ${\delta\eta\approx\mathrm{const.})}$ with the mode amplitude $\delta M$ which gives the change of the peak mass.
Integration of the continuity equation over the peak region thus yields
\begin{equation}\label{eq:relax-rate-int-cont-eq}
    \sigma_\mathrm{relax} \delta M \int_0^{b}\mathrm{d}x\, \partial_M^{} \rhostat\approx \sigma_\mathrm{relax} \frac{\delta M}{2} \approx D_v \partial_x\delta\eta{\big |}_{x = b}
    \, .
\end{equation}
This ``inner'' solution at the peak has to be matched to the ``outer'' solution in the plateau, Eq.~\eqref{eq:peak-relax-mode-plateau}.
This gives
\begin{equation}
\label{eq:relax-self-consistency}
    \sigma_\mathrm{relax} \frac{\delta M}{\delta \eta_\mathrm{peak}} \approx \frac{4 D_v \chi}{\Lambda}
    \tan\left(\chi\right)
    \, .
\end{equation}
Here, ${\delta\eta_\mathrm{peak}=\delta\eta|_{x=0}\approx\delta\eta|_{x=b}}$ denotes the approximately constant change of the stationary mass-redistribution potential in the peak region.
Because ${\chi = \chi(\sigma_\mathrm{relax})}$, Eq.~\eqref{eq:relax-self-consistency} is an implicit equation that fixes the rate $\sigma_\mathrm{relax}$ such that the matched approximation of the relaxation mode is self-consistent.\footnote{
Matching of the density profile $\delta\rho(x)$ does not need to be considered separately because the matching region $x \approx b$ lies in the tail region of the peak pattern ($\ell_\mathrm{int}\ll b\ll \Lambda$ by assumption), and the local equilibrium assumption ${\eta\approx\eta^*(\rho)}$ holds [see Eq.~\eqref{eq:plateau-local-equilibrium}].
Hence, $\delta \rho$ and $\delta \eta$ are slaved to one another.}
In the diffusion-limited regime, the stationary mass-redistribution potential at the peak can be approximated by its QSS.
This yields ${\delta M/\delta\eta_\mathrm{peak}=1/\partial_M^{}\etastat=\partial_\etastat M}$, which closes Eq.~\eqref{eq:relax-self-consistency}.

In contrast, outside the diffusion-limited regime, $\delta\eta_\mathrm{peak}$ deviates from its QSS $\delta M \partial_M^{}\etastat$ due to the finite rate of reactive conversion between the $u$ and $v$ species (cf.\ Secs.~\ref{sec:mc-growth-rates},~\ref{sec:growth-rate-nMC-discussion}).
To capture this deviation, we approximate the `conversion-rate integral' given by
\begin{equation}
\label{eq:relax-convRateInt}
    \int_0^b\mathrm{d}x \,
    \tilde{f}_\eta \,
    \delta\rho
    \, .
\end{equation}
Because ${\delta\rho \sim \partial_M^{}\rhostat}$ is localized to the peak, this integral describes an average reaction rate at the peak.
It can be expressed in two different ways using the eigenmode approximation at the peak and the linear eigenmode dynamics, Eq.~\eqref{eq:eigenvalue-problem-MC}.
The procedure of how this fixes $\delta M/\delta\eta_\mathrm{peak}$ is detailed in the derivation of the mass-competition rates (see Appendix~\ref{app:MC-mass-competition}); the integral will appear in different variants in the derivation of all growth rates, also for the mass-competition instability.
For the analyzed relaxation mode this calculation results in the expression 
\begin{equation}\label{eq:peak-relax-mode-peak}
     \frac{\delta M}{\delta \eta_\mathrm{peak}}\approx \frac{1}{\frac{1}{\partial_\etastat M}+\frac{(1+d)\sigma_\mathrm{relax}}{2\ell_\mathrm{int}\langle \tilde{f}_\eta\rangle_\mathrm{int}}}
     \, .
\end{equation}
The quantities $\ell_\mathrm{int}$ and $\langle\tilde{f}_\eta\rangle_\mathrm{int}$ are defined as for the mass-competition rate (see Sec.~\ref{sec:mc-growth-rates} and Appendix~\ref{app:MC-mass-competition}).
In the limit of fast reactive conversion ${\langle \tilde{f}_\eta\rangle_\mathrm{int}/\sigma_\mathrm{relax}\gg 1}$, we retrieve the result ${\delta M/\delta\eta_\mathrm{peak}=\partial_\etastat M}$ of the QSS approximation.

\begin{figure}
    \centering
    \includegraphics{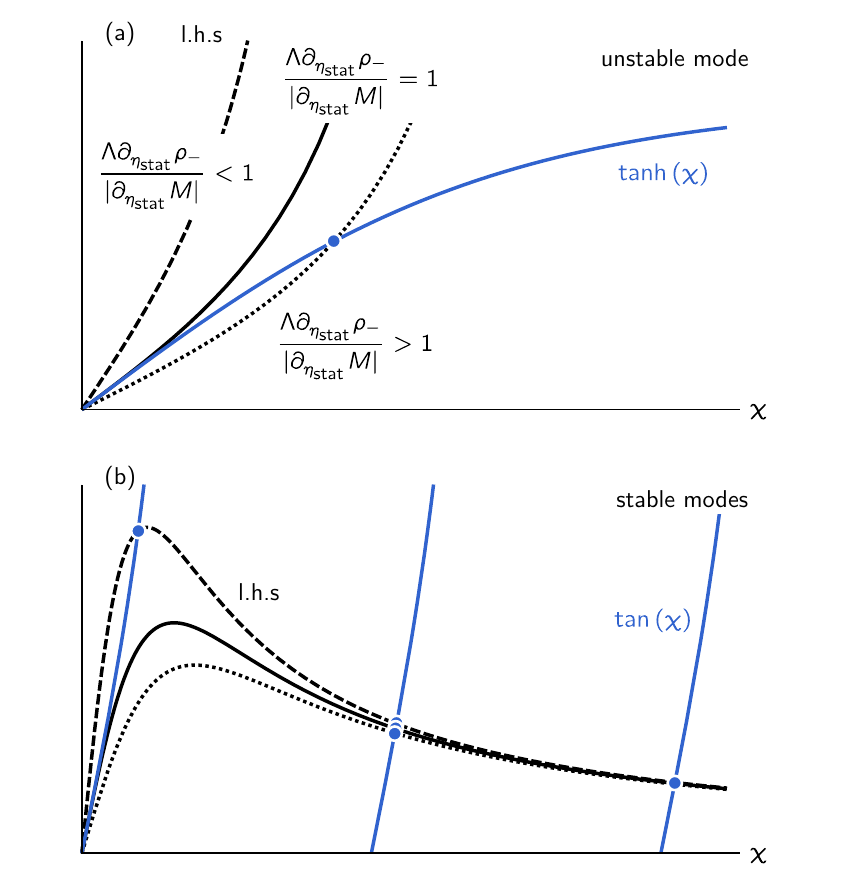}
    \caption{Graphical construction of the solutions to the matching conditions for the mass-redistribution modes of the stationary elementary peak pattern.
    (a) The growth rate of an unstable mode has to fulfill condition Eq.~\eqref{eq:consistency-relation-destabilizing-rate}.
    The solutions of this equation correspond to intersections of the left-hand side (black) and right-hand side [blue (dark gray) line].
    One unstable mode [(blue) dot)] arises if ${\Lambda\partial_\etastat^{}\rho_- > \partial_\etastat^{} M}$.
    (b) The relaxation rates have to solve Eq.~\eqref{eq:consistency-relation-relaxation-rate}.
    The left-hand side is shown in black (different dashing corresponds to the cases labeled in a), and the right-hand side is in blue (dark gray).
    Infinitely many relaxation modes are found due to the periodic branches of the tangent.
    A solution on the first branch exists if no solution is found for the unstable mode.
    }
    \label{fig:Stability-ConsistencyCondition}
\end{figure}

Combining Eq.~\eqref{eq:relax-self-consistency} with Eq.~\eqref{eq:peak-relax-mode-peak}, one retrieves the implicit (self-consistency) relation that determines $\chi$ and thus the relaxation rate $\sigma_\mathrm{relax}$ throughout both the diffusion- and reaction-limited regimes:
\begin{equation*}
    \chi \left[\frac{\Lambda \partial_\etastat \rho_-}{-\partial_\etastat M} + \frac{4 D_v (1+d)}{2\Lambda\ell_\mathrm{int}\langle \tilde{f}_\eta\rangle_\mathrm{int}}\chi^2\right]^{-1} \approx \tan(\chi)
    \, .
\end{equation*}
Recall that ${\partial_\etastat M < 0}$, such that the first term in the denominator is positive.
Comparing with the mass-competition rates [Eqs.~\eqref{eq:sigma-MC-peak-D-R}], we write more compactly
\begin{equation}\label{eq:consistency-relation-relaxation-rate}
    \chi \left(\frac{\Lambda \partial_\etastat\rho_-}{|\partial_\etastat M |} + \frac{\sigma_\mathrm{D}}{\sigma_\mathrm{R}}\chi^2\right)^{-1} \approx \tan(\chi)
    \, .
\end{equation}

Before calculating the slowest relaxation rate from the solutions $\chi(\sigma_\mathrm{relax})$ to Eq.~\eqref{eq:consistency-relation-relaxation-rate}, we have to ensure for the stability of the peak that no unstable mode exists.
The above analysis can be repeated under the assumption of a positive growth rate ${\sigma_\mathrm{relax} > 0}$.
If such a solution exists, the associated eigenmode renders the elementary stationary pattern unstable.
In this case, the mode profile [see Eqs.~\eqref{eq:plateau-profile-relax-mode}] in the plateaus is given by a hyperbolic cosine and the matching condition for the unstable mode(s) follows as
\begin{equation}\label{eq:consistency-relation-destabilizing-rate}
    \chi \left(\frac{\Lambda \partial_\etastat\rho_-}{|\partial_\etastat M |} - \frac{\sigma_\mathrm{D}}{\sigma_\mathrm{R}}\chi^2\right)^{-1} \approx \tanh(\chi)
    \, .
\end{equation}
This equation has at most one solution for ${\chi > 0}$ [see Fig.~\ref{fig:Stability-ConsistencyCondition}(a)].
It has no solution if
\begin{equation*}
    \Lambda \partial_\etastat \rho_- < -\partial_\etastat M
    \, ,
\end{equation*}
implying that there is no positive eigenvalue, and thus, that the peak is stable.
With this, we recover the heuristic stability argument introduced in the supplementary material of Ref.~\cite{Brauns.etal2021}.

If the peak is stable, Eq.~\eqref{eq:consistency-relation-relaxation-rate} yields solutions ${\chi > 0}$ on each branch of the tangent [see intersection points in Fig.~\ref{fig:Stability-ConsistencyCondition}(b)].
The solution on the first branch is lost and gives rise to the unstable mode if the stability condition is violated.
While the solution on the first branch yields redistribution of mass between the plateau and the peak, solutions on higher branches correspond to modes with several nodes in the cosine profile in the plateau.
Consequently, these describe in addition redistribution of mass within the plateau itself.
The slowest relaxation mode corresponds to the solution on the first branch because $\sigma_\mathrm{relax}$ increases monotonously with $\chi$: One has ${|\sigma_\mathrm{relax}|= 4\chi^2 D_-/\Lambda^2}$ by definition.
The higher modes have faster relaxation rates because the mass redistribution occurs on shorter scales in these modes.

We can now use Eq.~\eqref{eq:consistency-relation-relaxation-rate} to estimate the magnitude of the slowest relaxation rate $\sigma_\mathrm{relax}$.
If this slowest relaxation rate is fast compared to the rate of the mass-competition process, our assumption of fast (regional) relaxation of the elementary stationary patterns holds.
From the functional form of the tangent, we know that the first solution fulfills ${0  < \chi < \pi/2}$.
It follows that ${\chi \sim 1}$ if it is not particularly small.
Thus from the definition of $\chi$ we find, if ${\chi\sim 1}$,
\begin{equation}
\label{eq:relaxation-rate-diff-lim}
    \sigma_\mathrm{relax}^\mathrm{D}\sim -\frac{4D_v}{\Lambda^2}\partial_\rho\eta^*(\rho_-)
    \, .
\end{equation}
The relaxation rate shows different behavior if ${\chi \ll 1}$, a case which arises in two ways.
First, $\chi$ is small if the fraction $\Lambda \partial_\etastat \rho_-/| \partial_\etastat M|$ is very close to one, i.e., if ${|\Lambda \partial_\etastat \rho_-|-| \partial_\etastat M|\ll 1}$.
This is not generic in the 2cMcRD systems because the two terms in the latter condition scale differently with the peak mass or domain size (see Appendix~\ref{app:peak-scaling}), and we do not consider this case here.
Second, for ${\sigma_\mathrm{D} \gg \sigma_\mathrm{R}}$, that is, in the reaction-limited regime, we find ${\chi \approx \sqrt{\sigma_\mathrm{R}/\sigma_\mathrm{D}}\ll 1}$ and thus
\begin{equation}\label{eq:relaxation-rate-reac-lim}
    \sigma_\mathrm{relax}^\mathrm{R} \approx -\frac{2\ell_\mathrm{int}\langle\tilde{f}_\eta\rangle_\mathrm{int}}{(1+d)\Lambda} \, 
    \partial_\rho\eta^*(\rho_-)
    \, .
\end{equation}
Consequently, the first expression, Eq.~\eqref{eq:relaxation-rate-diff-lim}, describes the relaxation rate in the diffusion-limited regime while the second expression, Eq.~\eqref{eq:relaxation-rate-reac-lim}, gives the relaxation rate in the reaction-limited regime.

The above two expressions, Eqs.~\eqref{eq:relaxation-rate-diff-lim},~\eqref{eq:relaxation-rate-reac-lim}, resemble the growth rates of the mass-competition instability in the diffusion- and reaction-limited regime [see Eqs.~\eqref{eq:sigma-MC-peak-D-R}], with the crucial difference that there is the factor $\partial_\rho\eta^*(\rho_-)/\Lambda$, which is always positive, instead of $\partial_M \etastat$, which is negative for 2cMcRD systems (see Sec.~\ref{sec:stat-patterns}).
As a result, and in contrast to the rates $\sigma_\mathrm{D}$, $\sigma_\mathrm{R}$ of the mass-competition instability, here both rates are negative and thus stabilizing.
In addition, by comparison of the rate magnitudes, one finds that the mass-competition instability is indeed slow compared to this relaxation mode if [cf.\ Eq.~\eqref{eq:small-plateau-mass}]
\begin{equation}\label{eq:small-plateau-mass-app}
    \Lambda \partial_{\etastat} \rho_- \ll -\partial_\etastat M
    \, .
\end{equation}

The same analysis as above can be performed for mesa patterns.
The observed interface stability has been discussed as ``wave-pinning'' in Ref.~\cite{Mori.etal2011}.
Again, the relaxation mode has cosine profiles in the plateau regions.
For continuity, they have to meet at the interface. 
Mass incorporation or release at the interface during the relaxation of the mesa pattern toward its stationary profile creates a flux difference left and right of the interface, that is, a mismatch in the gradients left and right [integrate the continuity equation, Eq.~\eqref{eq:cont-eq} (${\varepsilon=0}$), over the interface region]
\begin{equation*}
    D_v \partial_x \eta |_{x_0 + b} - D_v \partial_x \eta |_{x_0 - b} 
    = 
    \partial_t M 
    \, ,
\end{equation*}
where $\pm b$ denotes a small offset from the interface position $x_0$ fulfilling ${\ell_\mathrm{int}\ll b\ll \Lambda/2}$.
This mass change corresponds to an interface shift
\begin{equation*}
    \partial_t M = \Delta\rho\, \partial_t x_0
    \, .
\end{equation*}
The reaction-rate integral is used to determine the value of the mass-redistribution potential at the interface consistent with the rate of mass uptake $\partial_t M$.
This inner solution is matched to the cosine profiles in the plateaus.
The procedure results in the same condition, Eq.~\eqref{eq:small-plateau-mass-app}.
For mesas, $|\partial_\etastat M|$ becomes exponentially large in the plateau size (see Sec.~\ref{sec:stat-patterns} and Appendix~\ref{app:stat-state-MC}) such that the above condition is clearly fulfilled for sufficiently narrow interfaces compared to the overall length $\Lambda/2$ of the elementary pattern.

\section{Growth rate of the mass-competition instability within the sharp-interface approximation}
\label{app:MC-mass-competition}
In this section, we perform the linear stability analysis (LSA) of two neighboring stationary peaks and mesas (peak/mesa competition) in the 2cMcRD system.
The result is the growth rate $\sigma$ of the mass-competition instability within a singular perturbation calculation.
The rate for mesa coalescence is found analogously as for mesa competition since the coalescence scenario can be interpreted as competition between the low-density plateaus of the mesa pattern (see Sec.~\ref{sec:coarsening-phenomenology}).
This rate also estimates the speed of the peak-coalescence scenario.

\subsection{Competition of two peaks}
We choose the system domain as ${\Omega = [-\Lambda/2,\Lambda/2]}$ and analyze the stability of the symmetric pattern containing two half peaks at the outer boundaries [see Fig.~\ref{fig:peak-MC-modeApprox}(a)].
This symmetry of the stationary pattern ensures that the eigenmodes of the linearized dynamics around this pattern are either symmetric or antisymmetric under the replacement ${x\to -x}$.
The mass-competition mode increases the mass of one peak while reducing the mass of the other such that it has to be antisymmetric around ${x = 0}$ [see Fig.~\ref{fig:peak-MC-modeApprox}(b)].
Assuming that mass competition is slow compared to the relaxation modes of the single peaks themselves (see Sec.~\ref{sec:approximations} and Appendix~\ref{app:stability}), the peak profiles adiabatically follow the stationary profiles as mass competition changes the peak mass.
All deviations from the stationary peak profile relax, by assumption, on a timescale fast compared to the timescale on which mass competition changes the peak masses.
Thus, we approximate the mode at each peak by the mass mode $(\partial_\etastat \rhostat,1)$, which describes the change of the stationary peak profile under mass change.
Because deviations from the stationary profile are due to the finite relaxation rate $\sigma_\mathrm{relax}$ onto the stationary profile in comparison to the rate $\sigma$ of the mass-competition dynamics, we expect deviations of the order ${\sim\sigma/\sigma_\mathrm{relax}}$.
This scaling of the eigenmode's deviations from its unperturbed form (here this is the mass mode of an isolated peak) can be supported by a more rigorous perturbation theory for the Jacobian operators, which we do not perform here.
Thus, one has
\begin{subequations}\label{eq:peak-mode-approx-mc}
\begin{align}
    \delta\rho(x) &= A \left[\operatorname{sgn}(x) \frac{2\partial_{\etastat}\rhostat(x)}{\partial_{\etastat}M} +\order{\sigma/\sigma_\mathrm{relax}}\right]
    \, ,
    \label{eq:rho-approx}\\
    \delta\eta(x) &= A\,
    \order{\sigma/\sigma_\mathrm{relax}}
    \, ,
\end{align}
\end{subequations}
introducing the mode amplitude $A$.
Using the chain rule in Eq.~\eqref{eq:rho-approx}, one obtains ${\delta\rho \approx 2 A\operatorname{sgn}(x)\, \partial_M^{}\rhostat(x)}$.
From this expression, one reads off that the discontinuity introduced in $\delta\rho$ at ${x=0}$ is only of order ${A\partial_M^{}\rho_- = A \, \order{\sigma/\sigma_\mathrm{relax}}}$ [using Eqs.~\eqref{eq:sigma-MC-peak-D-R},~\eqref{eq:relaxation-rate-diff-lim},~\eqref{eq:relaxation-rate-reac-lim}], and thus of the same order as the correction terms.
Hence, the continuity of the eigenmode profile can be ensured by the correction terms.
In contrast, no term larger than $\sim \sigma/\sigma_\mathrm{relax}$ can appear in $\delta\eta$.
Because $\eta$ is constant in the mass mode, such a term would induce a jump ${\sim A}$ at ${x=0}$, in contradiction with the continuity of the eigenmode profile.

We will now proceed in three steps. First, we show that $\delta\eta$ can be assumed to be linear between the peaks.
Second, integration of the continuity equation [first row of Eq.~\eqref{eq:eigenvalue-problem-MC}] yields a relationship between $\sigma$, $A$ and $\delta\eta|_{x=\Lambda/2}$.
Last, the equation for $\delta\eta$ [second row of Eq.~\eqref{eq:eigenvalue-problem-MC}] is used to find another equality based on approximating the conversion-rate integral [cf.\ Eq.~\eqref{eq:relax-convRateInt}] relating again $\sigma$, $A$ and $\delta\eta|_{x=\Lambda/2}$.
As the mode amplitude $A$ is arbitrary and can be scaled out (due to the linearity of the dynamics), the resulting two equations together determine the growth rate $\sigma$ and the ratio $\delta\eta|_{x=\Lambda/2}/A$.

\textit{The linear gradient in $\delta\eta$.\;---}
In the plateau between the peaks the eigenmode has to fulfill Eq.~\eqref{eq:eigenvalue-problem-plateaus} (${\varepsilon = 0}$):
\begin{equation}
\label{eq:diff-mass-red-plateau}
    \frac{\sigma}{D_v\partial_{\rho}\eta^*(\rho_-)} \, 
    \delta\eta 
    \approx 
    \partial_x^2\delta\eta
    \, .
\end{equation}
The relaxation rate of a single peak fulfills the inequality ${|\sigma_\mathrm{relax}| \lesssim \frac{4 D_v}{\Lambda^2}\partial_{\rho}\eta^*(\rho_-)}$ (Appendix~\ref{app:stability}), which implies
\begin{equation*}
    \frac{\sigma}{D_v\partial_{\rho}\eta^*(\rho_-)} \delta\eta
    \lesssim 
    \frac{\sigma}{\sigma_\mathrm{relax}} \frac{\delta\eta}{\Lambda^2}
    \, .
\end{equation*}
Consequently, the curvature of the $\delta\eta$ profile on the scale of the peak separation $\Lambda$ is of order $\order{\sigma/\sigma_\mathrm{relax}}$ and negligible:
\begin{equation*}
    \frac{\partial_x^2 \delta\eta}{\delta\eta/\Lambda^2}\lesssim \frac{\sigma}{\sigma_\mathrm{relax}}
    \, .
\end{equation*}
With Eq.~\eqref{eq:diff-mass-red-plateau} describing the diffusive mass redistribution within the plateau, this explicitly shows that mass redistribution \emph{in the plateau} proceeds fast compared to the mass competition process between the peaks.
Consequently, the mass-redistribution potential rapidly relaxes to fulfill the Laplace equation ${0 = \partial_x^2\delta\eta}$ between the peaks.
Using the antisymmetry of the mode it follows
\begin{equation}\label{eq:peak-eta-mode-approx-mc}
    \delta\eta(x) = 2 \delta\eta{\big|}_{x=\Lambda/2} \frac{x}{\Lambda}\, [1+\order{\ell_\mathrm{int}/\Lambda}]
    \, ,
\end{equation}
where the correction term is due to the finite peak width, and it may be neglected within the sharp-interface approximation [see Fig.~\ref{fig:peak-MC-modeApprox}(b)].

\textit{Mass redistribution from peak to peak.\;---}
Integration of the continuity equation [see Eq.~\eqref{eq:eigenvalue-problem-MC}] and using the approximation, Eq.~\eqref{eq:rho-approx}, of the density mode then gives
\begin{equation}\label{eq:peak-integrated-cont-eq}
    \sigma A\int_0^{\frac{\Lambda}{2}}\mathrm{d}x\, 2\partial_M^{}\rhostat(x) \approx \sigma A \approx -\frac{2 D_v}{\Lambda}\delta\eta{\big|}_{x=\Lambda/2}
    \, .
\end{equation}
Thus, the gradient in $\delta\eta$ between the peaks determines how fast the peaks grow/shrink.

\textit{The value $\delta\eta|_{x=\Lambda/2}$ at the peak.\;---}
The gradient in $\delta\eta$ is determined by the change of the mass-redistribution potential $\delta\eta|_{x=\Lambda/2}$ at the peak.
To determine a second condition on the change of the mass-redistribution potential, we will express the conversion-rate integral [cf.\ Eq.~\eqref{eq:relax-convRateInt}] in different ways.
Similar integrals will be central to the derivations of all other growth rates as well, and we already used the conversion-rate integral to determine the relaxation rates in appendix~\ref{app:stability}.
Why does it appear?
The mass-redistribution potential changes at the peak due to the reactive conversion of particles between the $u$ and $v$ states (see Sec.~\ref{sec:mc-growth-rates}).
The strength of the reactive conversion determines how strongly $\delta\eta|_{x=\Lambda/2}$ deviates from $\delta M \partial_M^{}\etastat$ (cf.\ Sec.~\ref{sec:growth-rate-nMC-discussion}).
This strength is captured by the conversion-rate integral (as we see below) while it does not enter in the mass-redistribution equation, Eq.~\eqref{eq:peak-integrated-cont-eq}.
On this note, we use the approximation of the density mode Eq.~\eqref{eq:rho-approx} to express the conversion-rate integral by
\begin{align}\label{eq:peak-reaction-rate-integral}
    \int_0^{\frac{\Lambda}{2}}\mathrm{d}x\, \tilde{f}_\eta\, \delta\rho \approx A \int_0^\frac{\Lambda}{2}\mathrm{d}x\, \tilde{f}_\eta\, 2 \partial_M^{}\rhostat = A \langle \tilde{f}_\eta\rangle_\mathrm{int}
    \, .
\end{align}
Here, $\langle \tilde{f}_\eta\rangle_\mathrm{int}$ represents the reaction rate $\tilde{f}_\eta$ averaged at the peak/interface.
The basis for this average is the distribution of mass inclusion at the peak/interface which can be defined
by the change of the density profile of the stationary elementary pattern due to mass increase:
\begin{equation*}
    P(x) = 2 \partial_{M}^{}\rhostat(x) = 2 \left(\partial_{M}^{}\etastat\right) \partial_{\etastat}^{}\rhostat(x)
    \, .
\end{equation*}
The factor 2 is introduced such that $P(x)$ is (approximately) normalized:
\begin{equation*}
    \int_0^\frac{\Lambda}{2} \mathrm{d}x \, P(x) = 2 \partial_{M}^{} \int_0^\frac{\Lambda}{2} \mathrm{d}x \, \rhostat(x) \approx \partial_{M}^{} M = 1
    \, .
\end{equation*}
Intuitively, the derivative $\partial_{M}^{}\rhostat(x)$ is localized at the peaks/interface.
Technically, this is ensured by the condition Eq.~\eqref{eq:small-plateau-mass}.
Thus, $P(x)$ defines a weighted average that is localized to the peak/interface by
\begin{equation}\label{eq:interface-average}
    \langle\cdot\rangle_\mathrm{int} = \int_0^\frac{\Lambda}{2}\mathrm{d}x\,\cdot P(x)
    \, .
\end{equation}
Moreover, we define the half-peak or interface width $\ell_\mathrm{int}$ as estimate of the width of $P(x)$:
Because $\langle P(x)\rangle_\mathrm{int}$ gives the average height of the distribution $P(x)$, a width estimate for $P(x)$ follows by distributing the area under the curve into a rectangle ${\ell_\mathrm{int} \langle P(x)\rangle_\mathrm{int} \approx 1}$, where we used that $P(x)$ is (approximately) normalized.
Thus, we define
\begin{equation}\label{eq:interface-width}
    \ell_\mathrm{int} = \big(\langle P\rangle_\mathrm{int}\big)^{-1}
    \, .
\end{equation}

Coming back to the conversion rate integral Eq.~\eqref{eq:peak-reaction-rate-integral} and using that the mass mode is a zero mode [specifically, Eq.~\eqref{eq:mass-mode-eq}], we find as second approximation of the conversion-rate integral
\begin{align}
    \int_0^{\frac{\Lambda}{2}}\mathrm{d}x\, \tilde{f}_\eta\, \delta\rho &\approx \int_0^{\frac{\Lambda}{2}}\mathrm{d}x\, \left(\partial_{\etastat}\rhostat\right) \mathcal{L}\,\delta\rho\nonumber\\
    &\approx \int_0^{\frac{\Lambda}{2}}\mathrm{d}x\,
    \left(\partial_{\etastat}\rhostat\right) \left[\tilde{f}_\eta\delta\eta - (1+d)\sigma \delta\rho \right]\nonumber\\
    &\approx \frac{1}{2}\delta\eta|_{x=\Lambda/2} \left(\partial_\etastat^{}M\right) \langle \tilde{f}_\eta\rangle_\mathrm{int}\nonumber\\
    &\quad - (1+d)\sigma \frac{A}{2} \left(\partial_{\etastat}^{}M\right) \frac{1}{\ell_\mathrm{int}}
    \, . \label{eq:peak-reaction-rate-integral-2}
\end{align}
In the first line, we neglected exponentially small boundary terms from the pattern tails approaching the low-density plateau. In the second step, we used the evolution equation of the eigenmode, Eq.~\eqref{eq:eigenvalue-problem-MC}, and ${\sigma/\tilde{f}_\eta\ll 1}$. 
In the last line, we applied the sharp-interface approximation in the first term to take $\delta\eta$ as constant across the peak.
In the second term, we used the mode approximation Eq.~\eqref{eq:rho-approx}.

Combining Eqs.~\eqref{eq:peak-integrated-cont-eq},~\eqref{eq:peak-reaction-rate-integral},~\eqref{eq:peak-reaction-rate-integral-2} one finally extracts the growth rate [cf.\ Eqs.~\eqref{eq:sigma-MC-massComp},~\eqref{eq:sigma-MC-peak-D-R}]
\begin{equation*}
    \sigma \approx -\frac{\partial_M^{}\etastat}{\frac{\Lambda}{4 D_v}+\frac{1+d}{2 \ell_\mathrm{int}\langle \tilde{f}_\eta\rangle_\mathrm{int}}}
    \, .
\end{equation*}

\begin{figure}
    \centering
    \includegraphics{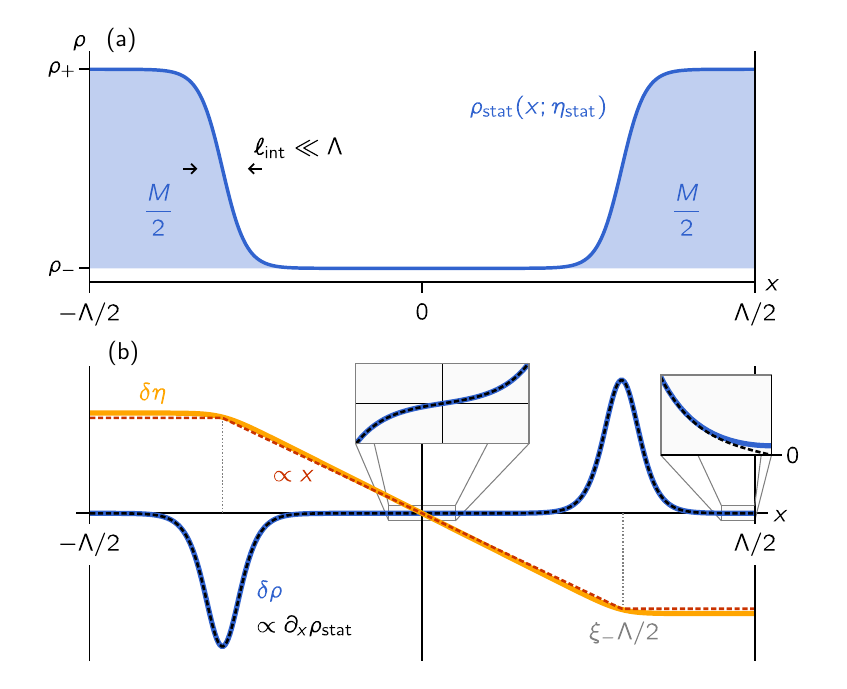}
    \caption{
    Mass competition of two mesas.
    (a) We consider the competition for mass between two stationary half mesas $\rhostat(x)$, each containing mass $M/2$.
    (b) The mass-competition mode [blue (dark gray), orange (light gray)] is antisymmetric.
    At the interfaces and in the inner plateau, the density mode $\delta\rho(x)$ [blue (dark gray)] is well approximated by the translation mode $\partial_x\rhostat$ (black, dashed) as mass-competition translates the trough between the two half mesas.
    Deviations appear at the boundaries ${x=\pm \Lambda/2}$ (inset).
    Within the sharp-interface approximation, the mass-redistribution potential $\delta\eta(x)$ [orange (light gray), strongly magnified] is linear in the plateaus [red (dashed, piecewise straight) line].
    In this figure, the stationary state and the mode approximations are exemplified for the cubic model ${\tilde{f} = \eta - \rho^3 + \rho}$ with parameters ${D_u = 1}$, ${D_v = 10}$, ${\Lambda = 40}$, and ${\bar{\rho} = -0.2}$ (see Appendix~\ref{app:cubic-model}).
    }
    \label{fig:mesa-MC-modeApprox}
\end{figure}

\subsection{Competition between mesas}
We can apply the same analysis as above to determine the growth rate $\sigma$ of the mass-competition instability of mesa patterns (mesa competition/coalescence).
However, the growth rate is exponentially small in the plateau lengths because the steady-state mass-redistribution potential $\etastat(M)$ is only varying due to exponentially small contributions from the pattern tails approaching the high- and low-density plateaus [recall that ${\partial_M^\pm{\etastat} = \order{\delta\rho_\pm^2}}$, see Sec.~\ref{sec:stat-patterns}].
Thus, one has to be careful where one can neglect boundary terms that arise due to these exponential tails of the pattern approaching the high- and low-density plateaus.
For peak competition, we neglected these contributions because the mass-redistribution potential changes much more strongly due to the changing peak heights (cf.\ Appendix~\ref{app:stat-state-MC}) than due to the exponential tails in the pattern plateau.

For concreteness, consider two high-density `half' mesas separated by a trough on the domain ${\Omega = [-\Lambda/2,\Lambda/2]}$ symmetrical about ${x = 0}$ (mesa competition).
Mass competition redistributes mass between the two mesas and shifts the interfaces---due to mass conservation---synchronously.
Thus, the trough separating the mesas is translated as a whole.
Accordingly, we approximate the mass-competition eigenmode by
\begin{subequations}
\begin{align}
    \delta\rho(x) &\approx A\, \partial_x\rhostat
    \, ,\\
    \delta\eta(x) &\approx \begin{cases}
    \frac{2\delta\eta_\mathrm{int}}{\xi_-\Lambda} x & x\in \left(-\frac{\xi_-}{2}\Lambda,\frac{\xi_-}{2}\Lambda\right)\\
    -\delta\eta_\mathrm{int} & x < -\frac{\xi_-}{2}\Lambda\\
    \delta\eta_\mathrm{int} & x > \frac{\xi_-}{2}\Lambda
    \end{cases}
    .\label{eq:mesa-ansatz-eta-mode}
\end{align}
\end{subequations}
We use the relative length $\xi_-$ of the lower plateau introduced for the stationary mesa patterns in Appendix~\ref{app:stat-state-MC}.
The derivative $\partial_x\rhostat$ describes the translation mode of the stationary density profile.
On the infinite line, it is an exact zero mode due to the translation invariance of the system.
In the finite system, deviations appear at the boundaries [see Fig.~\ref{fig:mesa-MC-modeApprox}(b)].
Moreover, as for the peak-forming system [see Eqs.~\eqref{eq:peak-mode-approx-mc}] this mode approximation neglects corrections ${\sim \sigma/\sigma_\mathrm{relax}}$ due to the finite relaxation rates of the elementary stationary pattern.

The piecewise linear approximation of the mass-redistribution potential $\delta\eta$ uses once more that the relaxation of the plateaus is fast compared to the mass redistribution between the two mesas (${\sigma_\mathrm{relax}\gg\sigma}$).
Thus, $\delta\eta$ has to fulfill the Laplace equation ${0 = \partial_x^2\delta\eta}$ in the plateaus up to corrections ${\sim\sigma/\sigma_\mathrm{relax}}$, that is, $\delta\eta$ has to be linear in the plateaus [cf.\ derivation for peak patterns, Eq.~\eqref{eq:peak-eta-mode-approx-mc}].
The antisymmetry of the eigenmode, continuity, and the no-flux boundary conditions then prescribe the chosen form, Eq.~\eqref{eq:mesa-ansatz-eta-mode}.
The finite interface width yields corrections that are smaller by a factor proportional to $\ell_\mathrm{int}/\Lambda$.

Integration of the continuity equation, first component of Eq.~\eqref{eq:eigenvalue-problem-MC}, then yields [analogously to Eq.~\eqref{eq:peak-integrated-cont-eq}]
\begin{equation}\label{eq:mesa-int-cont-eq}
    \sigma_+ A \int_0^\frac{\Lambda}{2}\mathrm{d}x\, \partial_x\rhostat(x) \approx \sigma_+ A \Delta\rho \approx - \frac{2 D_v}{\xi_-\Lambda}\delta\eta_\mathrm{int} \, ,
\end{equation}
with ${\Delta\rho = \rho_+-\rho_-}$.
The conversion rate integral yields analogously to Eqs.~\eqref{eq:peak-reaction-rate-integral},~\eqref{eq:peak-reaction-rate-integral-2} the two expressions [using Eq.~\eqref{eq:mass-mode-eq}]
\begin{subequations}\label{eq:mesa-reaction-rate-integral}
\begin{align}
    \int_0^\frac{\Lambda}{2}\mathrm{d}x\, \tilde{f}_\eta \delta\rho &\approx A \Delta\rho \langle \tilde{f}_\eta\rangle_\mathrm{int}
    \, , \\
    \int_0^\frac{\Lambda}{2}\mathrm{d}x\, \tilde{f}_\eta \delta\rho &= D_u{\big[}\left(\partial_{\etastat}\rhostat\right)\partial_x\delta\rho{\big]}_0^\frac{\Lambda}{2}\nonumber\\
    &\quad +\int_0^\frac{\Lambda}{2}\mathrm{d}x\, \left(\partial_{\etastat}\rhostat\right) \mathcal{L}\,\delta\rho\nonumber\\
    &\approx -D_u\partial_{\etastat}\rhostat{\big|}_{x=0}\partial_x\delta\rho{\big|}_{x=0}\nonumber\\
    &\quad + \frac{\delta\eta_\mathrm{int}}{2} \langle \tilde{f}_\eta\rangle_\mathrm{int} \left(\partial_{\etastat}^{}M\right)\nonumber\\
    &\quad - (1+d) \frac{\sigma_+ A \Delta\rho}{2\ell_\mathrm{int}} \left(\partial_{\etastat}^{}M\right),\label{eq:mesa-MC-convRateInt-BT}
\end{align}
\end{subequations}
where $[\,]_y^x$ denotes the boundary terms due to partial integration.
The conversion rate $\langle \tilde{f}_\eta\rangle_\mathrm{int}$ and the interface width $\ell_\mathrm{int}$ are defined as for peaks [see Eqs.~\eqref{eq:interface-average}, \eqref{eq:interface-width}] using ${P(x) = \partial_x\rhostat/\Delta\rho}$.
We can use the translation mode $\partial_x\rhostat$ to define the distribution $P(x)$ of mass inclusion at the mesa interface because, within the sharp-interface approximation, additional mass added to the mesa only translates the interface (see Sec.~\ref{sec:stat-patterns} and Appendix~\ref{app:stat-state-MC}).

The new aspect in Eq.~\eqref{eq:mesa-MC-convRateInt-BT} for mesa patterns compared to peak patterns is the boundary term that must not be dropped because it is of the same order as the other terms, as we see now.
We approximate the boundary term using the asymptotic result for the tails of the density profile in the lower plateau, Eqs.~\eqref{eq:mesa-exp-tails}.
This yields
\begin{align}\label{eq:mesa-boundary-term}
    -&\partial_{\etastat}\rhostat{\big|}_{x=0}\partial_x\delta\rho{\big|}_{x=0}\nonumber\\
    &\qquad\approx -\left(\partial_{\etastat}\delta\rho_-\right) A \partial_x^2\rhostat{\big|}_{x=0}\nonumber\\
    &\qquad\approx -\frac{A}{\ell_-^2}\frac{\partial_{\etastat}\delta\rho_-^2}{2}
    \, .
\end{align}
Combining Eqs.~\eqref{eq:mesa-int-cont-eq}, \eqref{eq:mesa-reaction-rate-integral}, \eqref{eq:mesa-boundary-term} the growth rate follows as [cf.\ Eqs.~\eqref{eq:sigma-MC-massComp},~\eqref{eq:sigma-MC-mesa-D-R}]
\begin{align}
    \sigma_+
    &\approx -\frac{1+\frac{\tilde{f}_\rho(\rho_-,\etastat)}{\Delta\rho \langle \tilde{f}_\eta\rangle_\mathrm{int}}\frac{\partial_{\etastat}\delta\rho_-^2}{2}}{\frac{\xi_- \Lambda}{4 D_v}+ \frac{1+d}{2\ell_\mathrm{int}\langle \tilde{f}_\eta\rangle_\mathrm{int}}}\partial_M{}\etastat\nonumber\\
    &= -\frac{\partial_{M}^+\etastat}{\frac{\xi_- \Lambda}{4 D_v}+ \frac{1+d}{\ell_\mathrm{int}\langle \tilde{f}_\eta\rangle_\mathrm{int}}}
    \, .
\end{align}
The second mass-competition scenario, that is, mesa coalescence may be treated analogously.
Repeating the above analysis yields the growth rate
\begin{equation*}
    \sigma_- \approx -\frac{\partial_{M}^-\etastat}{\frac{\xi_+ \Lambda}{4 D_v}+ \frac{1+d}{2\ell_\mathrm{int}\langle \tilde{f}_\eta\rangle_\mathrm{int}}}
    \, ,
\end{equation*}
describing the growth of one and simultaneous shrinking of the other trough [translation of the mesa in the middle of the domain, cf.\ Fig.~\ref{fig:coarsening-modes}(b)].
Because $\partial_{M}^\pm\etastat$ is exponentially small in $\xi_\pm\Lambda/\ell_\pm$, the relative size of the two growth rates $\sigma_\pm$ (mainly) depends on the relative lengths $\xi_\pm\Lambda/\ell_\pm$ of the upper and lower plateaus compared to the exponential tails.
Moreover, recall that the length ratio of the upper and lower plateaus, i.e., the interface position is tuned by the average density $\bar{\rho}$ (see Sec.~\ref{sec:stat-patterns} and Appendix~\ref{app:stat-state-MC}).
Consequently, at sufficiently low average density $\bar{\rho}$, we have ${\sigma_+ \gg \sigma_-}$, and coarsening proceeds via growth and shrinking of mesas [see Fig.~\ref{fig:coarsening-phenomenology}(c)].
In contrast, at high average density ${\sigma_-\gg \sigma_+}$ holds, and coarsening is mainly driven by the growth and shrinking of troughs, that is, coalescence of mesas [see Fig.~\ref{fig:coarsening-phenomenology}(e)].

In a periodic pattern containing several peaks/mesas, modes of mass competition between second-next and further neighbors exist as well.
Their growth rates are suppressed with respect to the nearest-neighbor competition as mass redistribution is slower over larger distances~\cite{Langer1971,Brauns.etal2021}.
Also in the reaction-limited regime, the rate of nearest-neighbor competition is a good estimate of the fastest competition rate in all cases.\footnote{
In Ref.~\cite{McKay.Kolokolnikov2012}, growth rates for next-neighbor and higher modes are calculated in the reaction-limited regime for mesa patterns, that is, if diffusive mass redistribution is not the limiting factor of the instability.
The growth rates for the higher modes are found to be the same as for nearest-neighbor competition if ${\sigma_+\gg\sigma_-}$ or ${\sigma_-\gg\sigma_+}$.
Only if the average density $\bar{\rho}$ is tuned such that ${\sigma_+\approx\sigma_-}$, the longest-wavelength mode is the most unstable with a rate ${\approx\sigma_+ +\sigma_-}$.
}
Hence, the full growth rates derived for 2cMcRD systems in this section can be used to estimate the dynamics of coarsening as described in Sec.~\ref{sec:coarsening-scaling}.
The rates for the mass-conserving CH and cAC models, agreeing with the limiting expressions in the diffusion- and reaction-limited regimes, can be obtained by the same methods.

\section{Modification of the stationary states by weak source terms}
\label{app:stat-state-nMC}
\begin{figure}
    \centering
    \includegraphics{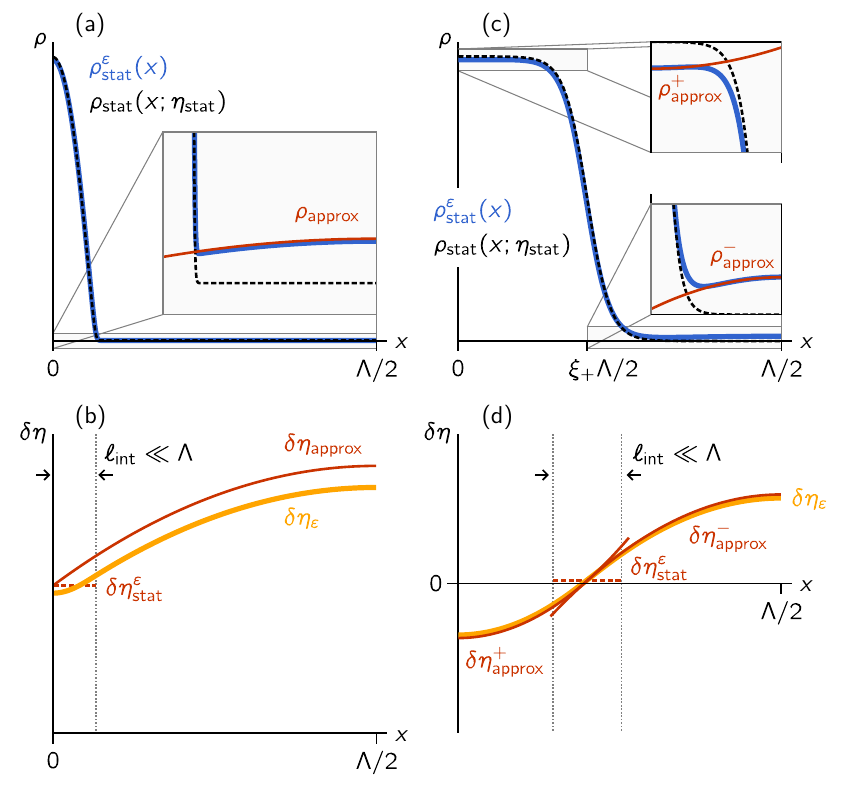}
    \caption{
    Modification of the stationary elementary pattern by weak source terms.
    (a) The peak profile of the mass-conserving system $\rhostat$ (black dashed) approximates the peak profile in the presence of source terms $\rhostateps$ [blue (dark gray)] up to corrections ${\sim \varepsilon}$.
    In the plateau, $\rhostateps$ can be approximated analytically by a parabolic profile [$\rho_\mathrm{approx}$, red (thin) line].
    (b) Due to the source terms, the stationary mass-redistribution potential ${\etastateps = \etastat + \delta\eta_\varepsilon}$ [$\delta\eta_\varepsilon$ shown in orange (light gray)] attains a spatially varying profile.
    In the plateau, it is also approximately parabolic [$\delta\eta_\mathrm{approx}$, red (dark gray)].
    The offset at the peak is approximated by $\delta\etastateps$ (red dashed).
    (c) The same construction as for the stationary peak profile is shown for the elementary stationary mesa pattern.
    Two distinct approximations [$\rho_\mathrm{approx}^\pm$, red (dark gray)] apply in the high- and low-density plateaus.
    (d) The change of the mass-redistribution potential $\delta\etastateps$ [orange (light gray)] induced by the source terms is approximated by matching the quadratic approximations in both plateaus $\delta\eta_\mathrm{approx}^\pm$ [red (dark gray)] at the interface position to the offset $\delta\etastateps$ (red dashed).
    We illustrate the construction for the peak-forming model ${\tilde{f}=\eta-10\rho/(1+\rho^2)}$, ${(s_1,s_2) = (p-\rho,0)}$ with parameters ${D_u=1}$, ${D_v=10^3}$, ${p=4}$, ${\varepsilon=10^{-3}}$, and ${\Lambda=200}$ (see Appendix~\ref{app:const-reac-rate-peak-model}).
    The cubic model ${\tilde{f}=\eta-\rho^3+\rho}$ with the same source terms is used with parameters ${D_u=1}$, ${D_v=10}$, ${p=-0.2}$, ${\varepsilon=10^{-2}}$, and $\Lambda=40$ to exemplify the construction for mesa patterns (see Appendix~\ref{app:cubic-model}).}
    \label{fig:LS-stat-Construction}
\end{figure}

Before we can analyze mass competition under the influence of weak source terms, we need to discuss the stationary state $[\rhostateps(x),\etastateps(x)]$ of the 2cRD system including weak source terms (see Fig.~\ref{fig:LS-stat-Construction}).
The effect of weak source processes is captured as perturbation $\varepsilon(\delta\rho_\varepsilon,\delta\eta_\varepsilon)$ to the stationary state of the mass-conserving system:
\begin{align*}
    \rhostateps(x) 
    &= \rhostat(x;\etastat) + \varepsilon\,\delta\rho_\varepsilon(x)
    \, , \\
    \etastateps(x) 
    &= \etastat + \varepsilon\,\delta\eta_\varepsilon(x)
    \, ,
\end{align*}
where the dependence of the stationary profile $\rhostat(x;\etastat)$ on the value of the stationary mass-redistribution potential of the mass-conserving system $\etastat$ is stated explicitly (see Sec.~\ref{sec:stat-patterns}).
As mass conservation is broken, the stationary mass-redistribution potential is no longer spatially constant under the influence of source terms.
This coincides with a deformation of the density profile.
Moreover, the source terms that break mass conservation fix the mass of the stationary pattern because the peak or mesa mass $M$ is no longer determined by the initial condition (see Sec.~\ref{sec:stat-patterns-nMC}).
Rather, the pattern mass, and equivalently the average density $\bar{\rho}$, evolves until global production and degradation balance [Eq.~\eqref{eq:dens-evol-source-terms}].

In the following, we determine the deformation of the stationary pattern and the selected pattern mass to first order in the source strength $\varepsilon$.
Within the same perturbative approach, we find conditions on the source terms necessary for the elementary stationary patterns to be stable.
One starts out from Eqs.~\eqref{eq:gen-stat-pattern}, which imply that at first order in $\varepsilon$ the steady-state correction has to fulfill the equations
\begin{subequations}\label{eq:nMC-stat-state}
\begin{align}\label{eq:stat-cont-eq-LS}
    0 = D_v\partial_x^2\delta\eta_\varepsilon + s_\mathrm{tot}(\rhostat,\etastat)\,
    ,
\end{align}
and
\begin{align}\label{eq:stat-prof-eq-LS}
    &s_1(\rhostat,\etastat) 
    + d\, s_2(\rhostat,\etastat)
    \nonumber\\
    &\qquad= 
    -D_u\partial_x^2\delta\rho_\varepsilon - \tilde{f}_\rho \delta\rho_\varepsilon - \tilde{f}_\eta \delta\eta_\varepsilon
    \nonumber\\
    &\qquad= 
    \mathcal{L}\delta\rho_\varepsilon - \tilde{f}_\eta \delta\eta_\varepsilon
    \, .
\end{align}
\end{subequations}
The linear operator $\mathcal{L}= -D_u\,\partial_x^2-\tilde{f}_\rho$ was already defined in Eq.~\eqref{eq:lin-op-MC} for the linear mass-conserving dynamics.
For concreteness, we analyze the modification of the stationary elementary pattern on the domain ${\Omega = \left[0,\Lambda/2\right]}$ with no-flux boundary conditions and the high-density region (peak or high-density plateau) located around ${x = 0}$ (see Fig.~\ref{fig:LS-stat-Construction}).

\textit{Source balance.\;---}
Integration of Eq.~\eqref{eq:stat-cont-eq-LS} over the domain $\Omega$ respecting no-flux boundary conditions yields an explicit equation selecting the stationary pattern $(\rhostat,\etastat)$ of the mass-conserving system which approximates the stationary pattern $(\rhostateps,\etastateps)$ under the influence of weak source terms (solvability condition). 
It reads
\begin{equation}
\label{eq:source-balance}
    0 
    = 
    \int_0^\frac{\Lambda}{2}\mathrm{d}x\, s_\mathrm{tot}(\rhostat,\etastat)
    \, .
\end{equation}
This equation enforces the overall balance of production and degradation in the stationary state and thereby selects a pattern mass $M$ out of the continuous family of stationary patterns in the mass-conserving system.
For mesa patterns, this condition simplifies within the sharp-interface approximation. Since in that limit production and degradation in the interface region can be neglected, and the source terms can be approximated in the plateaus by ${s_\mathrm{tot}^\pm = s_\mathrm{tot}(\rhostat^\pm,\etastat)}$, Eq.~\eqref{eq:source-balance} reduces to a balance between the overall production and degradation in the upper and lower plateaus:
\begin{equation}
\label{eq:source-balance-mesas}
    s_\mathrm{tot}^+ \, \xi_+
    \approx
    -s_\mathrm{tot}^- \, \xi_-
    \, .
\end{equation}

\textit{Stability of elementary stationary patterns.\;---}
Next, we exploit the requirement that a stationary peak or mesa must be stable.
This means that any change of the profile caused by the addition of a small amount of mass $\delta M$ must lead to increased degradation.
Since we consider systems with a small source strength $\varepsilon$, the induced dynamics of the mass change $\delta M$ is slow and the change of the pattern profile adiabatically follows the stationary profile.
Thus, we can approximate the change of the pattern profile by the change $\partial_M^{}\rhostat \delta M$ of the stationary profile.
Integration of the modified continuity equation [using that the boundary conditions ensure ${\partial_x\eta|_{x=0,\Lambda/2}=0}$ and employing the definition of the interface average, Eq.~\eqref{eq:interface-average}]
\begin{align*}
    \partial_t \frac{\delta M }{2}
    &\approx \int_0^\frac{\Lambda}{2}\mathrm{d}x\, {\big [} \left(\partial_\rho s_\mathrm{tot}\right) \partial_M^{}\rhostat\\
    &\qquad\qquad\quad + \left(\partial_\eta s_\mathrm{tot}\right) \partial_M^{}\etastat {\big ]} \delta M\\
    &\approx \langle \partial_\rho s_\mathrm{tot}\rangle_\mathrm{int}\frac{\delta M}{2}
    \, ,
\end{align*}
where we neglect the second term as $\partial_M^{}\etastat$ is small by assumption [Eq.~\eqref{eq:small-plateau-mass}].
We conclude from this analysis that peak stability in the presence of source terms demands that ${\langle \partial_\rho s_\mathrm{tot}\rangle_\mathrm{int} <0}$. 
For mesa patterns, the additional mass $\delta M$ only shifts the interface such that ${\partial_M^{}\rhostat \sim \partial_x\rhostat}$ and ${\langle \partial_\rho s_\mathrm{tot}\rangle_\mathrm{int}\approx (s_\mathrm{tot}^+-s_\mathrm{tot}^-)/\Delta\rho }$.
In addition, the source balance condition, Eq.~\eqref{eq:source-balance-mesas} ensures that the source terms $s_\mathrm{tot}^\pm$ have different signs. 
Taken together, the stability criterion ${(s_\mathrm{tot}^+-s_\mathrm{tot}^-)<0}$ [derived from ${\langle \partial_\rho s_\mathrm{tot}\rangle_\mathrm{int}<0}$] translates into the conditions ${s_\mathrm{tot}^- > 0}$ and ${s_\mathrm{tot}^+ < 0}$ \cite{Brauns.etal2021}.
They imply net degradation when the length of the upper plateau is increased, and vice versa net production when its length is decreased.
Similarly, the condition ${\langle \partial_\rho s_\mathrm{tot}\rangle_\mathrm{int} <0}$ for peak patterns ensures that the source terms lead to net degradation if the peak density is increased, i.e., if the peak mass increases.

\textit{Plateau profiles.\;---}
We will now show that the second effect of the source terms is a parabolic concentration profile in the plateaus (see Fig.~\ref{fig:LS-stat-Construction}).
This profile is due to a gradient in the stationary mass-redistribution potential $\etastateps$ which is necessary to redistribute particles from the production regions at low pattern density (low-density plateau) toward the degradation regions at high pattern densities (peak or high-density plateau).
This process is described by the modified continuity equation, Eq.~\eqref{eq:stat-cont-eq-LS}, which yields (up to corrections due to the exponential tails of the stationary pattern of the mass-conserving system)
\begin{equation}
\label{eq:etastat-profile-non-mc}
    \delta\eta_\varepsilon(x) = A + \frac{s_\mathrm{tot}^\pm}{2 D_v} (x-B)^2 \, ,
\end{equation}
where the constants $A$, $B$ remain to be determined to fulfill the boundary conditions and continuity of the profile across the pattern interface or at the peak [see Fig.~\ref{fig:LS-stat-Construction}(b,d)].
Because gradients are small in the plateaus [see Appendix~\ref{app:lin-dynamics-plateaus} and Fig.~\ref{fig:LS-stat-Construction}(a,c)], the density profile $\delta\rho_\varepsilon$ in the plateaus follows the profile $\delta\eta_\varepsilon$.
Hence, we make an ansatz for $\delta\rho_\varepsilon$ by linearization of the nullcline around the plateaus.
We set ${\delta\eta_\varepsilon=\delta\rho_\varepsilon\partial_{\rho_\pm}\eta^* + \mathrm{const.}}$ with a constant accounting for deviations from the nullcline.
From Eq.~\eqref{eq:stat-prof-eq-LS} one then finds
\begin{equation}\label{eq:tails-non-mc}
    \delta\rho_\varepsilon(x) = \delta\eta_\varepsilon(x) \,\partial_{\etastat}\rho_\pm - \frac{s_1^{\pm} + d \left(s_2^{\pm}+ s_\mathrm{tot}^\pm \partial_\etastat\rho_\pm\right)}{\tilde{f}_\rho^\pm}
    \, ,
\end{equation}
in the upper and lower plateau, respectively.
Here, we employed the implicit function theorem to write ${\partial_{\etastat}\rho_\pm=(\partial_{\rho_\pm}\eta^*)^{-1}}$.
With this, the first term in Eq.~\eqref{eq:tails-non-mc} states that the pattern profile is pinned to the nullcline in the plateaus (cf.\ discussion of mesa splitting in Ref.~\cite{Brauns.etal2021}).
However, the source terms induce an offset from the nullcline which gives rise to weak reactive conversion between the $u$ and $v$ species.
This offset is analogous to the offset of the stationary mass-redistribution potential $\delta\etastateps$ at a stationary peak [cf.\ Eq.~\eqref{eq:nMC-etastat-shift}].

\textit{Shift of the stationary mass-redistribution potential.\;---}
These shifts can be understood as follows:
In steady state, production (degradation) in the slow-diffusing species $u$ has to be balanced by the outflow (inflow) of particles toward (from) pattern regions where degradation (production) prevails.
In the (biologically relevant) limit $D_u\ll D_v$, redistribution of particles (in- and outflow) proceeds mainly through the fast-diffusing species $v$.
Hence, these redistributed particles need to be converted by the reactions $\tilde{f}$ to balance production or degradation in the slow species $u$. 
To allow for this net reactive flux, the concentrations in the plateaus must deviate from the nullcline where $\tilde{f} = 0$ [see Eq.~\eqref{eq:tails-non-mc}].
For the same reason, at a peak or a mesa interface, the stationary mass-redistribution potential will be offset by $\varepsilon\,\delta\etastateps$ from the value which fulfills reactive turnover balance, Eq.~\eqref{eq:reactive-turnover-balance} [cf.\ Eq.~\eqref{eq:nMC-etastat-shift}].
To determine the offset $\varepsilon\,\delta\etastateps$ at a peak, one needs that it is (approximately) constant in the peak region (discussion of mesa patterns below).
Therefore, one has to assume that ${D_v/\ell_\mathrm{int}^2 \gg \langle\tilde{f}_\eta\rangle_\mathrm{int}}$, that is, fast cytosolic diffusion on the length scale of peaks or interfaces compared to the average reaction rate [see Sec.~\ref{sec:growth-rate-nMC}].
As argued in the main text, this condition is naturally fulfilled in the biologically relevant limit ${D_u\ll D_v}$.
Apart from this condition of $\eta$ being constant in the peak region, the following analysis is independent of the condition ${D_u\ll D_v}$.
One can then determine the offset of the mass-redistribution potential at the peak ${\delta\eta_\varepsilon(0) = \delta\etastateps}$  via averaging of Eq.~\eqref{eq:stat-prof-eq-LS} over the interface [using the interface average Eq.~\eqref{eq:interface-average}], which gives
\begin{align}
    \langle s_1+ d s_2\rangle_\mathrm{int} &=\int_0^\frac{\Lambda}{2}\mathrm{d}x\, (s_1 + d\, s_2) 2 \partial_M^{}\rhostat\nonumber\\
    &\approx \int_0^\frac{\Lambda}{2}\mathrm{d}x \left(2\partial_M^{}\rhostat\right)\mathcal{L}\,\delta\rho_\varepsilon - \langle \tilde{f}_\eta\rangle_\mathrm{int} \delta\etastateps\nonumber\\
    &= 2\partial_M^{}\etastat \int_0^\frac{\Lambda}{2}\mathrm{d}x\, \tilde{f}_\eta \delta\rho_\varepsilon - \langle \tilde{f}_\eta\rangle_\mathrm{int} \delta\etastateps
    \, .\label{eq:delta-etastateps}
\end{align}
In the second line we used that ${\delta\eta_\varepsilon\approx\delta\etastateps}$ is approximately constant in the peak region, and the third line is obtained by partial integration and the property, Eq.~\eqref{eq:mass-mode-eq}, of the linear operator $\mathcal{L}$.
Finally, by assumption, Eq.~\eqref{eq:small-plateau-mass}, the change $\partial_M^{}\etastat$ of the stationary mass-redistribution potential as a function of the peak mass is small.
Thus, one can neglect the first term on the right-hand side of Eq.~\eqref{eq:delta-etastateps}, which yields [see Fig.~\ref{fig:LS-stat-Construction}(b)]
\begin{equation*}
    \delta\etastateps 
    \approx 
    -\frac{\langle s_1 + d\, s_2\rangle_\mathrm{int}}{\langle \tilde{f}_\eta\rangle_\mathrm{int}}
    \, .
\end{equation*}
This agrees with the expression, Eq.~\eqref{eq:nMC-etastat-shift}, given in the main text.
Moreover, one can read off that in the limit ${D_u\ll D_v}$ that we discussed above to interpret the shift, indeed only the source term $s_1$ in the slow-diffusing species contributes.
This underlines that the shift in the mass-redistribution potential is due to the additional reactive turnover necessary to balance particle production and degradation.

In the main text, we analyzed the shift $\delta\etastateps$ also from a different perspective.
For this, one modifies the reaction term as ${\tilde{f}\to\tilde{f}'=\tilde{f}+\varepsilon(s_1+d s_2)}$.
This redefinition cancels all source terms in the profile equation, Eq.~\eqref{eq:profile-equation}, and the stationary equations take the form [cf.\ Eqs.~\eqref{eq:gen-stat-pattern}]
\begin{subequations}\label{eq:gen-stat-pattern-mod}
\begin{align}
    0 
    &= D_v \partial_x^2 \etastateps + \varepsilon s_\mathrm{tot}^\varepsilon
    \, , \label{eq:cont-eq-mod}\\
    0 
    &= D_u \partial_x^2\rhostateps + \tilde{f}'(\rhostateps,\etastateps)
    \, .
\end{align}
\end{subequations}
One defines the stationary pattern ${[\rhostat'(x),\eta'_\mathrm{stat}]}$ of the mass-conserving system with the modified reaction term $\tilde{f}'$ [setting ${\varepsilon=0}$ in Eq.~\eqref{eq:cont-eq-mod} where it appears explicitly].
This stationary pattern deviates from the stationary pattern ${[\rhostat(x),\etastat]}$ of the original mass-conserving model due to the modified reaction term.
Expanding Eqs.~\eqref{eq:gen-stat-pattern-mod} (${\varepsilon s_\mathrm{tot}=0}$) to first order in $\varepsilon$, the deviations ${\varepsilon[\delta\rho'(x),\delta\eta']=[\rhostat'-\rhostat,\eta'_\mathrm{stat}-\etastat]}$ fulfill
\begin{equation}\label{eq:MC-stat-state-deviations}
    0 = -\mathcal{L} \delta\rho' + \tilde{f}_\eta \delta\eta' + s_1(\rhostat,\etastat)+ d s_2(\rhostat,\etastat).
\end{equation}
As both $\etastat$ and $\eta'_\mathrm{stat}$ are spatially uniform, Equation~\eqref{eq:cont-eq-mod} (${\varepsilon s_\mathrm{tot}=0}$) is trivially fulfilled.
Importantly, Eq.~\eqref{eq:MC-stat-state-deviations} agrees with Eq.~\eqref{eq:stat-prof-eq-LS}.
The only difference is that $\delta\eta'$ in Eq.~\eqref{eq:MC-stat-state-deviations} is exactly constant.
Performing the interfacial averaging, Eq.~\eqref{eq:delta-etastateps}, for Eq.~\eqref{eq:MC-stat-state-deviations} we can identify ${\delta\eta'=\delta\etastateps}$.
Thus, the shift ${\varepsilon\,\delta\etastateps=\eta'_\mathrm{stat}-\etastat}$ describes the change of the stationary mass-redistribution potential due to the modification of the reaction term $\tilde{f}\to\tilde{f}'$.
This builds the basis for the interpretation of the mathematical form of the growth rate $\sigma^\varepsilon$ in Sec.~\ref{sec:growth-rate-nMC-discussion}.

Now with a full understanding of the shift $\delta\etastateps$, we shortly note where the condition ${D_v/\ell_\mathrm{int}^2 \gg \langle\tilde{f}_\eta\rangle_\mathrm{int}}$ stems from.
One can derive this condition by considering a quadratic correction in $\delta\eta_\varepsilon$ in the peak region and using the average $\langle\tilde{f}_\eta\rangle_\mathrm{int}$ as a scale for the typical reaction rate $\tilde{f}_\eta$ in the peak region.
If the condition is fulfilled, the quadratic correction is negligible in comparison to the constant contribution.
Hence, $\delta\eta_\varepsilon$ is approximately constant at the peak if the condition ${D_v/\ell_\mathrm{int}^2 \gg \langle\tilde{f}_\eta\rangle_\mathrm{int}}$ is fulfilled.

We can determine the same quantity $\delta\etastateps$ for mesa patterns by averaging Eq.~\eqref{eq:stat-prof-eq-LS} over the interface region [again using the interface average, Eq.~\eqref{eq:interface-average}].
Different from peak patterns, in mesa patterns, most of the mass produced in the low-density plateau is transported through the interface into the high-density plateau where degradation prevails.
Production and degradation in the interface region only weakly contribute.
In contrast, for peak patterns all mass produced in the low-density plateau must be degraded in the peak region, requiring a stronger shift $\delta\etastateps$ than for mesa patterns.
At the mesa interface the gradient in $\delta\eta_\varepsilon$ is large to transport the particles from the low-density into the high-density plateau [compare the profiles in Figs.~\ref{fig:LS-stat-Construction}(b,d)].
Therefore, the linear gradient in $\delta\eta_\varepsilon$ at the interface cannot be neglected and ${\delta\eta_\varepsilon(x)\approx \delta\etastateps}$ cannot be assumed constant at the interface, even if ${D_v/\ell_\mathrm{int}^2 \gg \langle\tilde{f}_\eta\rangle_\mathrm{int}}$.
However, the offset ${\varepsilon\,\delta\etastateps\sim \varepsilon/\langle\tilde{f}_\eta\rangle_\mathrm{int}}$ of the mass-redistribution potential at the interface only becomes significant compared to the profile of the mass-redistribution potential in the plateaus (${\sim \varepsilon \Lambda^2/D_v}$) if ${D_v/\Lambda^2\gg \langle \tilde{f}_\eta\rangle_\mathrm{int}}$.
In this regime, a calculation analogous to Eq.~\eqref{eq:delta-etastateps} including the linear gradient in $\delta\eta_\varepsilon(x)$ at the interface shows that the gradient can be neglected for mesa patterns as well.
Thus, in the regime where the offset of the mass-redistribution potential at the interface is significant, it is again given by $\delta\etastateps$, Eq.~\eqref{eq:nMC-etastat-shift}.

In summary, we showed in this section that source balance fixes the stationary pattern ${[\rhostat(x),\etastat]}$ of the mass-conserving system that approximates the stationary pattern ${[\rhostateps(x),\etastateps(x)]}$ in the presence of source terms [see Eq.~\eqref{eq:source-balance}].
In the plateaus, the pattern profiles in the presence of source terms were found to be parabolic, and they are approximated by Eqs.~\eqref{eq:etastat-profile-non-mc},~\eqref{eq:tails-non-mc}.
Lastly, we discussed that the value of the stationary mass-redistribution potential $\etastateps$ at a peak or interface shifts compared to its value $\etastat$ in the corresponding mass-conserving system.
The shift is given by $\varepsilon\,\delta\etastateps$, Eq.~\eqref{eq:nMC-etastat-shift}.

\section{Mass-competition growth rate under the influence of weak source terms}
\label{app:nMC-mass-competition}
In this appendix, we now adapt the linear stability analysis of the stationary patterns in the mass-conserving system to include the effects of source terms and calculate the growth rate $\sigma^\varepsilon$ [cf.\ Eq.~\eqref{eq:sigma-eps}].
To address the regime in which the interplay between mass redistribution and the source terms is important, we assume that the strength of source processes $\varepsilon\sigma_\mathrm{S}$ [cf.\ Eq.~\eqref{eq:sigma-eps}] is of the same order as the growth rates $\sigma_\mathrm{D}$ and $\sigma_\mathrm{R}$.
We neglect all higher-order terms $\varepsilon^2$, $\sigma \varepsilon$, $\sigma^2$, as well as $\varepsilon\, \delta\rho_\pm$ [recall ${\delta\rho_\pm^2\lesssim\partial_M^{}\etastat}$ following from Eqs.~\eqref{eq:mesa-delMeta-tot}, and the assumption, Eq.~\eqref{eq:small-plateau-mass}].

\begin{figure}
    \centering
    \includegraphics{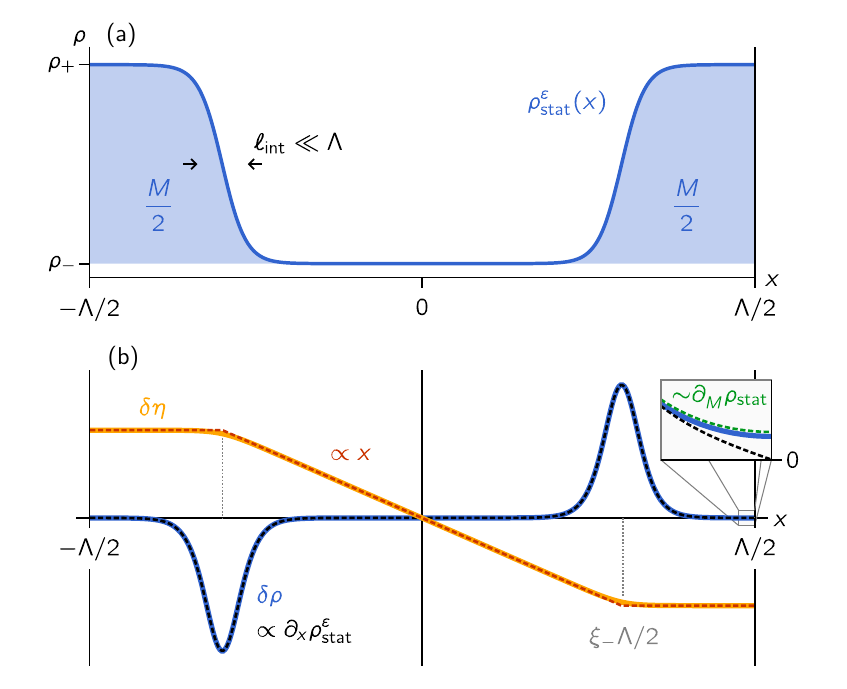}
    \caption{The mass-competition scenario under the influence of weak source terms for mesa patterns.
    (a) We consider the competition for mass of two stationary half mesas $\rhostateps(x)$ [blue (dark gray)].
    (b) The density mode $\delta\rho$ [blue (dark gray)] is approximated by the translation mode $\partial_x\rhostateps$ (black, dashed) in the inner plateau and at the interfaces.
    Because the mass-competition eigenmode changes the length $L_+$ of the outer (high-density) plateaus, the density mode has to be approximated by the corresponding change of the stationary profile ${\partial_{L_+}^{}\rhostateps\sim \partial_M^{}\rhostateps}$ at the domain boundaries (inset).
    The change of the mass-redistribution potential $\delta\eta$ [orange (light gray)] is again approximately linear in the plateaus (red, dashed).
    The construction is shown for the cubic model ${\tilde{f}=\eta-\rho^3+\rho}$ with parameters ${D_u=1}$, ${D_v=10^3}$, ${p=-0.2}$, ${\varepsilon = 10^{-3}}$, and ${\Lambda=40}$ (see Appendix~\ref{app:cubic-model}).
    }
    \label{fig:mesa-LS-modeApprox}
\end{figure}

\subsection{Mesa patterns}
We first consider the competition of two half mesas separated by a trough on the domain $\Omega= \left[-\frac{\Lambda}{2},\frac{\Lambda}{2}\right]$ with no-flux boundary conditions [see Fig.~\ref{fig:mesa-LS-modeApprox}(a)].
As in the mass-conserving case, mass competition redistributes mass between the two mesas, thereby translating the trough.
Due to the antisymmetry of the eigenmode (stationary pattern symmetric around ${x = 0}$), the two interfaces shift synchronously.
Due to this antisymmetry, the length of the trough stays constant as in the case with mass conservation although the conservation law is broken here.
Therefore, the change $\delta \rho (x)$ in the spatial profile can again be approximated by a displacement of the stationary profile, that is, by the (approximate) Goldstone mode 
\begin{equation}\label{eq:LS-mode-approx}
    \delta\rho(x) \approx A \frac{\partial_x\rhostateps}{\Delta\rho}
\end{equation}
in the inner plateau and at the interfaces [see Fig.~\ref{fig:mesa-LS-modeApprox}(b)].
Correction terms arise in the outer plateaus (the high-density plateaus in Fig.~\ref{fig:mesa-LS-modeApprox}) due to the boundaries which break translation invariance [see inset in Fig.~\ref{fig:mesa-LS-modeApprox}(b)].
Again, $A$ is the amplitude of the eigenmode and ${\Delta \rho=\rho_+-\rho_-}$.
Correction terms to Eq.~\eqref{eq:LS-mode-approx} are suppressed by ${\sim \sigma^\varepsilon/\sigma_\mathrm{relax}}$, analogously as in the mass-conserving system [see Appendix~\ref{app:MC-mass-competition}]. 

To approximate the mode in the outer plateaus (at ${x \gtrless \pm\xi_-\Lambda/2}$, see Fig.~\ref{fig:mesa-LS-modeApprox}), one observes that the translation of the interfaces changes the length of these plateaus as if the mesa mass $M$ is changed. 
Therefore, we approximate the eigenmode in the outer plateau by the mass mode $\partial_M^{}\rhostat(x)$ [cf.\ inset in Fig.~\ref{fig:mesa-LS-modeApprox}(b)].

As in the mass-conserving system, the mass-redistribution potential $\delta\eta(x)$ is linear in the plateau regions due to the fast relaxation of the plateau density compared to mass competition [see Fig.~\ref{fig:mesa-LS-modeApprox}(b) and Appendix~\ref{app:MC-mass-competition}].
We again employ for $\delta\eta(x)$ the ansatz given in Eq.~\eqref{eq:mesa-ansatz-eta-mode}.
Inserting the ansatz, Eqs.~\eqref{eq:LS-mode-approx},~\eqref{eq:mesa-ansatz-eta-mode}, into the modified continuity equation, Eq.~\eqref{eq:cont-eq-non-mc}, integrating over $\left[0,\Lambda/2\right]$, and using the interface average, Eq.~\eqref{eq:interface-average}, gives within the sharp-interface approximation
\begin{align}
\label{eq:mesa-int-cont-eq-non-mc}
    \frac{\delta\eta_\mathrm{int}}{A}
    &\approx -\frac{\xi_-\Lambda}{2 D_v} \left(\sigma_+^\varepsilon -\varepsilon\langle \partial_\rho s_\mathrm{tot}\rangle_\mathrm{int}\right)\nonumber\\
    &\approx -\frac{\xi_-\Lambda}{2 D_v} \left(\sigma_+^\varepsilon -\varepsilon \frac{s_\mathrm{tot}^+-s_\mathrm{tot}^-}{\Delta\rho}\right)
    .
\end{align}
Here, we used ${\delta\eta/A \sim \max(\sigma, \varepsilon)}$, following from this equation, Eq.~\eqref{eq:mesa-int-cont-eq-non-mc}, to drop the second term on the right-hand side of the modified continuity equation~\eqref{eq:cont-eq-non-mc} used in its derivation.

Again, we need a second condition to determine the two unknown quantities $\delta\eta_\mathrm{int}/A$ and $\sigma_+^\varepsilon$.
To this end, we now determine a (modified) conversion-rate integral [cf.~\eqref{eq:relax-convRateInt}].
First, we note that the translation mode of the stationary profile $(\rhostateps,\etastateps)$ fulfills [applying the derivative $\partial_x$ in the stationarity equations, Eqs.~\eqref{eq:gen-stat-pattern}]
\begin{align}
\label{eq:translation-mode-non-mc}
    &\left[ \tilde{f}_\eta^\varepsilon + \varepsilon\partial_\eta\left(s_1^\varepsilon + d\, s_2^\varepsilon\right)\right]
    \partial_x\etastateps\nonumber\\
   &\qquad =
   \left[\mathcal{L}^\varepsilon - \varepsilon \partial_\rho\left(s_1^\varepsilon + d\, s_2^\varepsilon\right) \right] \partial_x\rhostateps 
   \, .
\end{align}
Comparing this relation with Eq.~\eqref{eq:mass-mode-eq} and the conversion-rate integral used in the mass-conserving system, Eqs.~\eqref{eq:mesa-reaction-rate-integral}, we define here the modified conversion-rate integral
\begin{align}
\label{eq:mesa-reac-rate-int-non-mc}
    &\int_0^\frac{\Lambda}{2}\mathrm{d}x\, \left[ \tilde{f}_\eta^\varepsilon + \varepsilon \partial_\eta\left(s_1^\varepsilon + d\, s_2^\varepsilon\right)\right]
    \left(\partial_x\etastateps\right) \delta\rho\nonumber\\
    &\qquad \approx  
    A \partial_x\etastateps{\big|}_{x=\frac{\xi_-\Lambda}{2}} \langle \tilde{f}_\eta\rangle_\mathrm{int}
    \, .
\end{align}
The approximation holds because $\partial_x\etastateps$ is approximately constant across the interface in the sharp-interface approximation [see Fig.~\ref{fig:LS-stat-Construction}(d)].

Again we determine a second approximation for the conversion-rate integral.
Here, this is not done with the help of Eq.~\eqref{eq:mass-mode-eq}--which only holds for the mass-conserving system---but with the identity, Eq.~\eqref{eq:translation-mode-non-mc}, for the translation mode.
Using in addition, the dynamic equation, Eq.~\eqref{eq:lin-eta-eq-non-mc}, one finds
\begin{align}\label{eq:mesa-reac-rate-int-non-mc-2}
    &\int_0^\frac{\Lambda}{2}\mathrm{d}x \left[ \tilde{f}_\eta^\varepsilon + \varepsilon \partial_\eta\left(s_1^\varepsilon + d\, s_2^\varepsilon\right)\right] \partial_x\etastateps \delta\rho\nonumber\\
    &\qquad = -D_u \delta\rho{\big|}_{x=\Lambda/2}\, \partial_x^2\rhostateps{\big|}_{x=\Lambda/2}\nonumber\\
    &\qquad\quad + \int_0^\frac{\Lambda}{2}\mathrm{d}x\left[\tilde{f}_\eta \delta\eta - (1+d)\sigma_+^\varepsilon \delta\rho\right]\partial_x\rhostateps.
\end{align}
For the boundary term, we first use the approximation of the non-mass-conserving stationary states Eq.~\eqref{eq:tails-non-mc} to find
\begin{align*}
    D_u \partial_x^2\rhostateps{\big|}_{x=\Lambda/2} &= D_u\partial_x^2\rhostat{\big|}_{x=\Lambda/2} + D_u\partial_x^2\delta\rho_\varepsilon{\big|}_{x=\Lambda/2}\\
    &\approx  -|\tilde{f}_\rho^+|\delta\rho_+ + \order{\varepsilon}
    \, ,
\end{align*}
where $\delta\rho_+$ describes the offset from the upper plateau of the stationary pattern in the mass-conserving system [see Eqs.~\eqref{eq:mesa-exp-tails}].
Then, we approximate the eigenmode in the outer plateau using the change of the stationary profile due to a change of the plateau length, that is, due to a change of the plateau mass [cf.\ inset in Fig.~\ref{fig:mesa-LS-modeApprox}(b)]:
\begin{equation*}
    \delta\rho{\big|}_{x=\Lambda/2} \approx -2 A \partial_M^{}\delta\rho_+{\big|}_{x=\Lambda/2}  +\order{\varepsilon}
    \, .
\end{equation*}
Neglecting terms of order ${\sim \varepsilon \delta\rho_\pm}$ [one has ${\delta\rho_\pm^2\lesssim \partial_M^{}\etastat}$ and the assumption, Eq.~\eqref{eq:small-plateau-mass}], the two approximations lead to the boundary term
\begin{align*}
    -D_u \delta\rho{\big|}_{x=\Lambda/2} \, \partial_x^2\rhostateps{\big|}_{x=\Lambda/2} 
    &\approx 
    -2 A \frac{|\tilde{f}_\rho^+|}{\partial_{\etastat}M} \frac{\partial_{\etastat}\delta\rho_+^2}{2}\\
    &= -2 A  \Delta\rho \langle \tilde{f}_\eta\rangle_\mathrm{int} \partial_M^+\etastat
    \, .
\end{align*}
The second line follows from the definition of $\partial_M^+\etastat$, Eq.~\eqref{eq:mesa-delMeta-formula}.

We are left to determine the gradient ${\partial_x\etastateps|_{x=\xi_-\Lambda/2}}$ in Eq.~\eqref{eq:mesa-reac-rate-int-non-mc}, which is readily found from Eq.~\eqref{eq:etastat-profile-non-mc} as
\begin{equation*}
    \partial_x\etastateps{\big|}_{x=\frac{\xi_-\Lambda}{2}} \approx -\varepsilon\frac{|s_\mathrm{tot}^+|\xi_+ \Lambda}{2 D_v}
    \, .
\end{equation*}
Altogether, we then find from Eqs.~\eqref{eq:mesa-int-cont-eq-non-mc},~\eqref{eq:mesa-reac-rate-int-non-mc},~\eqref{eq:mesa-reac-rate-int-non-mc-2} analogously to the mass-conserving case [cf.\ 
Eq.~\eqref{eq:sigma-eps}]:
\begin{equation*}
    \sigma_\pm^\varepsilon \approx \frac{\sigma_\mathrm{R}^\pm}{\sigma_\mathrm{D}^\pm+\sigma_\mathrm{R}^\pm}\left(\sigma_\mathrm{D}^\pm - \varepsilon \frac{|s_\mathrm{tot}^\pm|}{\Delta\rho}\right).
\end{equation*}
The rates $\sigma_\mathrm{D,R}^\pm$ are the diffusion- and reaction-limited growth rates of the mass-competition instability in the mass-conserving system.
We performed the derivation for mesa competition, that is, $\sigma_+^\varepsilon$.
The derivation proceeds analogously for the growth rate $\sigma_-^\varepsilon$ describing mesa coalescence, i.e., the competition of two troughs.

\subsection{Peak patterns}
At last, we analyze the competition between two half peaks situated at the boundaries of the domain ${\Omega=[-\Lambda/2,\Lambda/2]}$ symmetric around ${x=0}$ with no-flux boundary conditions (see Fig.~\ref{fig:peak-MC-modeApprox}).
Neglecting corrections of order $\varepsilon$ and corrections due to relaxation modes [${\sim \sigma^\varepsilon/\sigma_\mathrm{relax}}$, see Appendix~\ref{app:MC-mass-competition}], we will approximate the competition mode again by the mass mode, that is, we use the same ansatz, Eqs.~\eqref{eq:peak-mode-approx-mc}, as in the mass-conserving case.

Integration of the modified continuity equation, Eq.~\eqref{eq:cont-eq-non-mc}, over $\left[0,\frac{\Lambda}{2}\right]$ yields [using the no-flux boundary conditions and the interface average, Eq.~\eqref{eq:interface-average}]
\begin{equation}
\label{eq:peak-int-cont-eq-non-mc}
    \frac{\delta\eta{\big|}_{x=\Lambda/2}}{A} \approx -\frac{\Lambda}{2 D_v} \left(\sigma^\varepsilon -\varepsilon \langle \partial_\rho s_\mathrm{tot}\rangle_\mathrm{int}\right) \, .
\end{equation}
Following the same line of argument as in the previous cases, one needs a second relation between ${\delta\eta\big|_{x=\Lambda/2}/A}$ and $\sigma^\varepsilon$ to determine these two unknown quantities.
As a second condition, we again write down approximations for the conversion-rate integral. 
To find two expressions analogous to Eqs.~\eqref{eq:peak-reaction-rate-integral},~\eqref{eq:peak-reaction-rate-integral-2} in the mass-conserving system, one needs to construct a mode for the system including source terms that resembles the mass mode.
We cannot directly use the mass mode as in Eq.~\eqref{eq:peak-reaction-rate-integral-2} because the source terms break the conservation law and there does not exist a family of stationary peak patterns of different masses.
Thus, no mass mode ``$\partial_M^{}(\rhostateps,\etastateps)$'' exists in the system including weak source terms.
Also, we cannot use the translation mode as in the above derivation for mesa patterns because the peak interfaces are not only translated but the whole peaks grow or shrink.

Instead, we need a different approach.
Recall that source balance, Eq.~\eqref{eq:source-balance}, fixes the average density of the stationary pattern.
Thus, we can emulate the mass mode by modifying the source terms in a way that changes the average density of the stationary pattern.
To this end, we introduce an auxiliary parameter $p$ inducing this shift of the source terms, $p=0$ recovering the original source terms.
We might introduce $p$ by changing ${s_\mathrm{tot} \to s_\mathrm{tot} + p}$ or ${s_1(\rho,\eta)\to s_1(\rho -p, \eta)}$, ${s_2(\rho,\eta)\to s_2(\rho -p, \eta)}$ [cf.\ Eq.~\eqref{eq:source-balance}].
The only condition is that $p$ parametrizes a family of stable elementary stationary patterns which fulfill ${\partial_p M \neq 0}$, i.e., that the peak mass indeed changes with $p$.
The final result for the growth rate will be independent of the choice of $p$.

From this new parameter $p$, one then gets the auxiliary mode $\partial_p(\rhostateps,\etastateps)$, which fulfills
\begin{align*}
    &-\varepsilon \left(\partial_\rho s_\mathrm{tot}^\varepsilon\right) \partial_p\rhostateps - \varepsilon  \left(\partial_p s_\mathrm{tot}^\varepsilon\right)\\
    &\qquad= \left(D_v\partial_x^2 + \varepsilon \partial_\eta s_\mathrm{tot}^\varepsilon\right) \partial_p\etastateps\, ,\\
    &{\Big[}\tilde{f}_\eta^\varepsilon +\varepsilon \partial_\eta ( s_1^\varepsilon + d\, s_2^\varepsilon){\Big]}\partial_p\etastateps + \varepsilon \partial_p\left(s_1^\varepsilon + d\, s_2^\varepsilon\right)\\
    &\qquad= \left[\mathcal{L}^\varepsilon - \varepsilon \partial_\rho\left(s_1^\varepsilon + d\, s_2^\varepsilon\right)\right] \partial_p\rhostateps\, .
\end{align*}
One may now calculate a modified conversion-rate integral for peak patterns, analogously to Eqs.~\eqref{eq:mesa-reac-rate-int-non-mc},~\eqref{eq:mesa-reac-rate-int-non-mc-2}, using the auxiliary mode, and one finds
\begin{align}
    &\int_0^\frac{\Lambda}{2}\mathrm{d}x \left(\left[\tilde{f}_\eta^\varepsilon + \varepsilon \partial_\eta\left(s_1^\varepsilon + d\, s_2^\varepsilon\right)\right] \partial_p\etastateps\right) \delta\rho\nonumber\\
    &\qquad\approx A\langle \tilde{f}_\eta \rangle_\mathrm{int} \partial_p \left(\etastat + \varepsilon\,\delta\etastateps\right),\nonumber\\
    &\qquad\approx \int_0^\frac{\Lambda}{2}\mathrm{d}x{\bigg(}\left[ \tilde{f}_\eta \delta\eta - (1+d) \sigma^\varepsilon \delta\rho\right] \partial_p\rhostateps \nonumber\\
    &\qquad\qquad\qquad\qquad - \varepsilon \left[\partial_p\left(s_1^\varepsilon + d\, s_2^\varepsilon\right]\right) \delta\rho{\bigg)}
    \, . \label{eq:peak-nMC-convRateInt}
\end{align}
The boundary term arising during the calculation is exponentially small in the plateau length and can be safely neglected for the competition between peaks due to their strong mass competition (cf.\ Appendix~\ref{app:MC-mass-competition}).
Using that ${\rhostateps-\rhostat=\order{\varepsilon}}$ implies ${\partial_p\rhostateps = \partial_p M\, \partial_M^{}\rhostat + \order{\varepsilon}}$, and one finds from  Eq.~\eqref{eq:peak-nMC-convRateInt} 
\begin{align*}
    &A\left(\partial_p M\right) \langle\tilde{f}_\eta \rangle_\mathrm{int} \left(\partial_M^{}\etastat - \varepsilon \partial_M^{}\frac{\langle s_1 + d\, s_2\rangle_\mathrm{int}}{\langle\tilde{f}_\eta\rangle_\mathrm{int}} {\Bigg |}_{p = 0}\right) \\
    &\qquad\qquad - \varepsilon A\langle \partial_p\left(s_1^\varepsilon + d\, s_2^\varepsilon\right)\rangle_\mathrm{int} \\
    &\qquad\approx \frac{\delta\eta{\big |}_{x = \Lambda/2}}{2}\left(\partial_p M\right) \langle\tilde{f}_\eta\rangle_\mathrm{int} - \frac{A}{2}\left(\partial_p M\right) \sigma^\varepsilon \frac{1+d}{ \ell_\mathrm{int}} \\
    &\qquad\qquad - \varepsilon A\langle \partial_p\left(s_1^\varepsilon + d\, s_2^\varepsilon\right)\rangle_\mathrm{int}
    \, .
\end{align*}
Applying the modified continuity equation, Eq.~\eqref{eq:peak-int-cont-eq-non-mc}, we finally derive the growth rate, Eq.~\eqref{eq:sigma-eps}, which we restate here for the reader's convenience
\begin{align*}
    \sigma^\varepsilon &\approx\frac{\sigma_\mathrm{R}}{\sigma_\mathrm{D} + \sigma_\mathrm{R}} \\
    & \quad\cdot\left[- \frac{4 D_v}{\Lambda}\partial_{M}^{} \left(\etastat +\varepsilon\, \delta\eta_\mathrm{stat}^\varepsilon\right) + \varepsilon \langle \partial_\rho s_\mathrm{tot}\rangle_\mathrm{int}\right].
\end{align*}
The rates $\sigma_\mathrm{D,R}$ are the diffusion- and reaction-limited growth rates of the mass-competition instability in the mass-conserving system [given in Eqs.~\eqref{eq:sigma-MC-peak-D-R}].
Any dependence on the auxiliary variable $p$ cancels out.

Finally, we use a (rough) scaling argument to show that the additional effect of the shift $\varepsilon\,\delta\eta_\mathrm{stat}^\varepsilon$ only becomes significant in the reaction-limited regime.
This argument justifies the simple expression, Eq.~\eqref{eq:sigma-eps-diffLim}, for the growth rate $\sigma^\varepsilon$ in the diffusion-limited regime, which was also verified numerically (see Sec.~\ref{sec:numerics}).
The growth rate $\sigma^\varepsilon$ [cf.\ Eq.~\eqref{eq:sigma-eps}] shows that the term including $\varepsilon\,\delta\eta_\mathrm{stat}^\varepsilon$ cannot be neglected as compared to the term $\varepsilon \sigma_\mathrm{S}$ if
\begin{equation*}
    \frac{4 D_v}{\Lambda}\left|\partial_M^{} \delta\eta_\mathrm{stat}^\varepsilon\right| \gtrsim \left|\sigma_\mathrm{S}\right|.
\end{equation*}
With the expressions for the shift $\delta\etastateps$, Eq.~\eqref{eq:nMC-etastat-shift}, and the rate $\sigma_\mathrm{S}$, Eq.~\eqref{eq:sigma-stab-peak}, this yields
\begin{equation}
    \frac{D_v}{\Lambda}\left|\partial_M^{} \frac{\langle s_1 + d\, s_2 \rangle_\mathrm{int}}{\langle \tilde{f}_\eta\rangle_\mathrm{int}}\right| \gtrsim \left|\langle\partial_\rho s_\mathrm{tot}\rangle_\mathrm{int}\right|. \label{eq:shift-relevance-estimate}
\end{equation}
To make further progress, we use that a change $\delta \bar{\rho}$ of the average density $\bar{\rho}$ mainly changes the peak mass [by assumption, Eq.~\eqref{eq:small-plateau-mass}].
Thus, the average local change $\delta\rho_\mathrm{peak}$ of the density at the peak can be estimated as ${\delta\bar{\rho}\sim \frac{\ell_\mathrm{int}}{\Lambda}\delta\rho_\mathrm{peak}} $.
As a consequence, we have ${\partial_M^{} = \Lambda^{-1}\partial_{\bar{\rho}}\sim \ell_\mathrm{int}^{-1} \partial_{\rho_\mathrm{peak}}}$.
Building then on the observation that the interface average, Eq.~\eqref{eq:interface-average}, is localized at the peak one may estimate
\begin{align}\label{eq:eta-shift-relevance-scaling}
    \left|\partial_M^{} \frac{\langle s_1 + d\, s_2 \rangle_\mathrm{int}}{\langle \tilde{f}_\eta\rangle_\mathrm{int}}\right|&\sim \frac{1}{\ell_\mathrm{int}}\left|\partial_{\rho_\mathrm{peak}}^{} \frac{\langle s_1 + d\, s_2 \rangle_\mathrm{int}}{\langle \tilde{f}_\eta\rangle_\mathrm{int}}\right|\nonumber\\
    &\sim \frac{\left|\langle\partial_\rho s_\mathrm{tot}\rangle_\mathrm{int}\right|}{\ell_\mathrm{int} \langle \tilde{f}_\eta\rangle_\mathrm{int}}
    \, .
\end{align}
In the second line, we additionally assumed that derivatives with respect to $\rho$ are of the same order when acting on the average source term or the average reaction rate.
This estimate is expected to hold if the patterns are (approximately) scale-free because all expressions are then given by power laws (cf.\ Appendix~\ref{app:peak-scaling}).
Moreover, the strength of the source term ${s_1+ d\, s_2}$ can be estimated by $s_\mathrm{tot}$ if the source terms are not restricted to the fast-diffusing species $v$, that is, if one has ${s_1 \gtrsim s_2}$.
Otherwise, the effect of the shift $\varepsilon\, \delta\etastateps$ is even smaller.

If the scaling given in Eq.~\eqref{eq:eta-shift-relevance-scaling} holds (approximately), one may use it in the above inequality, Eq.~\eqref{eq:shift-relevance-estimate}, and one finds that the shift in the steady-state mass-redistribution potential only yields a significant contribution to the growth rate $\sigma^\varepsilon$ if
\begin{equation*}
    \frac{D_v}{\Lambda \ell_\mathrm{int}\langle \tilde{f}_\eta\rangle_\mathrm{int}}\gtrsim 1
    \, ,
\end{equation*}
which agrees with the condition for the reaction-limited regime [cf.\ Eq.~\eqref{eq:diff-reac-crossover}].
The significance of the shift in the stationary mass-redistribution potential in the reaction-limited regime underlines our heuristic explanation that this shift is induced by the limited rate of reactive conversion between the slow and fast species at the peak (see Appendix~\ref{app:stat-state-nMC}).
In the diffusion-limited regime, this conversion is fast in comparison to the mass-redistribution process between the peaks as well as between a peak and its plateau, and the shift is negligible.

\section{Example systems}
\label{app:examples}
The numerical analysis of example systems was implemented using Mathematica v12.2.
The simulations of coarsening and its interruption (see Figs.~\ref{fig:coarsening-phenomenology},~\ref{fig:interrupted-coarsening-phenomenology},~\ref{fig:interrupted-coarsening}) were performed using Comsol Multiphysics, Version 5.6 \cite{Comsol}.
The Mathematica scripts and Comsol setup files are available under \url{https://github.com/henrikweyer/2cRD-wavelength-selection}.
We now give details on the analysis of the example models studied in detail in Sec.~\ref{sec:numerics}.

\subsection{The cubic model}
\label{app:cubic-model}
The cubic model with source terms, which serves as a prototypical model for mesa-forming 2cRD systems, is defined in Sec.~\ref{sec:cubic-model}.
The profile equation, Eq.~\eqref{eq:profile-equation}, which determines the stationary pattern of the mass-conserving cubic model reads
\begin{equation}\label{eq:profile-eq-cubic-model}
    0 
    = 
    D_u \partial_x^2 \rhostat(x) 
    + 
    \etastat 
    - 
    \big[\rhostat(x)\big]^3 
    + 
    \rhostat(x)
    \, .
\end{equation}
It is equivalent to the stationary equation for the classical CH equation.
The interface profile on the infinite line is thus given by
\begin{equation*}
    \rhostat^\infty(x) 
    = 
    \tanh
    \left(
    \frac{x}{\sqrt{2 D_u}}
    \right)
    ,
\end{equation*}
and ${\eta_\mathrm{stat}^\infty = 0}$.
Asymptotically, at $x\to\pm\infty$, the interface profile can be approximated as
\begin{equation*}
    \rhostat^\infty(x)\to \pm 1 \mp 2 \exp\left(-|x|/\ell\right)
    ,
\end{equation*}
with the diffusion lengths ${\ell_\pm =\ell = \sqrt{D_u / 2}}$.\footnote{
Note that we set the unit of time to $1$ by defining the reaction term as $\tilde{f}=\eta-\rho^3+\rho$ with the reaction rate $\tilde{f}_\eta$ set to $1$.}
This length scale is related to the interface width, Eq.~\eqref{eq:interface-width}, which reads in the cubic model
\begin{equation}\label{eq:cubic-model-interface-width}
    \ell_\mathrm{int} \approx 4\left( \int_{-\infty}^\infty\mathrm{d}x\, (\partial_x \rhostat^\infty)^2\right)^{-1} = 6\, \ell
    \, .
\end{equation}
The approximation neglects that we analyze the pattern on a finite domain, not on the infinite line.
Hence, it holds within the sharp-interface approximation because then the interface is well separated from the system boundaries.

To construct the stationary plateau profile on the finite domain of length $\Lambda/2$, we have to account for the boundaries.
To this end, we exploit that we know the pattern profile in the plateaus from the linearization around the plateau densities [see Eqs.~\eqref{eq:mesa-exp-tails}].
These $\cosh$-profiles quickly become close to exponential a distance ${\sim\ell}$ away from the boundaries [see Fig.~\ref{fig:MC-stat-construction}(d)].
Thus, in the sharp-interface limit, we can match these  pattern tails to the interface profile calculated on the infinite line [${6\ell=\ell_\mathrm{int}\ll\Lambda}$, see Eq.~\eqref{eq:cubic-model-interface-width}].
This procedure is called \emph{asymptotic matching} \cite{Kolokolnikov.etal2006,Pismen2006,Wei.Winter2014} and determines the offsets $\delta\rho_\pm$ as
\begin{equation*}
    \delta\rho_\pm 
    = 
    4 
    \exp\left(-\frac{L_\pm}{\ell}\right)
    ,
\end{equation*}
which yields [using Eq.~\eqref{eq:mesa-delMeta-formula}]
\begin{equation*}
    \partial_M^\pm\etastat 
    = 
    - \frac{4}{\ell}\exp\left(-2\frac{L_\pm}{\ell}\right)
    .
\end{equation*}

Thus, inserting the above terms into the general growth-rate expressions, Eqs.~\eqref{eq:sigma-MC-mesa-D-R},~\eqref{eq:sigma-stab-mesas}, one finds for the cubic model introduced in Sec.~\ref{sec:cubic-model} 
\begin{align*}
    \sigma_\mathrm{D}^\pm &= \frac{16 D_v}{\xi_\mp \Lambda \ell}\exp\left(-\frac{\xi_\pm\Lambda}{\ell}\right),\\
    \sigma_\mathrm{R}^\pm &= \frac{48}{1+d}\exp\left(-\frac{\xi_\pm\Lambda}{\ell}\right),\\
    \sigma_\mathrm{S}^\pm &= \frac{|p\mp 1|}{2}
    \, .
\end{align*}

\textit{Approximation including pattern deformation by the source term $s_1$.\;---}
Beyond the above `standard' approximation of the growth rates, we also employed in Sec.~\ref{sec:cubic-model} an improved approximation that incorporates the deformation of the stationary pattern by source terms in the slowly diffusing species $u$.
This modified approximation for the growth rates is obtained by shifting source terms into the fast-diffusing species $v$ via replacement ${\tilde{f}\to\tilde{f}'=\tilde{f}+\varepsilon (p-\rho)}$ [for the source terms chosen for Fig.~\ref{fig:numerics-eigenvalueMapping-cubicModel}(b)].
For this choice of the source terms, the profile equation, Eq.~\eqref{eq:profile-equation}, for the stationary pattern $\rhostat'(x)$ of the modified mass-conserving system takes the same form as in the original system, Eq.~\eqref{eq:profile-eq-cubic-model}, but with $\varepsilon$-dependent coefficients:
\begin{equation}
    0 
    = 
    D_u \partial_x^2\rhostat' + \varepsilon p + \eta_\mathrm{stat}'-\left(\rhostat'\right)^3+(1-\varepsilon)\rhostat'
    \, .
\end{equation}
The solution on the infinite line is therefore changed to
\begin{equation}\label{eq:mod-inf-profile-cubic-model}
    \rhostat'^\infty(x) 
    = 
    \sqrt{1-\varepsilon} \, 
    \tanh\left(x \Bigg/ \sqrt{\frac{2 D_u}{1-\varepsilon}}\right)
    .
\end{equation}
The threshold of interrupted coarsening for mesa competition is then obtained from numerically solving the implicit expression for $\varepsilon_\mathrm{stop}$ [cf.\ Eq.~\eqref{eq:eps-threshold-mesas}]:
\begin{equation}
    \frac{16 D_v (1-\varepsilon_\mathrm{stop})}{\xi'_- \Lambda \ell'}\exp\left(-\frac{\xi'_+\Lambda}{\ell'}\right)= \varepsilon_\mathrm{stop} \frac{\rho'_+-p}{2\sqrt{1-\varepsilon_\mathrm{stop}}}
    \, ,
\end{equation}
where ${\ell' = \sqrt{\frac{D_u}{2(1-\varepsilon)}}}$ and $\xi'_\pm$ are evaluated using ${\rho'_\pm = \pm \sqrt{1-\varepsilon}}$ from Eq.~\eqref{eq:mod-inf-profile-cubic-model} and $\varepsilon=\varepsilon_\mathrm{stop}$.

\textit{Plateau splitting.\;---}
In the numerical simulations discussed in Sec.~\ref{sec:cubic-model} we observed splitting of the lower pattern plateau [see Fig.~\ref{fig:numerics-eigenvalueMapping-cubicModel}(a,b)].
In our previous publication, Ref.~\cite{Brauns.etal2021}, a criterion for the onset $\varepsilon_\mathrm{split}$ of plateau splitting is derived.
We can apply this threshold here to determine the parameter region of plateau splitting in the cubic model.
We briefly outline the construction given in Ref.~\cite{Brauns.etal2021}:
Splitting occurs if the amplitude of the (approximately) parabolic profile of the high- or low-density plateaus [see Eqs.~\eqref{eq:etastat-profile-non-mc},~\eqref{eq:tails-non-mc}] becomes large and its minimum or maximum, respectively, enters into the density regime of lateral instability [${-\sqrt{1/3}<\rho<\sqrt{1/3}}$ in the cubic model, Eq.~\eqref{eq:cubic-model}], that is, the nullcline slope becomes negative at the minimum or maximum of the plateau profiles (see Sec.~\ref{sec:HSS-instability}).
Because the lower plateau is longer than the upper plateau for the chosen source terms [see insets in Fig.~\ref{fig:numerics-eigenvalueMapping-cubicModel}(a,b)], the maximum of the lower plateau enters the regime of lateral instability first, and plateau splitting occurs first in the lower plateau.
Because in the plateaus the local equilibrium approximation holds [Eq.~\eqref{eq:plateau-local-equilibrium}], the density profile $\rhostateps(x)$ is slaved to the profile $\etastateps(x)$ given by [see Eq.~\eqref{eq:etastat-profile-non-mc}]
\begin{equation*}
    \etastateps(x) = \eta^*(\rho_-) - \frac{\varepsilon s_\mathrm{tot}^-}{2 D_v} \left[ x^2-\left(\frac{\xi_-\Lambda}{2}\right)^2\right].
\end{equation*}
The profile enters the regime of lateral instability when the maximum of $\etastateps$ equals the maximum of the nullcline ${\eta_\mathrm{max} = 2/(3\sqrt{3})}$, because the nullcline slope becomes negative at its maximum.
This yields the threshold for mesa splitting
\begin{equation}\label{eq:cubicModel-splitting}
    \varepsilon_\mathrm{split} \approx \frac{64 D_v}{3\sqrt{3}(1+p)(1-p)^2\Lambda^2}
    \, ,
\end{equation}
where we neglected $\eta^*(\rho_-)$ as it is exponentially small in the plateau length [cf.\ Sec.~\ref{sec:stat-patterns}].
In Fig.~\ref{fig:numerics-eigenvalueMapping-cubicModel}(a,b), this threshold is depicted as a purple (top-most, diagonal) line.

\subsection{The Brusselator model}
\label{app:Brusselator}
The Brusselator model was introduced in Sec.~\ref{sec:Brusselator} as a mass-conserving core system supplemented by source terms.
Here we use the mesa shape of the patterns at slow cytosolic diffusion $D_v$ and the asymptotic peak shape at large cytosolic diffusion ${D_v\gg 1}$ to obtain explicit analytic approximations for the growth rates in these two limiting regimes [cf.\ inset in Fig.~\ref{fig:numerics-eigenvalueMapping-Brusselator}(a)]. 
Afterward, we discuss how the growth rates can be determined throughout the crossover from mesa toward peak patterns by numerically solving the stationary pattern profile.

\textit{Mesa-forming regime.\;---}
For the mesa-forming regime, the calculation of the stationary interface pattern on the infinite domain and the asymptotic matching to the pattern on the finite domain was already performed in the supplementary material of Ref.~\cite{Brauns.etal2021}.
This gives the change of the stationary mass-redistribution potential with the mesa mass as
\begin{equation*}
    \partial_M^\pm\etastat 
    = 
    - \frac{6 \ell}{(1-d) D_v}\exp\left(-\frac{\xi_\pm\Lambda}{\ell}\right)
    ,
\end{equation*}
where ${\ell_\pm = \ell = \sqrt{D_u}}$.
Because the interfacial profile on the infinite line is given---as for the cubic model---by an appropriately scaled hyperbolic tangent function (see Ref.~\cite{Brauns.etal2021}), the interface width is again given by
\begin{equation*}
    \ell_\mathrm{int} 
    = 
    6 \, \ell
    \, .
\end{equation*}
Moreover, using the definition of the interface average, Eq.~\eqref{eq:interface-average}, the average conversion rate is found as
\begin{equation*}
    \langle\tilde{f}_\eta\rangle_\mathrm{int} 
    = 
    \frac{2}{3 d}
    \, .
\end{equation*}
Using the stationary profile and inserting the above terms into the general growth-rate expressions, Eqs.~\eqref{eq:sigma-MC-mesa-D-R},~\eqref{eq:sigma-stab-mesas}, one finds for the mesa-forming regime of the Brusselator model introduced in Sec.~\ref{sec:Brusselator}
\begin{align*}
    \sigma_\mathrm{D}^\pm 
    &= \frac{12 \ell}{(1-d)\xi_\mp \Lambda}\exp\left(-\frac{\xi_\pm\Lambda}{\ell}\right)
    , \\
    \sigma_\mathrm{R}^\pm 
    &= \frac{24}{1-d^2}\exp\left(-\frac{\xi_\pm\Lambda}{\ell}\right)
    , \\
    \sigma_\mathrm{S}^\pm 
    &= \frac12 \left|p-\frac{1}{\sqrt{2 d}}\mp \frac{1}{\sqrt{2 d}}\right|
    .
\end{align*}

\textit{Peak-forming regime.\;---}
As $D_v$ is increased, the upper plateau density ${\rho^+ = 3 \sqrt{d/2}+ (1-d)\sqrt{2/d}}$ increases and a transition into peak patterns occurs (if the average density $\bar{\rho}$ is fixed).
The peak height is then limited by the total mass in the system and its profile approaches an asymptotic form as $\rho_+$ moves to much higher densities, that is, as ${D_v\to\infty}$.
The asymptotic peak profile $\rho_\mathrm{peak}(x)$ [${u = u_\mathrm{peak}(x)}$ and ${\eta = \eta_\mathrm{peak}=v_\mathrm{peak}}$] can be calculated from the profile equation, Eq.~\eqref{eq:profile-equation}, in the limit ${D_v\to\infty}$, which reads
\begin{equation*}
    0 = D_u \partial_x^2 u_\mathrm{peak} +(u_\mathrm{peak})^2 \eta_\mathrm{peak} - u_\mathrm{peak}
    \, .
\end{equation*}
It is solved on the infinite line by (cf.\ the second model in Ref.~\cite{Otsuji.etal2007})
\begin{equation}\label{eq:Brusselator-peak-profile}
    u_\mathrm{peak}(x) = \frac{M}{4\ell} \operatorname{sech}^2\left(\frac{x}{2\ell}\right),
\end{equation}
and the stationary mass-redistribution potential $\eta_\mathrm{peak}$ 
\begin{equation*}
    \eta_\mathrm{peak} 
    = 
    \frac{6\ell}{M}
    \, .
\end{equation*}
The peak mass is denoted by $M$.
Using this stationary solution and the definition of the interface width and average, Eqs.~\eqref{eq:interface-width},~\eqref{eq:interface-average}, one finds the interface (half-peak) width ${\ell_\mathrm{int}^\mathrm{peak} = 3 \ell}$ and the average conversion rate
\begin{equation*}
    \langle \tilde{f}_\eta\rangle_\mathrm{int} 
    = 
    \frac{M^2}{30 \ell^2}
    \, .
\end{equation*}
Collecting the above results and using the general growth-rate expressions, Eqs.~\eqref{eq:sigma-MC-peak-D-R},~\eqref{eq:sigma-stab-peak}, one has in the peak-forming regime of the Brusselator model introduced in Sec.~\ref{sec:Brusselator}
\begin{equation*}
    \sigma_\mathrm{D} 
    = \frac{24 \ell D_v}{\Lambda^3 p^2} 
    \, , \quad
    \sigma_\mathrm{R} 
    = \frac{6}{5}
    \, , \quad
    \sigma_\mathrm{S} 
    = 1
    \, .
\end{equation*}
To calculate the full growth rate $\sigma^\varepsilon$ from these rates, one also needs the shift of the stationary mass-redistribution potential for the case ${(s_1,s_2) = (p-u, 0)}$.
With the stationary peak profile, Eq.~\eqref{eq:Brusselator-peak-profile}, the shift given by Eq.~\eqref{eq:nMC-etastat-shift} follows as
\begin{equation*}
    \delta\eta_\mathrm{stat}^\varepsilon \approx 
    -\frac{p - \frac{M}{6 \ell}}{\frac{M^2}{30 \ell^2}} 
    \approx 
    \frac{5\ell}{M}
    \, ,
\end{equation*}
where the change in the density of the lower plateau was neglected, and the second approximation holds for large peaks (${M/\ell\gg p}$).
The last term ${\delta\eta_\mathrm{stat}^\varepsilon \approx 5\ell/M}$ is used in Fig.~\ref{fig:numerics-eigenvalueMapping-Brusselator}(b) to calculate the dashed, black line.

\textit{Mass-competition instability at the crossover from mesa to peak patterns.\;---}
Next to these analytic approximations in the mesa- and peak-forming regimes, we can numerically determine the growth rate $\sigma^\varepsilon$ throughout the crossover from mesa to peak patterns.
For this, we need to find an expression for $\sigma^\varepsilon$ that interpolates in the crossover region in which the pattern is neither purely peak- nor mesa-like.
For the considered scenario of mesa/peak competition, we guessed the following expression based on the basic physics underlying the mass-competition instability:
\begin{align}
\label{eq:sigma-eps-crossover}
    \sigma^\varepsilon = \frac{\sigma_\mathrm{R}^+}{\sigma_\mathrm{R}^+ + \sigma_\mathrm{D}^+}{\Bigg[}&- \frac{4 D_v}{\xi_-\Lambda}\left(\partial_{M}^+ \etastat +\varepsilon\,\partial_{M}^{}  \delta\eta_\mathrm{stat}^\varepsilon\right) \nonumber\\
    &+\varepsilon \left(\langle \partial_\rho s_\mathrm{tot}\rangle_\mathrm{int} + \frac{s_\mathrm{tot}^-}{\Delta\rho}\right){\Bigg]}
    \, .
\end{align}
Let us analyze this expression in detail.
To this end, we first focus on the first term describing the mass exchange between the peaks.
Within the sharp-interface approximation, one has ${\partial_{M}^{}  \delta\eta_\mathrm{stat}^\varepsilon\approx 0}$ in the mesa-forming regime as discussed in Sec.~\ref{sec:growth-rate-nMC} [see Eq.~\eqref{eq:nMC-etastat-shift}].
Hence, we recover the correct competition term in the mesa-forming regime [cf.\ Sec.~\ref{sec:growth-rate-nMC}]. 
In the peak-forming regime, one finds ${\partial_{M}^+ \approx\partial_{M}^{}}$, and also $[\sigma_\mathrm{D,R}^+\approx\sigma_\mathrm{{D,R}}]$, by neglecting terms due to the pattern tails in the low-density plateau.
These correction terms are exponentially small in the plateau length and do not contribute significantly for peak patterns because the change of the peak height induces much stronger changes of $\etastat$ (see Appendix~\ref{app:peak-scaling}).
Furthermore, in the peak-forming regime we have ${\xi_-\approx 1}$ (up to corrections of the order of ${\ell_\mathrm{int}/\Lambda\ll 1}$.
Thus, the chosen expression, Eq.~\eqref{eq:sigma-eps-crossover}, also correctly describes the peak-forming regime.
In the crossover region, the use of $\partial_M^+\etastat$ and $\sigma_\mathrm{D,R}^+$ ensures that no contributions from the pattern tails in the inner, merely translated, trough are wrongly accounted for in the mass-competition rate.
Introducing $\xi_-$ accounts for the reduced distance---and thus the increased gradient---between the mesa/peak interfaces due to the finite mesa/peak width.

Second, also the source contribution, the second term in Eq.~\eqref{eq:sigma-eps-crossover}, has to be modified to describe the crossover from mesa to peak patterns.
Again, one has to ensure that no contributions from the inner, low-density plateau are included.
As the trough is only shifted during the mass-competition process, total production in the low-density plateau remains constant.
To honor the shifting of the trough, let $-\delta x$ describe the shift of the interface position if the small amount of mass $\delta M$ is transferred from the left to the right (half-)mesa or (half-)peak, i.e., $\delta x \approx \delta M / (\Delta\rho)$ for mesa patterns. 
The source terms then induce additional degradation at the right mesa (peak) leading to a relaxation of the mass difference by [cf.\ Eq.~(63) in the supplementary material of Ref.~\cite{Brauns.etal2021} and use Eqs.~\eqref{eq:source-derivation},~\eqref{eq:source-prod-heuristic}]
\begin{align*}
    \partial_t^\mathrm{(source)}\delta M &= -\varepsilon \sigma_\mathrm{S}\delta M\\
    &=\varepsilon \int_{-\delta x}^{\frac{\Lambda}{2}}\mathrm{d}x\, s_\mathrm{tot}\\
    &\approx \varepsilon \left(\frac{1}{2}\langle\partial_\rho s_\mathrm{tot}\rangle_\mathrm{int} + s_\mathrm{tot}^- \partial_M^{}\delta x\right) 2 \delta M \, .
\end{align*}
The translation of the lower integration boundary accounts for the translation of the inner plateau and keeps the length of the lower plateau within the integration interval constant.
For mesa patterns we have ${\partial_M^{}\delta x = 1/(2 \Delta\rho)}$ and ${\varepsilon\langle\partial_\rho s_\mathrm{tot}\rangle_\mathrm{int} =(s_\mathrm{tot}^+-s_\mathrm{tot}^-)/\Delta\rho}$ within the sharp-interface approximation (see Sec.~\ref{sec:growth-rate-nMC-discussion}).
Therefore, we recover ${\sigma_\mathrm{S} = \left| s_\mathrm{tot}^+\right|/\Delta\rho}$ in the mesa-forming regime [cf.\ Eq.~\eqref{eq:sigma-stab-mesas}].
In fact, $\partial_M^{}\delta x \approx 1/(2 \Delta\rho)$ is a good estimate as long as the pattern amplitude does not change strongly with the mesa/peak mass $M$.
In the peak limit where this is no longer the case, ${\partial_M^{}\delta x \lesssim 1/[2(\hat{\rho}-\rho_-)]}$ is an upper bound because the peak height increases with the peak mass as well, letting the peak grow less strongly in width than ${\sim 1/(\hat{\rho}-\rho_-)}$.
Since the peak height is large in the sharp-peak limit [${(\hat{\rho}-\rho_-)\sim (\bar{\rho}-\rho_-)\Lambda/\ell_\mathrm{int}}$ due to the definition of the peak mass, Eq.~\eqref{eq:mesa-peak-mass}], it follows that the additional term $s_\mathrm{tot}^-\partial_M^{}\delta x$ is negligible in comparison to $\langle\partial_\rho s_\mathrm{tot}\rangle_\mathrm{int}$ in the peak-forming regime, and one recovers the correct rate $\sigma_\mathrm{S}$ for peak patterns, Eq.~\eqref{eq:source-prod-heuristic}.
Moreover, one can use the approximation $\partial_M^{}\delta x \approx 1/(2 \Delta\rho)$ to express the corresponding term in Eq.~\eqref{eq:sigma-eps-crossover}.
Taken together, Equation~\eqref{eq:sigma-eps-crossover} correctly accounts for the basic physics of mass competition under the influence of weak source terms in the crossover regime between peak and mesa patterns.

For the analysis in Fig.~\ref{fig:numerics-eigenvalueMapping-Brusselator}, we apply Eq.~\eqref{eq:sigma-eps-crossover}.
For this, the phase-space construction yields ${s_\mathrm{tot}^-/\Delta\rho = p \sqrt{d/2}/(1-d)}$ and the simple choice of the source terms gives ${\langle\partial_\rho s_\mathrm{tot}\rangle_\mathrm{int}=-1/(1-d)}$.
In contrast, one has to numerically calculate $\partial_M^+ \etastat$ in the crossover region.
To this end recall that it holds ${\partial_M^{}\etastat = \partial_M^+ \etastat + \partial_M^- \etastat}$ where for plateaus $\partial_M^\pm \etastat$ are exponentially suppressed in the relative plateau length $2 L_\pm/\ell = \xi_\pm \Lambda/\ell$ [see Eqs.~\eqref{eq:mesa-delMeta-tot}].
Given the length of the upper plateau $L_+$ for which we want to determine $\partial_M^+ \etastat$, the value of $\partial_M^{} \etastat$ is a good estimate for $\partial_M^+ \etastat$ if ${L_-\gg L_+}$.
Then, one can use the value of $\partial_M^{} \etastat$ to approximate $\partial_M^+ \etastat$.
In addition, if the plateau lengths fulfill ${L_-\lesssim L_+}$ on the simulation domain, one can construct the elementary stationary pattern on an enlarged domain of length $\tilde{\Lambda}/2$ such that, for example, ${\xi_-\geq 0.6}$ while keeping $L_+$ fixed (the precise threshold value of $\xi_-$ is arbitrary).
Within the sharp-interface approximation, the length of the upper plateau is kept constant by changing the average density $\bar{\rho}\to \tilde{\rho}$ on the enlarged domain by
\begin{equation*}
   \tilde{\rho}-\rho_- 
   = 
   \frac{2 L_+(\rho_+-\rho_-)}{\tilde{\Lambda}}
   = 
   (\bar{\rho}-\rho_-)\frac{\Lambda }{\tilde{\Lambda}}
   \, .
\end{equation*}
The value of $\partial_M^{} \etastat$ calculated on the enlarged domain is again a good estimate for $\partial_M^+ \etastat$ at the given half-mesa length $L_+$ because $\partial_M^+ \etastat$ only depends on the upper plateau.

\subsection{The constant-reaction-rate peak model}
\label{app:const-reac-rate-peak-model}
As the third model system, we introduce a system that always forms peaks in the mass-conserving case (see Sec.~\ref{sec:const-reac-rate-peak-model}).
The peak profile is independent of the fast diffusion constant $D_v$ but the profile is not known analytically.
The quantities appearing in $\sigma^\varepsilon$ are determined from the numerically determined stationary peak profile.

\textit{Approximation including pattern deformation by the source term $s_1$.\;---}
Moreover, for the analysis of the source term $s_1$ in Sec.~\ref{sec:const-reac-rate-peak-model} we again used the improved approximation obtained by the reaction-term modification ${\tilde{f}\to\tilde{f}'=\tilde{f}+ \varepsilon s_1}$ [see Fig.~\ref{fig:numerics-eigenvalueMapping-constReacRateModel}(b)]:
\begin{equation*}
    \tilde{f}' 
    = 
    \eta - a\frac{\rho}{1+\rho^2} + \varepsilon(p-\rho)
    \, .
\end{equation*}
Consequently, the nullcline attains a (highly asymmetric) $\mathsf{N}$-shape for finite source strengths $\varepsilon$.
This explains the observed transition from peak into mesa patterns [see inset in Fig.~\ref{fig:numerics-eigenvalueMapping-constReacRateModel}(b)].
The profile equation for the stationary profile (of the modified mass-conserving system) turns into
\begin{equation*}
    0 
    = 
    D_u \partial_x^2\rhostat' +\eta_\mathrm{stat}' - a\frac{\rhostat'}{1+\rhostat'^2} + \varepsilon(p-\rhostat')
    \, ,
\end{equation*}
which has to be solved numerically for all values of $\varepsilon$.
The properties of the stationary pattern entering $\sigma^\varepsilon$ are then calculated from the stationary patterns for the corresponding value of $\varepsilon$.

\textit{Plateau splitting.\;---}
As for the cubic model, we observe plateau splitting in the simulation of this model [see Fig.~\ref{fig:numerics-eigenvalueMapping-constReacRateModel}(a,b)].
Again, we use the approximation described in the supplementary material of Ref.~\cite{Brauns.etal2021}, to find the threshold of plateau splitting as (calculation shown for the cubic model in Appendix~\ref{app:cubic-model})
\begin{equation}\label{eq:constReacRateModel-splitting}
    \varepsilon_\mathrm{split} 
    \approx 
    \frac{4 D_v a}{p\Lambda^2}
    \, ,
\end{equation}
where we neglected $\eta^*(\rho_-)$ as well as $\rho_-$ because $\rho_-\approx 0$ is small for large peaks.
In Fig.~\ref{fig:numerics-eigenvalueMapping-constReacRateModel}, this threshold is depicted as a purple (top-most, diagonal) line.

\bibliography{literature}

\end{document}